\documentclass[floats,floatfix,showpacs,amssymb,prd,twocolumn,superscriptaddress,nofootinbib,nolongbibliography,reprint]{revtex4-2}

\usepackage{amssymb,amsmath,verbatim,mathtools,needspace,enumitem,etoolbox,graphicx,physics,microtype,afterpage,bigints,gensymb,tabularx,mathtools,siunitx, scalerel}
\usepackage[dvipsnames, usenames]{xcolor}
\definecolor{linkcolor}{rgb}{0.0,0.3,0.5}
\definecolor{dodgerblue}{HTML}{1E90FF}
\usepackage{hyperref}
\hypersetup{unicode, colorlinks=true, linkcolor=linkcolor, citecolor=linkcolor, filecolor=linkcolor,urlcolor=linkcolor, pdfusetitle}
\usepackage[all]{hypcap}
\usepackage[T1]{fontenc}
\usepackage[utf8]{inputenc}
\usepackage{orcidlink}
\usepackage [english]{babel}
\usepackage [autostyle, english = american]{csquotes}
\MakeOuterQuote{"}
\usepackage{soul}
\usepackage{bbm}
\usepackage{siunitx}
\usepackage{booktabs}

\newcommand{\Msol}{\rm \,M_{\odot}}

\newcommand{\umani}{\affiliation{Department of Physics and Astronomy \& Winnipeg Institute for Theoretical Physics, University of Manitoba, Winnipeg, Manitoba, R3T 2N2, Canada}}

\newcommand{\cmu}{\affiliation{McWilliams Center for Cosmology and Astrophysics, Department of Physics, Carnegie Mellon University, Pittsburgh, PA 15213, USA}}

\newcommand{\utrgv}{\affiliation{Department of Physics and Astronomy, University of Texas Rio Grande Valley, Brownsville, TX 78520, USA}
\affiliation{South Texas Space Science Institute, University of Texas Rio Grande Valley, Brownsville, TX 78520, USA}}

\newcommand{\ttu}{\affiliation{Department of Physics \& Astronomy, Texas Tech University, Box 41051, Lubbock, TX, 79409-1051, USA}}

\newcommand{\cca}{\affiliation{Center for Computational Astrophysics, Flatiron Institute, 162 Fifth Ave, New York, NY, 10010, USA}}

\begin{document}

\title{A new model for the continuum spectra of AM CVn binaries and \\ multi-messenger inference with normalizing flows} 

\author{Nathan Steinle$\,$\orcidlink{0000-0003-0658-402X}}
\email{nathan.steinle@umanitoba.ca}
\umani

\author{Samar Safi-Harb$\,$\orcidlink{0000-0001-6189-7665}}
\umani

\author{Austin MacMaster$\,$\orcidlink{0009-0002-3696-7339}}
\umani

\author{Liliana Rivera Sandoval$\,$\orcidlink{0000-0002-9396-7215}}
\utrgv

\author{Katelyn Breivik$\,$\orcidlink{0000-0001-5228-6598}}
\cmu

\author{Tyrone E. Woods$\,$\orcidlink{0000-0003-1428-5775}}
\umani

\author{Tom Maccarone$\,$\orcidlink{0000-0003-0976-4755}}
\ttu

\author{Tom Wagg$\,$\orcidlink{0000-0001-6147-5761}}
\cca

\begin{abstract} 
Future electromagnetic telescopes, such as \textit{NewAthena}, \textit{CASTOR}, and an \textit{AXIS}-like mission, along with milli-Hz gravitational-wave (GW) detectors such as \textit{LISA}, are expected to unearth the population of Galactic ultra-compact binaries (UCBs). Joint multi-messenger detections will probe the uncertain formation, evolution, and observables of mass-transferring UCBs such as AM CVns, but theoretical tools need to be advanced to anticipate future data challenges. Motivated by this, we present a new forward model for the continuum emission of AM CVn binaries that connects source binary parameters to X-ray, optical, and ultraviolet observables. The model assumes GW-driven mass transfer with physically motivated prescriptions for accretion energetics, emission geometry, absorption, and instrumental response. 
Combining this with \textit{LISA} observations and the output of binary population synthesis enables exploration of the multi-messenger properties of AM CVns. 
Although uncertain, our model predicts that approximately one per $7000$ AM CVn binaries will permit a joint multi-messenger detection with \textit{LISA}, \textit{CASTOR}, and \textit{AXIS}. 
We also develop a framework for inferring binary parameters from the inverse model with a convolutional neural net and normalizing flows. 
Testing the trained flow with our synthetic AM CVn population, we find mean absolute fractional error on the inferred accretor mass of $0.05 \Msol$, donor mass of $0.26 \Msol$, orbital period of $0.1$ s, and distance of $0.2$ pc, while Spearman's rank shows strongly correlated true and predicted distributions except for the donor mass. 
These efforts lay a foundation for follow-up studies that will explore detailed binary astrophysics and observational requirements for effective multi-messenger scientific discovery in the coming decade. 
\end{abstract}

\maketitle

\section{Introduction}
\label{Sec:Intro}

The forthcoming era of space-based gravitational-wave (GW) astronomy, led by the Laser Interferometer Space Antenna (\textit{LISA}), will enable the systematic study of compact stellar-mass binaries in the Milky Way \cite{2017arXiv170200786A,2023LRR....26....2A}. Operating in the millihertz GW band, \textit{LISA} is uniquely sensitive to ultra-compact binaries (UCBs) with orbital periods shorter than approximately one hour \cite{2017MNRAS.470.1894K,2018MNRAS.480..302K,2013ASPC..467...27N, 2026arXiv260313813W}. These systems are expected to dominate the astrophysical foreground in the \textit{LISA} band while simultaneously offering a rich arena for multi-messenger astrophysics, binary evolution, and compact-object physics under extreme conditions through their electromagnetic (EM) emission \cite{2017MNRAS.470.1894K,2018MNRAS.480..302K}.

Galactic UCBs have been observed electromagnetically for decades across the X-ray, ultraviolet (UV), optical, and infrared bands \cite{2005ASPC..330...27N,2010PASP..122.1133S,2005A&A...440..675R}. The principal classes include AM CVn systems composed of a white dwarf (WD) accretor and a helium-rich WD donor; ultra-compact X-ray binaries (UCXBs) containing a neutron star or black hole accretor; detached and interacting double-WD (DWD) systems; and hot subdwarfs with compact companions (sdBs) \cite{2010PASP..122.1133S,2018MNRAS.480..302K}. Their EM properties vary widely with orbital period, mass-transfer rate, and accretion geometry. In AM CVn systems, short-period binaries in the direct-impact regime typically exhibit hard X-ray and extreme-UV emission with luminosities of $\gtrsim10^{31}$--$10^{33}\,\mathrm{erg\,s^{-1}}$, while longer-period disk-accreting systems are dominated by UV emission with often comparatively weaker X-ray flux \cite{2002MNRAS.331L...7M,2005ASPC..330...27N,2010PASP..122.1133S,2005A&A...440..675R,2023JAVSO..51..227B}. UCXBs are generally brighter X-ray sources, reaching luminosities  $\gtrsim 10^{36}\,\mathrm{erg\,s^{-1}}$ in persistent systems, whereas detached DWDs are usually electromagnetically faint or undetectable \cite{2013ApJ...768..184H,2006csxs.book..623T}.

The scientific return from \textit{LISA} will be substantially enhanced by contemporaneous observations with next-generation EM facilities \cite{2019BAAS...51g.107M}. In the X-ray band, a telescope similar to the Advanced X-ray Imaging Satellite (\textit{AXIS}) \cite{2019BAAS...51g.107M} would combine sub-arcsecond angular resolution with high sensitivity and adequate timing and target-of-opportunity response, making it particularly effective for identifying faint UCBs in crowded Galactic environments. For example, its Galactic Plane Survey would reach flux limits of $\sim10^{-16}\,\mathrm{erg\,cm^{-2}\,s^{-1}}$, enabling detections of accreting DWDs out to several kiloparsecs.
The planned future Advanced Telescope for High Energy Astrophysics (\textit{NewAthena}) \cite{2013arXiv1306.2307N,2025NatAs...9...36C} will provide high-resolution spectroscopy. Missions such as the Cosmological Advanced Survey Telescope for Optical and UV Research (\textit{CASTOR}) \cite{2025JATIS..11d2202C} would probe thermal emission from accretion disks and heated WD photospheres.

Multi-wavelength EM observations have historically been important for revealing the underlying physical processes governing AM CVn binaries \cite{2005ASPC..330...27N,2010PASP..122.1133S,2005A&A...440..675R,2005A&A...440..675R,2010PASP..122.1133S}.
This provides strong motivation for combining such data with GW observations to extract the most information possible for probing uncertain astrophysical processes.  
A key subset of \textit{LISA} sources are so-called verification binaries (VBs): compact binaries already known from EM observations whose GW signals are guaranteed to be detected by \textit{LISA} \cite{2018MNRAS.480..302K,2024ApJ...963..100K}. The current census of candidate VBs is dominated by short-period AM CVn systems and detached DWDs identified through optical, UV, and X-ray surveys \cite{2018MNRAS.480..302K}. Population synthesis models predict that several dozen such systems will be detected with high signal-to-noise by \textit{LISA}, with many more individually resolvable sources emerging over the mission lifetime \cite{2017MNRAS.470.1894K,2017ApJ...846...95K,2013ASPC..467...27N}. These binaries will play a critical role in instrument validation and in characterizing the Galactic DWD foreground that sets a noise floor near the peak of \textit{LISA} sensitivity \cite{2017MNRAS.470.1894K,2019MNRAS.482.5222L}.

From a multi-messenger perspective, Galactic AM CVn systems can be divided into three observational regimes. The first consists of the intrinsic Galactic population distributed throughout the Milky Way, with characteristic distances on the order of the Galactic solar radius ($\sim8$~kpc). These systems trace stellar mass and binary evolution but are largely accessible only through EM observations and their cumulative GW signal \cite{2004MNRAS.349..181N,2017MNRAS.470.1894K,2018ApJ...854L...1B,2019MNRAS.482.5222L}. The second regime includes systems detectable jointly by \textit{LISA} and future EM facilities, typically within $\sim1$--$10$~kpc, and is strongly shaped by GW and EM selection effects \cite{2018MNRAS.480..302K}. The third regime comprises nearby systems within $\lesssim1$~kpc that have already been identified electromagnetically, including most current verification binary candidates. Understanding how these regimes relate to one another is essential for interpreting future joint EM--GW datasets.

While expanding the sample of verification binaries is an important goal, the broader astrophysical significance of AM CVn systems lies in their ability to probe fundamental uncertainties in compact binary evolution \cite{2001A&A...375..890N,2010PASP..122.1133S}. These systems test models of common-envelope evolution, angular momentum loss, donor star structure, and mass-transfer stability under extreme conditions \cite{2001A&A...375..890N,2007MNRAS.382..685R,2018MNRAS.480..302K}. In contrast to detached DWDs, whose GW signals can often be modeled with relatively few assumptions, mass-transferring systems pose a greater challenge: both their EM and GW observables depend sensitively on the mass-transfer rate, which itself is governed by poorly constrained physics \cite{2001A&A...375..890N,2006csxs.book..623T,2010PASP..122.1133S}. Consequently, much of the existing \textit{LISA} literature has focused on detached binaries, for which population modeling and GW predictions are simpler \cite{2017MNRAS.470.1894K,2019MNRAS.482.5222L}.

The scientific opportunities enabled by joint EM--GW observations of AM CVn systems are substantial. Combined modeling can break degeneracies inherent in EM-only analyses, such as those between inclination, distance, and accretion luminosity \cite{2018MNRAS.480..302K}. It enables direct tests of mass-transfer prescriptions and accretion geometries, facilitates classification of UCBs by their combined signatures, and improves estimates of the contribution of interacting binaries to the Galactic GW foreground \cite{2017MNRAS.470.1894K,2019MNRAS.482.5222L}. Realizing this potential requires forward models that are both physically grounded and computationally efficient.

\begin{figure*}
\centering
\includegraphics[width=\textwidth]{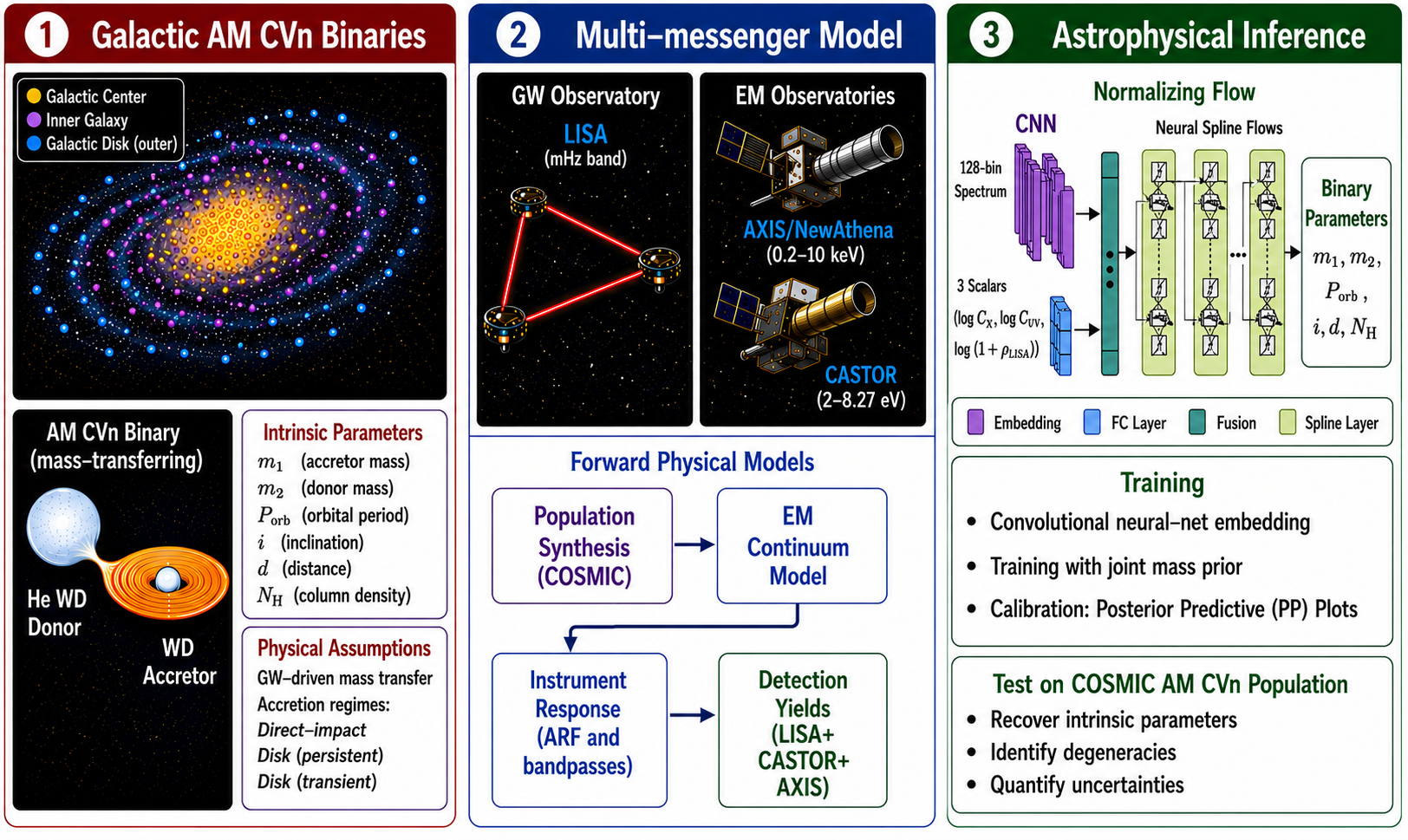}
\caption{
Overview of the multi-messenger AM CVn framework, illustrating the connection between Galactic AM CVn populations, forward physical modeling, and simulation-based Bayesian inference of individual source parameters.
\textit{Stage~1 (Galactic AM CVn Binaries):} The left panel shows schematics of Galactic AM CVn systems and of the mass-transferring binary. The EM continuum model is parameterized by the intrinsic and extrinsic binary properties $(m_1,\,m_2,\,P_{\rm orb},\,i,\,d,\,N_{\rm H})$, assumes GW-driven mass transfer, and determines the accretion state as direct-impact, persistent-disk, or transient-disk regimes. 
\textit{Stage~2 (Multi-messenger Model):} The middle panel summarizes the multi-messenger pipeline. AM CVn populations generated with \texttt{COSMIC} are evolved and passed through an analytic continuum-emission model; the resulting spectra are folded through EM instrument responses of the future observatories \textit{AXIS}/\textit{NewAthena} in X-rays and \textit{CASTOR} in optical/UV. GW detection is modeled for \textit{LISA} with \texttt{LEGWORK}. The observables (spectrum, count rates, and GW signal-to-noise ratio) are used to estimate joint EM--GW detection yields. 
\textit{Stage~3 (Astrophysical Inference):} The right panel depicts the simulation-based inference framework, where the multi-messenger observation vector, consisting of a 128-bin continuum spectrum together with X-ray and UV count rates and the \textit{LISA} signal-to-noise ratio, is compressed by a convolutional neural-net embedding architecture and transformed into a latent representation which conditions a neural spline flow that learns the inverse model from training data generated by the forward model, enabling efficient inference of the AM CVn parameters. The framework is calibrated using posterior predictive diagnostics and independently applied to a \texttt{COSMIC} population to test parameter recovery and evaluate the information content of future joint EM--GW observations. 
}
 \label{F:ProjectOverview}
\end{figure*}

The primary aim of this study is to develop such a forward model for AM CVn binaries and apply it to multi-messenger analysis, as shown in the overview of Figure~\ref{F:ProjectOverview}. We construct an analytic framework that maps intrinsic and extrinsic binary parameters---component masses, orbital period, inclination, distance, sky location, and interstellar absorption---to observable EM spectra in the X-ray, optical, and UV bands, with explicit applicability to joint analyses with \textit{LISA}. The model assumes GW--driven mass transfer and self-consistently assigns each system to one of three accretion regimes depending on the orbital period: direct-impact accretion, persistent disk accretion, or transient disk accretion.

For a given binary configuration, the EM spectrum is built as a sum of continuum components associated with physically distinct emission regions, including accretion disks, boundary layers or spreading layers on the accretor surface, and localized hot spots in the direct-impact regime \cite{2005ASPC..330...27N,2010PASP..122.1133S,2005A&A...440..675R,2005A&A...440..675R,2010PASP..122.1133S}. Luminosities are normalized by enforcing energy conservation relative to the accretion power, with inclination-dependent geometric effects treated consistently. Interstellar absorption and instrumental responses for EM missions are incorporated to yield observable spectra. This model provides an instantaneous snapshot of the AM CVn spectrum, and line emission, winds, and explicit time dependence are left for future extensions. In the X-ray band, we fold model spectra through realistic instrumental response using the energy-dependent effective area file (ARF) of the \textit{AXIS} detector to predict photon count rates, and we leave a more detailed study of \textit{NewAthena} to forthcoming work. 
In the UV, u, and g bands, we model \textit{CASTOR} observations by integrating intrinsic spectral energy distributions over instrument-specific bandpasses\cite{2024AJ....167..178C}, accounting for WD heating and disk emission components that dominate the optical/UV output of AM CVns \cite{2005A&A...440..675R}. 
This approach enables rapid evaluation of individual systems, making it suitable for survey forecasting with population synthesis and Bayesian inference. By providing a physically motivated mapping between binary parameters and EM observables, the model offers a flexible tool for interpreting future multi-messenger observations of Galactic UCBs. 

We test and validate the model in two complementary ways. First, we explore the intrinsic binary parameter space, varying component masses, orbital period, inclination, distance, and column density to see how spectral shapes, luminosities, and observational regimes change across physically relevant domains. This allows us to identify dominant trends, regime transitions, and parameter degeneracies inherent to the forward model. 
Second, we apply the model to a population of AM CVns drawn from the \texttt{COSMIC} population synthesis framework \cite{2017ApJS..230...15B}, assigning Galactic positions using an analytic Milky Way stellar density model and computing line-of-sight column densities using the \textit{Bayestar~2019} three-dimensional dust maps \cite{2019ApJ...887...93G,2018JOSS....3..695G}.
The multi-messenger framework shows that systems jointly detectable by \textit{LISA}, \textit{CASTOR}, and \textit{AXIS} are dominated by short-period, high mass-transfer-rate binaries with strong optical, UV and X-ray continuum emission and large GW signal-to-noise ratios. Although the intrinsic Galactic AM CVn population extends to long orbital periods and low accretion luminosities, only a small subset remains observable in both EM and GW observations. 
Our estimated detection yields suggest that approximately 1 out of every 7000 Galactic AM CVns can be observed jointly by \textit{LISA}, \textit{CASTOR}, and \textit{AXIS}, though this depends on important astrophysical and progenitor-population uncertainties, full exploration of which we leave to a future study. 
Our results are consistent with the expectation that the primary limitation on the joint multi-messenger sample is not EM sensitivity, but rather the comparatively restrictive subset of systems individually detectable by \textit{LISA}. 

Further, we apply conditional normalizing flows using the \texttt{SBI} framework \cite{2021JOSS....6.2505T} to perform likelihood-free Bayesian inference on the combined EM and GW observables. Training on arbitrary binaries generated directly from the forward model shows that the learned posterior distributions remain well calibrated while accurately recovering the dominant structure of the underlying parameter distributions. Testing on independent \texttt{COSMIC} populations, we find that the orbital period, inclination, distance, and absorption column density are inferred robustly from the combined multi-messenger observables, while the accretor mass is recovered with comparatively small systematic bias despite residual degeneracies associated with the accretion physics. In contrast, the donor mass remains substantially degenerate, reflecting overlap in the observable signatures of physically distinct systems. Restricting the population to shorter orbital periods further improves inference performance due to the stronger EM emission and larger \textit{LISA} signal-to-noise ratios of these systems. These results show that Bayesian inference combined with forward modeling provides a promising pathway for extracting astrophysical information from future multi-messenger populations of interacting UCBs. 

This paper is organized as follows. In Sections~\ref{Sec:ForwardEM}, \ref{Sec:Cosmic}, \ref{Sec:LISA}, \ref{Sec:GalacticLocation} and \ref{Sec:NormalFlows} we describe the new physical assumptions and formulation of the EM continuum model, the generation of the synthetic AM CVn population with \texttt{COSMIC}, the model for \textit{LISA} detections, the model for the AM CVn Galactic locations, and the new framework utilizing normalizing flows for binary parameter inference, respectively. 
In Section~\ref{Sec:Results} we present the results, and in Section~\ref{Sec:Conclusions} we summarize our conclusions and discuss implications for multi-messenger studies.

\section{Model for the continuum spectrum of an AM CVn binary}
\label{Sec:ForwardEM}

This section describes the construction of the forward EM model used in this study. The model provides a physically grounded yet computationally efficient mapping between the intrinsic parameters of AM CVn binaries and their observable EM spectra in the X-ray and ultraviolet bands, with explicit applicability to joint analyses with GW observations by \textit{LISA}. The approach is deliberately analytic: the orbital evolution and mass-transfer rate are determined self-consistently from GW angular momentum losses, and the resulting accretion power is partitioned among physically motivated emission components whose spectral forms admit closed-form or efficiently normalized expressions. This design choice ensures that evaluating the model remains computationally efficient, enabling systematic parameter-space exploration and population-level applications while retaining direct interpretability of how individual physical assumptions influence the predicted observables.

Early AM CVn population studies employed prescriptions for the EM emission, typically modeling the optical/UV and X-ray bands using approximate blackbody components, average disk temperatures, or phenomenological luminosity scalings in order to estimate detectability across surveys and in combination with GW observations \cite{2004MNRAS.349..181N,2007ApJ...655.1010G}. In comparison, the present framework constructs the continuum spectrum explicitly as a superposition of physically distinct emission components, including a multi-temperature accretion disk, optically thick and optically thin boundary-layer emission, direct-impact hotspot emission, and UV WD heating. Each component is normalized such that its bolometric luminosity reproduces the accretion energetics self-consistently before geometric projection and absorption effects are applied.

The observational studies of \citet{2014ApJ...785..157S} emphasized the empirical connection between orbital period, accretion state, and long-term photometric behavior, distinguishing persistent high-state systems from transient outbursting systems primarily through observed variability properties. Similarly, the spectral analyses and atmosphere/disk modeling discussed by \citet{2015ApJ...806...76K} focused on fitting observed spectra of individual AM CVn systems using helium-rich atmosphere or accretion-disk models. In contrast, the framework developed here is not intended as a phenomenological fitting model for individual sources, but rather as a predictive forward model in which the UV, optical, and X-ray continuum spectra follow directly from the binary parameters, accretion energetics, and regime-dependent emissivity assumptions. In particular, the present model treats direct-impact accretion, persistent disk accretion, and transient disk accretion using distinct luminosity partitions and emission geometries tied explicitly to the mass-transfer rate and accretion state.

\subsection{AM CVn binary model}

AM CVn binaries consist of a WD accretor of mass $M_1$ and a hydrogen-deficient donor of mass $M_2$, transferring mass via Roche-lobe overflow. The fully degenerate helium donor is assumed to be helium-rich and obeys a zero-temperature mass--radius relation,
\begin{equation}\label{Eq:DonorRadius}
R_2 \propto M_2^{-1/3},
\end{equation}
appropriate for cold, degenerate helium WDs \cite{1969ApJ...158..809Z}. The donor’s structural response to mass loss is characterized by the logarithmic derivative \cite{1997A&A...327..620S}, 
\begin{equation}
\zeta_2 \equiv \frac{d\ln R_2}{d\ln M_2},
\end{equation}
which plays a central role in determining the mass-transfer rate.

The free parameters of the binary are:
\begin{itemize}
    \item accretor mass $M_1$,
    \item donor mass $M_2$,
    \item orbital period $P_{\rm orb}$,
    \item distance $d$,
    \item inclination $i$,
    \item sky position (right ascension $\alpha$ and declination $\delta$), 
    \item column density along the line of sight $N_{\rm H}$. 
\end{itemize}

The accretor radius $R_1$ is computed using the Nauenberg \cite{1972ApJ...175..417N} mass--radius relation for cold WDs,
\begin{equation}
R_1 = 0.0112 \left[\left(\frac{M_{\rm Ch}}{M_1}\right)^{2/3} - \left(\frac{M_1}{M_{\rm Ch}}\right)^{2/3}\right]^{1/2} {\rm R}_\odot\,\,,
\end{equation}
where $M_{\rm Ch}=1.44 \Msol$ is the Chandrasekhar mass.

\subsection{Accretion state classification}

The model assumes three distinct accretion states for AM CVn systems:
\begin{enumerate}
\item Direct-impact accretion,
\item Persistent disk accretion,
\item Transient disk accretion.
\end{enumerate} 
The classification is determined self-consistently from the binary parameters and the resulting mass-transfer rate. Direct-impact accretion occurs when the ballistic mass-transfer stream from the donor impacts the accretor directly, without forming an accretion disk. This condition is evaluated by comparing the circularization radius $R_{\rm circ}$ of the mass-transfer stream to the accretor radius $R_1$ \cite{1975ApJ...198..383L}.

The circularization radius is defined as the radius of a Keplerian orbit around the accretor that has the same specific angular momentum as the ballistic stream at its point of closest approach. Following the analytic treatment of Lubow \& Shu (1975) \cite{1975ApJ...198..383L}, the minimum distance of the stream from the accretor, $r_{\rm min}$, is approximated as a function of the mass ratio $q = M_2/M_1$ by
\begin{equation}
\frac{r_{\rm min}}{a} = 0.0488 + 0.27\, q^{1/3},
\end{equation}
where $a$ is the binary separation computed from the orbital period $P_{\rm orb}$ and binary masses. The corresponding circularization radius is then given by
\begin{equation}
R_{\rm circ} = (1+q)\left(\frac{r_{\rm min}}{a}\right)^4 a .
\end{equation}

Direct-impact accretion is assumed to occur when the stream cannot circularize outside the WD surface,
\begin{equation}
R_{\rm circ} \le R_1,
\end{equation}
in which case no accretion disk forms and the accretion energy is dissipated in a localized impact region on the WD. If $R_{\rm circ} > R_1$, the stream circularizes and an accretion disk is assumed to form. This criterion is intended only as a first-order approximation to the direct-impact boundary.

For disk-forming systems, the accretion state is further classified as either persistent or transient based on whether the GW-driven mass transfer rate exceeds a critical threshold $\dot{M}_{\rm crit}$. Persistent systems maintain a hot, stable helium accretion disk, while transient systems are assumed to occupy a low state dominated by optically thin emission. This classification reflects the observed phenomenology of AM\,CVn systems and is consistent with disk instability arguments adapted to helium disks \cite{2008A&A...486..523L,1997PASJ...49...75T}.

In practice, the critical mass-transfer rate $\dot{M}_{\rm crit}$ is implemented as a fixed threshold motivated by helium disk instability models, such that systems with $\dot{M} > \dot{M}_{\rm crit}$ are classified as persistent and those below as transient.         
Although the model allows one to compute the critical threshold as the geometric mean of the thin and thick mass transfer rates, for simplicity we set $\dot{M}_{\rm crit} = 10^{15}$ g/s $\approx \num{1.6e-11}\Msol$/yr throughout this work. 
This prescription is not intended to capture detailed physics of outburst cycles, but rather to provide a phenomenological separation between observationally distinct accretion states.

\subsection{Gravitational-wave--driven mass transfer}
\label{SubSec:GW_masstransfer}

The mass-transfer rate is computed assuming that angular momentum losses are dominated by gravitational radiation and that the binary evolves through stable Roche-lobe overflow. The orbital separation $a$ is obtained from Kepler’s third law,
\begin{equation}\label{Eq:Keplers3rd}
a^3 = \frac{G(M_1+M_2)}{(2\pi)^2} P_{\rm orb}^2.
\end{equation}

The orbital angular momentum of the binary is denoted by $J$. The total angular momentum evolution includes both GW losses and changes induced by mass transfer. We assume that gravitational radiation is the only external angular-momentum loss mechanism, $\dot J = \dot J_{\rm GW}$. 
The fractional angular momentum loss rate due to GW emission for a circular binary is \cite{1964PhRv..136.1224P},
\begin{equation}\label{Eq:Jdot}
\frac{\dot{J}_{\rm GW}}{J} =
-\frac{32}{5}\frac{G^3}{c^5}
\frac{M_1 M_2 (M_1+M_2)}{a^4}.
\end{equation}

Under the assumption of stable Roche-lobe overflow, the donor and Roche-lobe radii evolve together,
\begin{equation}\label{Eq:dRdt}
\frac{d\ln R_2}{dt}
=
\frac{d\ln R_L}{dt}.
\end{equation}
With the response of the donor to mass loss parameterized by the donor mass--radius exponent, 
\begin{equation}
\zeta_2 \equiv \frac{d\ln R_2}{d\ln M_2}\,,
\end{equation}
Eq.~(\ref{Eq:dRdt}) becomes
\begin{equation}\label{Eq:dRLdt}
\zeta_2
\frac{\dot M_2}{M_2}
=
\frac{d\ln R_L}{dt}\,.
\end{equation} 
The Roche-lobe response is, 
\begin{equation}
\zeta_L \equiv \left.\frac{\partial\ln R_L}{\partial\ln M_2}\right|_{\rm P_{orb}}\,.
\end{equation} 
This definition isolates the explicit mass dependence of the Roche-lobe response while orbital evolution is handled via 
$\dot{J}/J$, and approximates the full derivative along a conservative mass-transfer sequence which implicitly includes orbital effects \cite{1998ASPC..137..174K,2001A&A...375..890N,2004MNRAS.350..113M,2004MNRAS.349..181N}. 
Using the chain rule for $R_L = R_L(M_2,\,J)$,
\begin{equation}
\frac{d\ln R_L}{dt}
=
\left.\frac{\partial \ln R_L}{\partial \ln M_2}\right|_{J}
\frac{\dot M_2}{M_2}
+
\left.\frac{\partial \ln R_L}{\partial \ln J}\right|_{M_2}
\frac{\dot J}{J}\,,
\end{equation}
and substituting into Eq.~(\ref{Eq:dRLdt}) yields
\begin{equation}
(\zeta_2-\zeta_L)
\frac{\dot M_2}{M_2}
=
\left.\frac{\partial\ln R_L}{\partial\ln J}\right|_{M_2}
\frac{\dot J}{J}.
\end{equation}
For a circular binary, the Roche-lobe radius scales with orbital separation, $R_L \propto a$, and the orbital angular momentum scales as $J \propto a^{1/2}$, implying that, 
\begin{equation}
\left.\frac{\partial\ln R_L}{\partial\ln J}\right|_{M_2} \approx 2,
\end{equation}
yielding, 
\begin{equation}
(\zeta_2-\zeta_L)
\frac{\dot M_2}{M_2}
=
2\frac{\dot J_{\rm GW}}{J}\,, 
\end{equation}
where we have also used $\dot J = \dot J_{\rm GW}$. 
Solving for the mass-transfer rate gives, 
\begin{equation}\label{Eq:Mdot}
\frac{\dot{M}_2}{M_2} =
\frac{2(\dot{J}_{\rm GW}/J)}{\zeta_2 - \zeta_L}.
\end{equation}
The mass-transfer rate obtained from Eq.~(\ref{Eq:Mdot}) is treated internally in our model as signed, with $\dot M_2 < 0$ corresponding to donor mass loss. The absolute value is returned for computing and modeling accretion luminosities in Sections~\ref{SubSec:SpectralComponents} and \ref{SubSec:SpectralModeling}, respectively.

The donor structural response exponent is fixed to $\zeta_2 = -1/3$ for the degenerate WD stars we consider. The Roche-lobe response exponent $\zeta_L$ is evaluated numerically at fixed orbital period (so that orbital evolution effects are captured separately through the $\dot{J}/J$ term) using a symmetric finite-difference scheme. The Roche-lobe radius is approximated using the Eggleton relation \cite{1983ApJ...268..368E},
\begin{equation}\label{Eq:RocheRadius}
\frac{R_{\rm L}}{a} =
\frac{0.49 q^{2/3}}
{0.6 q^{2/3} + \ln(1+q^{1/3})}.
\end{equation}
where the orbital separation $a$ is obtained from Kepler’s law (Eq.~\ref{Eq:Keplers3rd}) given the period. The logarithmic derivative is then evaluated using
\begin{equation}
\zeta_L =
\frac{
\ln R_L(M_2+\Delta M_2)
-
\ln R_L(M_2-\Delta M_2)
}
{
\ln(M_2+\Delta M_2)
-
\ln(M_2-\Delta M_2)
}.
\end{equation}
This procedure self-consistently accounts for the dependence of the Roche-lobe radius on both mass ratio and orbital separation. A small fractional perturbation in donor mass is used to ensure numerical stability across the UCB parameter space. 

Next, we compute the rate of change of the orbital period. 
Taking the logarithmic derivative of the orbital angular momentum of a circular binary $J =
(M_1 M_2/(M_1+M_2))
(G(M_1+M_2)a)^{1/2}$ yields, 
\begin{equation}\label{Eq:Jdotstep1}
\frac{\dot J}{J} =
\frac{\dot M_1}{M_1}
+
\frac{\dot M_2}{M_2}
+
\frac{1}{2}\frac{\dot a}{a}
-
\frac{1}{2}\frac{\dot M_1 + \dot M_2}{M_1+M_2}.
\end{equation}
Assuming conservative mass transfer,
\begin{equation}
\dot M_1 = -\dot M_2,
\end{equation}
Eq.~(\ref{Eq:Jdotstep1}) simplifies to, 
\begin{equation}\label{Eq:Jdotstep2}
\frac{\dot a}{a}
=
2\frac{\dot J}{J}
-
2\dot M_2\left(\frac{1}{M_1}-\frac{1}{M_2}\right)\,.
\end{equation}
Taking the logarithmic time derivative of Kepler’s third law (Eq.~\ref{Eq:Keplers3rd}) yields, 
\begin{equation}
3\frac{\dot a}{a}
=
\frac{\dot M_1 + \dot M_2}{M_1+M_2}
+
2\frac{\dot P_{\rm orb}}{P_{\rm orb}} \,.
\end{equation}
Assuming conservative mass transfer ($\dot M_1 + \dot M_2 = 0 $) and substituting the resulting expression for $\dot a/a$ into Eq.~(\ref{Eq:Jdotstep2}) gives, 
\begin{equation}\label{Eq:Pdot}
\frac{\dot P_{\rm orb}}{P_{\rm orb}} =3\frac{\dot J}{J}+3\dot M_2\left(\frac{1}{M_1}-\frac{1}{M_2}\right).
\end{equation}

Additionally, we can consider whether the sources can be considered monochromatic sources of GW emission. 
The GW frequency and its time derivative are computed assuming emission dominated by the quadrupole harmonic, 
\begin{equation}\label{Eq:fdot} 
f_{\rm GW} = \frac{2}{P_{\rm orb}}, \qquad \dot{f}_{\rm GW} = -f_{\rm GW}\frac{\dot{P}_{\rm orb}}{P_{\rm orb}}. 
\end{equation} 
For an observation time $T_{\rm obs}$, the frequency drift during the observation is \begin{equation} \Delta f \simeq \dot f \, T_{\rm obs}. \end{equation} The Fourier frequency resolution is \begin{equation} \delta f \simeq \frac{1}{T_{\rm obs}}. \end{equation} If \begin{equation} \Delta f \ll \delta f, \end{equation} or equivalently 
\begin{equation} 
|\dot f| \ll \frac{1}{T_{\rm obs}^2}, 
\end{equation} the source is effectively monochromatic and its power remains within a single frequency bin. Thus, a source can be considered monochromatic when it satisfies the condition 
\begin{equation}\label{Eq:Monochromaticity} 
|\dot f| \, T_{\rm obs}^2 \ll 1\,. 
\end{equation}

\subsection{Spectral components}
\label{SubSec:SpectralComponents}

The EM continuum spectrum produced by the model is constructed as a superposition of physically distinct emission components, each associated with a specific accretion geometry and radiative regime. Which components are present, and how the accretion luminosity is partitioned among them, depends on whether the system undergoes direct-impact accretion or disk-mediated accretion, and---if a disk is present---on the mass-transfer rate and resulting accretion state. In all cases, each component is normalized such that integration of its energy reproduces the luminosity assigned by the accretion physics, before geometric and absorption effects are applied.

In addition to the intrinsic binary parameters that determine the mass-transfer rate and accretion geometry, the continuum spectrum depends on a set of regime-dependent emissivity parameters that specify how accretion power is converted into radiation. These parameters describe physically meaningful quantities such as fractional luminosity partitions between emission components, fractional surface areas covered by optically thick emitting regions, characteristic temperatures of optically thin plasma, and radiative efficiency factors that account for energy advection into the WD. Within the model, these emissivity parameters are treated as fixed for each of the three observable accretion regimes and are not tuned on a source-by-source basis. Consequently, variations in the predicted spectra across a DWD population arise primarily from differences in the binary parameters and how they are mapped to the observational regimes, rather than from changes in the assumed microphysics of accretion. On the other hand, in a future work we will explore the impact of varying specific accretion assumptions on the observables and the possibility of probing such assumptions with an extended version of our framework. 
Inclination effects enter the model exclusively through geometric projection factors applied after bolometric normalization. The intrinsic spectral shapes and component temperatures are independent of inclination, and no relativistic, limb-darkening, or atmosphere-induced anisotropies are included.

\subsubsection{Disk-mediated accretion components}\label{subsubSec:DiskComponents}

The total accretion luminosity, $L_{\rm acc}$, is computed from the gravitational potential energy released as mass is accreted onto the WD surface. For a compact object of mass $M_1$ and radius $R_1$, the available gravitational binding energy per unit mass is approximately $GM_1/R_1$, where $G$ is the Newtonian gravitational constant. In the steady accretion limit, essentially all of this potential energy is liberated as radiation rather than being advected or stored, leading to the canonical accretion luminosity
\begin{align}\label{Eq:Lacc}
L_{\rm acc} = \frac{GM_{1}\dot{M}}{R_{1}}\,.    
\end{align}
This corresponds to the total power available from accretion in a Newtonian potential well in the absence of significant kinetic energy storage at the stellar surface. In the context of accretion disks around non-relativistic stars such as WDs, classical steady-state disk theory states that this total gravitational energy release is partitioned roughly equally between radiation from the extended disk and radiation from the boundary layer (BL) where the Keplerian disk flow transitions to corotation with the stellar surface \cite{1974MNRAS.168..603L,1981ARA&A..19..137P}. 
We adopt this canonical partitioning, where the disk radiates
$$
L_{\rm disk} = \frac{1}{2}\frac{GM_{1}\dot{M}}{R_{1}}\,,
$$
and the remaining half is dissipated in the boundary layer as the gas relinquishes its residual kinetic energy upon reaching the WD surface \cite{1981ARA&A..19..137P}. This partitioning is implemented in the model to assign the bolometric luminosities of individual components, with the disk and boundary layer (or direct-impact hotspot in the appropriate regime) receiving fixed fractions of the total accretion power set by these energetics. All component spectra are then normalized so that their integrated luminosity matches the assigned $L_{\rm disk}$ and $L_{\rm BL}$, respectively.

When the mass-transfer stream circularizes outside the WD radius, the model assumes the formation of a geometrically thin, optically thick accretion disk extending from an inner radius $R_{\rm in}$, taken to be the WD radius $R_1$, to an outer radius $R_{\rm out}$ set by tidal truncation. 
The disk emission is modeled using an analytic multi-temperature blackbody prescription that is physically consistent with the commonly used \texttt{diskbb} approximation, rather than the exact thin-disk temperature profile.

The parameters controlling the disk-mediated emission are therefore fully determined by the binary masses and orbital period through their influence on the mass-transfer rate, together with the assumption of a steady-state, radiatively efficient disk. No additional free parameters are introduced to modify the disk luminosity or spectral shape within a given accretion regime, ensuring that the disk contribution is uniquely specified once the accretion state is determined.

In this framework, the disk is treated as a superposition of concentric annuli extending from an inner radius $R_{\rm in}=R_1$ to an outer radius $R_{\rm out}$ set by tidal truncation, each radiating locally as a blackbody with effective temperature 
\begin{equation}\label{Eq:MultiDisk1}
T(r) = \left(\frac{3 G M_1 \dot{M}}{8\pi \sigma_{\rm SB} r^3}\right)^{1/4},
\end{equation}
which follows from local energy dissipation in a Keplerian disk under the assumption of zero torque at the inner boundary \cite{1973A&A....24..337S,1981ARA&A..19..137P}. The hottest annulus occurs at the inner disk edge, defining the characteristic inner disk temperature
\begin{equation}\label{Eq:MultiDisk2}
    T_{\rm in} \equiv T(R_{\rm in}) = \left(\frac{3 G M_1 \dot{M}}{8\pi \sigma_{\rm SB} R_1^3}\right)^{1/4}.
\end{equation}

The total disk spectrum is then constructed by integrating the blackbody emission from all annuli,
\begin{equation}\label{Eq:MultiDisk3}
L_E^{\rm disk} = \int_{R_{\rm in}}^{R_{\rm out}} 2\pi r B_E[T(r)] \, dr,
\end{equation}
where $B_E(T)$ is the Planck function expressed per unit energy. This produces a continuum spectrum with the well-known asymptotic form $L_E \propto E^{1/3}$ at energies below the spectral peak, a defining characteristic of optically thick, multi-temperature accretion disks.

The disk emission contributes primarily in the ultraviolet band for typical AM CVn parameters, with a weak extension into the soft X-ray band at the highest accretion rates and smallest inner disk radii. Inclination-dependent foreshortening is applied after bolometric normalization, reflecting the projected emitting area of the disk surface.

The remaining accretion power,
\begin{equation}
L_{\rm BL} = \frac{1}{2} L_{\rm acc},
\end{equation}
is released in the boundary layer between the disk and the WD surface. This relation defines the total boundary-layer luminosity prior to any partitioning into optically thick or optically thin emission components. 
The radiative character of the boundary layer depends on the mass-transfer rate and defines the accretion state. 

Throughout this work, accretion onto the WD is assumed to be radiatively efficient in the sense that all gravitational potential energy released outside the stellar interior is either radiated promptly or explicitly accounted for via an efficiency factor. No energy is assumed to be lost to large-scale outflows or jets.

\subsubsection{Optically thick boundary layer emission}

At sufficiently high mass-transfer rates (persistent or high-state systems), the boundary layer is assumed to be optically thick. In this regime, most of $L_{\rm BL}$ is emitted as thermal radiation from a belt-like region on the WD surface. This emission is modeled as a single-temperature blackbody with emitting area
$$
A_{\rm BL} = f_{\rm BL}4\pi R_1^2,
$$
where $f_{\rm BL}$ is a free parameter representing the fractional surface coverage of the boundary layer. 
The covering factor $f_{\rm BL}$ controls the effective temperature of the optically thick boundary-layer emission at fixed luminosity, but does not alter the total radiated power, which is fixed by $L_{\rm soft}$.

The effective temperature is determined self-consistently via the Stefan-Boltzmann relation,
$$
T_{\rm BL} = \left(\frac{L_{\rm soft}}{\sigma_{\rm SB} A_{\rm BL}}\right)^{1/4},
$$
where $L_{\rm soft}$ is the fraction of $L_{\rm BL}$ assigned to the optically thick component. 

In the model, this fraction is treated as a fixed, regime-dependent emissivity parameter that specifies how the boundary-layer luminosity is partitioned between optically thick and optically thin emission. 
The boundary-layer luminosity is decomposed as
\begin{equation}
L_{\rm BL} = L_{\rm soft} + L_{\rm thin} + L_{\rm hard},
\end{equation}
where the individual components are defined by fixed fractional parameters,
\begin{align}
L_{\rm soft} &= (1 - f_{\rm thin} - f_{\rm hard})\,L_{\rm BL}, \\
L_{\rm thin} &= f_{\rm thin}\,L_{\rm BL}, \\
L_{\rm hard} &= f_{\rm hard}\,L_{\rm BL}.
\end{align}

The resulting spectrum peaks in the extreme ultraviolet or soft X-ray band and is explicitly normalized so that its bolometric integral equals $L_{\rm soft}$. Inclination-dependent projection and partial occultation by the disk rim are applied to this component.

\subsubsection{Optically thin boundary layer emission}

At low mass-transfer rates (transient or low-state systems), the boundary layer is assumed to be optically thin, and a significant fraction of $L_{\rm BL}$ is radiated as hard X-rays. This emission is modeled as thermal bremsstrahlung from a hot plasma with characteristic temperature set by the virial temperature at the WD surface,
$$
kT_{\rm BL} \sim \frac{3}{8}\frac{G M_1 \mu m_p}{R_1},
$$
where $\mu$ is the mean molecular weight. 
The adopted temperature should be interpreted as a characteristic scale rather than a precise thermodynamic description of the plasma, encapsulating the expectation that the optically thin boundary layer reaches near-virial temperatures close to the WD surface. 
Throughout this work, the mean molecular weight is fixed to a helium-dominated composition appropriate for AM CVn systems. As a result, $\mu$ is not treated as a free parameter, and variations in the optically thin boundary-layer temperature arise solely from differences in the WD mass and radius.

As no closed-form expression exists for the bolometric integral of the adopted bremsstrahlung approximation, the spectrum is constructed as a shape function and normalized numerically over a wide energy range to ensure convergence. The luminosity assigned to the optically thin boundary-layer component is fixed by the emissivity parameter $f_{\rm thin}$,
\begin{equation}
L_{\rm thin} = f_{\rm thin}\,L_{\rm BL},
\end{equation}
with the remaining boundary-layer power distributed between soft thermal and hard X-ray emission as specified by the accretion regime. 
Intermediate accretion regimes allow for coexistence of optically thick and optically thin boundary-layer emission, with the relative luminosity fractions controlled by model parameters. 
In transient disk systems, an additional fraction $f_{\rm hard}$ of the boundary-layer luminosity is emitted as hard X-rays,
\begin{equation}
L_{\rm hard} = f_{\rm hard}\,L_{\rm BL},
\end{equation}
representing high-temperature, optically thin plasma emission characteristic of low accretion states.

\subsubsection{Direct-impact accretion emission}

In sufficiently compact systems, the mass-transfer stream impacts the WD directly before circularizing. In this direct-impact regime, the accretion disk and boundary layer are absent. Instead, a compact impact region near the WD equator radiates a fraction of the accretion luminosity,
$$
L_{\rm soft} = \epsilon_{\rm rad} L_{\rm acc},
$$
where $\epsilon_{\rm rad} \ll 1$ accounts for the fact that much of the dissipated energy may be advected into the WD envelope rather than radiated promptly.

The radiating region is modeled as a soft blackbody hotspot with emitting area
$$
A_{\rm DI} = f_{\rm DI}4\pi R_1^2,
$$
where $f_{\rm DI} \ll f_{\rm BL}$ reflects the highly localized nature of the impact. The effective temperature follows directly from the Stefan-Boltzmann law,
$$
T_{\rm DI} = \left(\frac{L_{\rm soft}}{\sigma_{\rm SB} A_{\rm DI}}\right)^{1/4}.
$$
This component is spectrally distinct from the optically thick boundary layer and dominates the soft X-ray and extreme ultraviolet emission of the shortest-period AM CVn systems.

In this regime, the emissivity parameters $\epsilon_{\rm rad}$ and $f_{\rm DI}$ fully determine the luminosity and temperature of the direct-impact component, and no additional partitioning of the accretion power is applied.

\subsubsection{Band-specific contributions}

In the X-ray band (0.1--10 keV), the spectrum may include optically thin boundary-layer emission, soft blackbody emission from either a boundary layer or direct-impact hotspot, and (optionally) magnetic column emission. In the UV and optical bands (which we restrict to 2.0--8.27 eV in our model), the dominant contributors are the accretion disk and the heated WD surface, with negligible contribution from optically thin plasma components. In all cases, spectral components are combined linearly before absorption and instrumental effects are applied.

\subsection{Spectral modeling}
\label{SubSec:SpectralModeling}

In the X-ray band ($0.1$--$10$~keV), the model includes:
\begin{itemize}
\item optically thin bremsstrahlung emission from the boundary layer,
\item soft blackbody emission from an optically thick boundary layer or direct-impact spot,
\item multi-temperature disk emission approximated by a \texttt{diskbb}-like spectrum.
\end{itemize}

For our model in the UV and optical bands ($2.0$--$8.27$~eV), the dominant components are:
\begin{itemize}
\item the accretion disk,
\item thermal emission from the heated WD surface.
\end{itemize}

The WD effective temperature in the UV includes a parameterized contribution from long-term compressional heating due to accretion, scaled by an efficiency factor. 
In practice, this contribution is implemented as a fixed fraction of the long-term accretion power that is assumed to be thermalized in the WD envelope and re-radiated from the stellar surface. This compressional heating efficiency is treated as a global model parameter rather than a regime-dependent emissivity parameter, and is held fixed across the population to isolate the effects of binary parameters on the emergent ultraviolet spectra.

\begin{table*}[th!]
\centering
\caption{Summary of accretion regimes, dominant emission components, and observational characteristics implemented in the AM CVn binary spectral model.}
\begin{tabular}{p{3cm}||p{5cm}|p{3cm}|p{5cm}}

Accretion Regime & Emission Components & Dominant Bands & Physical Characteristics \\
\hline
\hline
Direct-impact &
Soft blackbody hotspot &
soft X-ray &
No disk; stream impacts WD surface directly; low radiative efficiency \\
\hline
Persistent disk &
Multi-temperature disk + optically thick BL &
UV + soft X-ray &
Stable, hot helium disk; high $\dot{M}$; BL largely optically thick \\
\hline
Transient disk &
Disk (low state) + optically thin BL &
Hard X-ray &
Low $\dot{M}$; disk instability; BL dominated by bremsstrahlung \\

\end{tabular}
\label{Tab:Regimes}
\end{table*}

All spectral components are normalized such that their integrals reproduce the assigned bolometric luminosities exactly. Concretely, for each emission component $i$ with assigned bolometric luminosity $L_i$, the model first constructs a dimensionless or arbitrarily normalized spectral \textit{shape} $S_i(E)$ appropriate to the physical emission process (e.g.\ Planck emission for optically thick components, thermal bremsstrahlung for optically thin components, or a multi-temperature disk spectrum for the accretion disk). The shape function is then integrated over energy to compute
$$
I_i = \int_0^{\infty} S_i(E) dE ,
$$
where the integration is performed either analytically (when a closed-form solution exists) or numerically over a sufficiently wide energy grid to capture the full emitted power. The observer-frame spectral energy flux is then defined as
$$
F_{E,i}(E) = \frac{L_i}{4\pi d^2}\frac{S_i(E)}{I_i},
$$
where $d$ is the source distance. This construction guarantees that
$$
\int_0^{\infty} F_{E,i}(E) dE = \frac{L_i}{4\pi d^2},
$$
independent of the specific energy grid used to evaluate the spectrum.

For optically thick thermal components (the boundary layer and direct-impact hot spot), the shape function $S_i(E)$ is taken to be the Planck function expressed in energy units, $S(E)\propto E^3/[\exp(E/kT)-1]$, and the normalization integral is evaluated analytically using
$$
\int_0^{\infty} \frac{E^3}{\exp(E/kT)-1} dE = \frac{\pi^4}{15}(kT)^4,
$$
ensuring exact bolometric normalization. The temperature $T$ is determined self-consistently from the Stefan--Boltzmann relation,
$$
L_i = \sigma_{\rm SB} A_i T^4,
$$
where $A_i$ is the physically motivated emitting area (e.g.\ a fixed fraction of the WD surface).

For optically thin components, the model constructs a thermal bremsstrahlung photon spectrum and converts it to an energy spectrum. Because no closed-form bolometric integral exists for the adopted bremsstrahlung approximation, the normalization integral $I_i$ is computed numerically over a logarithmically spaced energy grid spanning many decades in energy, ensuring convergence of the total emitted power. The resulting spectrum is then scaled so that its integrated flux matches $L_i/(4\pi d^2)$ exactly. 

Multi-temperature disk emission is treated using an analytic prescription corresponding to a steady-state, optically thick, geometrically thin accretion disk. As detailed in Sec.~\ref{subsubSec:DiskComponents}, 
the disk is modeled as a superposition of concentric annuli. 
Rather than evaluating the radial integral in Eq.~(\ref{Eq:MultiDisk3}) explicitly at runtime, the model adopts the analytic spectral shape associated with this temperature profile, which exhibits the characteristic low-energy asymptotic behavior $L_E \propto E^{1/3}$. This shape is normalized analytically such that the bolometric disk luminosity satisfies
\begin{equation}
\int_0^{\infty} L_E^{\rm disk}\, dE = L_{\rm disk} = \frac{1}{2}\frac{G M_1 \dot{M}}{R_1},
\end{equation}
ensuring exact consistency with the accretion energetics. 

After bolometric normalization, inclination-dependent geometric factors and absorption are applied, preserving the intrinsic luminosity accounting while modifying the observer-frame spectrum. 
All spectral components are normalized using integrals over an effectively infinite energy range prior to the application of instrumental bandpass limits. The \textit{AXIS} ($0.1$--$10$ keV) and \textit{CASTOR} ($2.0$--$8.27$ eV) energy ranges therefore act only to truncate the observer-frame spectra after bolometric normalization, and do not affect the intrinsic luminosity accounting of any emission component.

\subsubsection{Regime-dependent emissivity parameterization}

The observational regime of the AM CVn binary in our model depends on how the computed mass transfer rate $\dot{M}$ compares with the critical mass-transfer rate $\dot{M}_{\rm crit}$ which we fix to $\dot{M}_{\rm crit} = 10^{15}$ g/s $\approx \num{1.6e-11}\Msol$/yr throughout this work. 
Systems with $\dot{M} > \dot{M}_{\rm crit}$ are classified as persistent (i.e., high state dominated by optically thick boundary layer) and are otherwise classified as transient (i.e., low state dominated by optically thin boundary layer). 
                                
To connect the intrinsic accretion energetics of the binary to observable continuum spectra, the model adopts a regime-dependent emissivity parameterization that specifies how the accretion luminosity is distributed among physically distinct emission components. This parameterization is defined separately for the three observable accretion regimes—direct-impact accretion, persistent disk accretion, and transient disk accretion—and reflects differences in accretion geometry, radiative efficiency, and optical depth rather than differences in the underlying gravitational energy release. Within a given regime, the emissivity parameters are held fixed and are intended to represent a characteristic or typical accretion configuration for that class of systems, while variations in the emergent spectrum across the population arise primarily from differences in the binary parameters that determine the mass-transfer rate and accretion state.

For disk-mediated accretion, the total accretion luminosity is partitioned equally between the disk and the boundary layer, consistent with steady-state thin-disk theory. The boundary layer luminosity is further divided between optically thick (soft thermal) and optically thin (hard X-ray) emission, with the relative fractions determined by the accretion regime. Persistent systems at high mass-transfer rates are assumed to host predominantly optically thick boundary layers, while transient or low-state systems radiate a larger fraction of their boundary-layer power via optically thin plasma emission. The fractional surface area covered by optically thick emitting regions is parameterized through fixed covering factors, which control the effective temperatures of the resulting blackbody components.

In the direct-impact regime, the absence of an accretion disk and boundary layer motivates a distinct treatment in which only a small fraction of the total accretion power is assumed to be radiated promptly from a localized impact region on the WD surface, with the remainder advected into the stellar envelope. This reduced radiative efficiency and highly localized emitting area naturally produce a soft, compact emission component characteristic of the shortest-period AM CVn systems. Across all regimes, the adopted emissivity parameters are chosen to be physically plausible, energetically self-consistent, and sufficiently simple to enable efficient forward modeling, while capturing the dominant qualitative differences between accretion states.

\begin{table*}[th!]
\centering
\caption{Default regime-dependent emissivity parameters adopted in the model. These parameters specify how the accretion luminosity is distributed among spectral components for each observable accretion regime. Numerical values correspond to the fiducial choices. 
All other quantities affecting the spectral modeling, including the white dwarf compressional heating efficiency, plasma composition, instrumental bandpasses, and extinction prescriptions, are treated as fixed global assumptions and are described in the text. 
}\label{Tab:RegimeParameters}
\begin{tabular}{lccc}
\hline\hline
Spectral Free Parameter & Direct Impact & Persistent Disk & Transient Disk \\
\hline
Disk luminosity fraction $L_{\rm disk}/L_{\rm acc}$ 
& ---
& 0.5 
& 0.5 \\

Boundary layer luminosity fraction $L_{\rm BL}/L_{\rm acc}$ 
& ---
& 0.5 
& 0.5 \\

Radiated fraction of $L_{\rm acc}$ ($\epsilon_{\rm rad}$) 
& 0.01 
& 1.0 
& 1.0 \\

Optically thick BL fraction $L_{\rm soft}/L_{\rm BL}$ 
& --- 
& $1 - f_{\rm thin}$ 
& $1 - f_{\rm hard}$ \\

Optically thin BL residual fraction ($f_{\rm thin}$) 
&---
& 0.01 
& 0.0 \\

Optically thin hard X-ray fraction ($f_{\rm hard}$) 
&---
& 0.0 
& 0.3 \\

BL surface covering factor ($f_{\rm BL}$) 
&---
& 0.10 
& 0.03 \\

Direct-impact spot covering factor ($f_{\rm DI}$) 
& $1\times10^{-4}$ 
&---
&---\\

Characteristic thin-plasma temperature $kT_{\rm BL}$ 
& --- 
& Virial 
& Virial \\

\hline
\end{tabular}
\end{table*}

\subsection{Inclination-dependent effects}

The observed EM spectrum of an AM CVn system depends not only on its intrinsic accretion luminosity and emission geometry, but also on the inclination angle $i$ between the orbital angular momentum vector and the line of sight. The forward model incorporates inclination-dependent effects in a component-specific manner, reflecting the differing geometries and optical depths of the emitting regions, while deliberately neglecting higher-order effects that are poorly constrained or observationally subdominant.

The inclination angle $i$ is defined such that $i=0^\circ$ corresponds to a face-on system and $i=90^\circ$ to an edge-on configuration. Throughout the model, inclination effects are treated as purely geometric and time-independent; orbital modulation, eclipses, and phase-dependent self-occultation are not modeled explicitly.

\subsubsection{Optically thick disk emission}

For disk-mediated accretion, the accretion disk is assumed to be geometrically thin and optically thick. The specific intensity emitted by such a disk is assumed to be locally isotropic, so that the observed flux scales with the projected emitting area. As a result, the disk contribution to the observed spectral energy flux is multiplied by a simple foreshortening factor,
\begin{align}\label{Eq:DiskInclination}
F_{E,\rm disk}(i) = F_{E,\rm disk}(i=0)\cos i.
\end{align}
This prescription follows standard thin-disk treatments and captures the leading-order inclination dependence of disk emission. No additional limb-darkening corrections are applied by default, although the formalism allows for such modifications if desired. At high inclinations, the model does not include disk rim obscuration or vertical disk structure; therefore, the $\cos i$ scaling is applied uniformly for all $i < 90^\circ$.

\subsubsection{Optically thick boundary layer and direct-impact emission}

Thermal emission from optically thick regions on the WD surface—including the boundary layer in disk-accreting systems and the hotspot in direct-impact systems—is modeled as blackbody radiation from a localized surface area. For these components, the inclination dependence arises from both geometric projection and partial self-occultation by the accretion disk.

In the absence of occultation, the projected emitting area scales as $\cos i$, leading to
$$
F_{E,\rm soft}(i) = F_{E,\rm soft}(i=0)\cos i.
$$

However, in disk-mediated systems, the inner edge of the accretion disk can partially obscure the boundary-layer emission at moderate to high inclinations. To account for this effect, the model applies an inclination-dependent suppression factor $f_{\rm occ}(i)$, such that
$$
F_{E,\rm BL}(i) = F_{E,\rm BL}(i=0)\cos i,f_{\rm occ}(i),
$$
where $0 \le f_{\rm occ}(i) \le 1$. This factor is parameterized phenomenologically and represents the fraction of the emitting region that remains visible above the disk plane. In direct-impact systems, where no disk is present, $f_{\rm occ}(i)=1$ by construction, and only geometric projection affects the observed flux.

The model does not include relativistic light bending or rotational Doppler effects, which are expected to be negligible for WD surface emission.

\subsubsection{Optically thin emission components}

Optically thin emission components—including thermal bremsstrahlung from an optically thin boundary layer or from a magnetic accretion column—are assumed to be emitted isotropically. These components are therefore taken to be independent of inclination,
$$
F_{E,\rm thin}(i) = F_{E,\rm thin}(i=0),
$$
reflecting the expectation that the emitting plasma occupies an extended, quasi-spherical or vertically distributed region. No angular dependence of emissivity or absorption within the plasma is modeled.

\subsubsection{Ultraviolet emission from the WD surface}

In the ultraviolet band, emission from the heated WD surface is treated as optically thick thermal radiation. As with boundary-layer emission, the inclination dependence is modeled via geometric projection alone,
$$
F_{E,\rm WD}(i) = F_{E,\rm WD}(i=0)\,\cos i.
$$
This component is not subject to disk occultation in the current model, implicitly assuming that the ultraviolet-emitting region lies sufficiently high above the disk midplane or that the disk is optically thin in the relevant wavelength range.

\subsubsection{Modeling assumptions and limitations}

The inclination treatment in the model is intentionally simplified. It neglects eclipses, orbital modulation, disk warping, flaring, and azimuthal asymmetries, as well as relativistic effects such as Doppler boosting and gravitational lensing. These effects are expected to be subdominant for the majority of AM CVn systems considered here. The adopted prescription captures simple inclination-dependent effects, but including others is an important direction for future work.

\subsection{Absorption and extinction}
\label{subSec:AbsorpExtinct}

X-ray absorption is modeled using a simple exponential attenuation,
\begin{equation}\label{Eq:XrayAbsorption}
T(E) = \exp[-N_{\rm H}\,\sigma(E)],
\end{equation}
where $N_{\rm H}$ is the equivalent neutral hydrogen column density and $\sigma(E)$ is an approximate photoelectric cross section. This treatment is intended to capture the dominant energy-dependent attenuation in the soft X-ray band and is applied to the intrinsic source spectrum.

UV extinction is treated using a physically motivated hybrid extinction model that combines standard prescriptions from the literature. 
The interstellar hydrogen column density $N_{\rm H}$ is converted to a visual extinction via the linear relation
\begin{equation}
A_V = \frac{N_{\rm H}}{2.21 \times 10^{21} \ \mathrm{cm}^{-2}\,\mathrm{mag}^{-1}},
\label{eq:Av_NH}
\end{equation}
where the conversion factor $2.21 \times 10^{21} \ \mathrm{cm}^{-2}\,\mathrm{mag}^{-1}$ is a fixed canonical value. 
The value of $N_{\rm H}$ is either treated as a free parameter or computed from three‑dimensional dust maps (e.g., Bayestar 2019; \cite{2018JOSS....3..695G}). 
The color excess is $E(B-V) = A_V / R_V$, where $R_V$ is the total‑to‑selective extinction ratio. 

The wavelength-dependent extinction curve $A(\lambda)/A_V$ is implemented piecewise. In UV ($x \equiv 1/\lambda \ge 3.3~\mu{\rm m}^{-1}$), the model adopts the formulation of \citep{1999PASP..111...63F}, including the 2175~\AA\ Drude bump and far-UV curvature terms with coefficients that depend explicitly on $R_V$. In the optical regime ($1.1 \le x < 3.3~\mu{\rm m}^{-1}$), the extinction curve follows the polynomial update \cite{1994ApJ...422..158O} to the original law \cite{1989ApJ...345..245C}. 
To avoid an artificial discontinuity at the UV-optical transition, the extinction curve is blended smoothly across the narrow interval $3.0 \le x \le 3.4~\mu{\rm m}^{-1}$ using a cosine interpolation between the optical and ultraviolet prescriptions. This ensures numerical stability and physical smoothness when applying extinction to broadband UV spectra. 

For a given photon energy $E$, the corresponding wavelength is computed via $\lambda = hc/E$, and the extinction transmission is applied as
\begin{align}\label{Eq:UVExtinction}
T(\lambda) = 10^{-0.4\,A(\lambda)}.
\end{align}
This transmission factor is applied to the energy-based spectral flux prior to instrumental response modeling.

\subsection{Instrument response and count-rate modeling}

To convert the intrinsic, absorption-modified source spectra into observable quantities, the model incorporates instrument-specific response functions for both X-ray and ultraviolet detectors. In all cases, the detector response is applied after the construction of the intrinsic spectral energy distribution and the inclusion of interstellar absorption, and before any comparison to observational sensitivities. The model computes photon or count rates by explicitly folding the source spectrum through the appropriate effective area or passband response, thereby reproducing the physical operations performed by standard spectral analysis tools while retaining full control over the forward-modeling pipeline.

\subsubsection{X-ray detector response}
\label{Subsubsec:XrayCounts}

For X-ray observations, the model implements a physically detailed response treatment for \textit{AXIS}. Its response is specified through a publicly available\footnote{See the \href{https://axis.umd.edu/researchers/simulation-resources}{\textit{AXIS} Simulation Resources} site.} ancillary response file (ARF), which encodes the energy-dependent effective collecting area of the telescope and detector system, including mirror reflectivity, detector quantum efficiency, and geometric throughput.

Given an absorption-modified source energy flux spectrum $F_E(E)$, the model first converts this to a photon flux spectrum,
\begin{equation}
\Phi(E) = \frac{F_E(E)}{E},
\end{equation}
expressed in units of photons~cm$^{-2}$~s$^{-1}$~keV$^{-1}$. The expected detector count rate is then computed by explicitly folding this photon spectrum through the \textit{AXIS} effective area,
\begin{equation}\label{Eq:XrayCountRate}
C_{\rm X} = \int \Phi(E)\, A_{\rm eff}^{\rm AXIS}(E)\, dE ,
\end{equation}
where $A_{\rm eff}^{\rm AXIS}(E)$ is taken directly from the ARF. The integration is performed numerically over the full energy range of interest ($0.1$--$10$~keV), using an energy grid sufficiently fine to resolve both spectral features and variations in the effective area.

This procedure captures the effective area weighting used in standard forward-folding analyses, such as with the spectral fitting package \texttt{XSPEC} \cite{1996ASPC..101...17A,1999ascl.soft10005A}, while omitting the response distribution between detector channels. We therefore fold the photon spectrum through the ARF and do not apply the response matrix file (RMF), an approximation that enables rapid broadband count rate predictions across a large parameter space without invoking external software, and is appropriate for estimating total 0.1--10 keV count rates. We note however that the RMF effects would be needed for channel-resolved spectral fitting or hardness ratio predictions, which is beyond the scope of this work.

\subsubsection{Ultraviolet and optical detector response}
\label{Subsubsec:CASTORCounts}

For UV and optical observations, the model adopts a band-integrated response treatment appropriate for \textit{CASTOR} imaging data. The intrinsic source spectrum is first expressed as an energy flux per unit energy, $F_E(E)$, after absorption and inclination effects are applied. This is converted to a spectral flux density per unit frequency using the standard relation
\begin{equation}
F_\nu(\nu) = F_E(E)\,\frac{dE}{d\nu} = h\,F_E(E),
\end{equation}
where $E = h\nu$ and $h$ is Planck’s constant. The resulting $F_\nu(\nu)$ is expressed in cgs units of erg~cm$^{-2}$~s$^{-1}$~Hz$^{-1}$.
 
For each \textit{CASTOR} filter $b$, characterized by a dimensionless transmission curve $T_b(\nu)$ that encodes the wavelength-dependent throughput of the telescope, optics, filter, and detector, the model computes a band-averaged flux density,
\begin{equation}\label{Eq:UVAverageFlux}
\langle F_\nu \rangle_b =
\frac{\int F_\nu(\nu)\, T_b(\nu)\, d\nu}
     {\int T_b(\nu)\, d\nu}.
\end{equation}
This definition ensures that the effective flux density corresponds to the mean flux transmitted by the instrument across the passband. 
The dimensionless passband transmission curves for the u, g, and UV filters are obtained from the \texttt{FORECASTOR} source code \cite{2024AJ....167..178C}.

The band-averaged flux density is converted into an AB magnitude using
\begin{equation}\label{Eq:UVMag}
m_{\rm AB} = -2.5 \log_{10}\!\left(\frac{\langle F_\nu \rangle_b}{3631~{\rm Jy}}\right),
\end{equation}
where the AB zero point of 3631~Jy is adopted. Finally, the expected \textit{CASTOR} count rate is obtained by applying the band-specific zero-point calibrations $ZP_{\rm AB}$ \cite{2024AJ....167..178C} for each filter, 
\begin{align}
C_{UV} = 10^{-0.4\,(m_{\rm AB} - ZP_{\rm AB})}\,,
\end{align} 
and summing over the counts for each filter for a total count rate. This procedure yields ultraviolet observables that are directly comparable to \textit{CASTOR} sensitivity estimates and exposure-time calculations, while avoiding the need for an explicit wavelength-resolved effective area. All extinction corrections are applied prior to bandpass integration using the extinction model described in Section~\ref{subSec:AbsorpExtinct}, ensuring consistency between the intrinsic spectral energy distribution and the \textit{CASTOR} count rate.

\section{Population Synthesis of AM CVns}
\label{Sec:Cosmic}

To construct a physically motivated population of AM CVn binaries suitable for multi-messenger inference, we employ the \texttt{COSMIC} binary population synthesis framework \cite{2020ApJ...898...71B} (version 3.6.0). This approach allows us to generate compact binary progenitors from zero-age main sequence (ZAMS) initial conditions, evolve them through common-envelope and mass-transfer phases, and identify systems that emerge as ultra-compact helium-transferring binaries. Because AM CVn systems evolve significantly after the onset of Roche-lobe overflow (RLOF), we supplement the \texttt{COSMIC} output with a forward evolution model that integrates each system through the GW--driven mass-transfer phase. This hybrid approach yields a physically consistent population spanning the full range of orbital periods relevant for multi-messenger observations.

\subsection{Binary Population Synthesis with COSMIC}

The \texttt{COSMIC} framework extends the Binary Stellar Evolution  formalism \cite{1998MNRAS.298..525P,2000MNRAS.315..543H,2002MNRAS.329..897H} to include updated prescriptions for processes such as stellar evolution, mass transfer, and compact object formation. Each binary system is initialized at the ZAMS with a set of parameters $\left( M_{1,\rm ZAMS}, M_{2,\rm ZAMS}, P_{\rm ZAMS}, e_{\rm ZAMS},  \alpha_{\rm CE}, Z, \eta_{\rm acc} \right)$, where $M_{1,\rm ZAMS}$ and $M_{2,\rm ZAMS}$ are the primary and secondary masses, $P_{\rm ZAMS}$ is the orbital period, $e_{\rm ZAMS}$ is the eccentricity, $\alpha_{\rm CE}$ is the common-envelope efficiency parameter, $Z$ is the metallicity fixed at solar abundance ($Z=0.014$), and $\eta_{\rm acc}$ is the mass accretion efficiency which we set to unity for all binaries.

Rather than drawing initial conditions from assumed distributions, we construct a structured grid spanning the physically relevant initial parameter space,
\begin{equation}
M_{1,\rm ZAMS} \in [0.7,\,10]\,M_\odot,\quad \nonumber
M_{2,\rm ZAMS} \in [0.7,\,10]\,M_\odot,
\end{equation}
\begin{equation}\label{Eq:COSMICGrid}
P_{\rm ZAMS} \in [15,\,1000]\,{\rm days}, \quad 
e_{\rm ZAMS} \in [0,\,0.7],
\end{equation}
and with $\alpha_{\rm CE} = 8.6$. 
Each grid point is evolved independently with \texttt{COSMIC}. 

Binary evolution proceeds through a sequence of evolutionary phases, including mass transfer, common-envelope evolution, and compact object formation. The outcome of common-envelope evolution is determined using the standard energy formalism
\begin{equation}\label{Eq:CommonEnvelope}
\alpha_{\rm CE} \left( \frac{GM_{\rm core,1}M_2}{2a_f} - \frac{GM_1M_2}{2a_i} \right)
= \frac{GM_1M_{\rm env,1}}{\lambda R_1},
\end{equation}
where $a_i$ and $a_f$ are the initial and final separations of the episode, $M_{\rm core}$ and $M_{\rm env}$ are the stellar core and envelope masses of the donor, and $\lambda$ parameterizes the envelope binding energy \cite{1984ApJ...277..355W,1990ApJ...358..189D,2002MNRAS.329..897H}. This standard prescription determines whether a system survives common-envelope evolution and emerges as a compact binary, and is a major uncertainty in AM CVn progenitor evolution. 
AM CVn systems are produced primarily through two pathways: either each star initiates separate CEE episodes to lose their envelopes, or a single CEE episode initiated by the primary star causes both stars to lose their envelopes.  
In these pathways, the short period of the surviving DWD binary allows for the less massive WD to fill its Roche lobe and begin transferring helium-rich material to the more massive WD companion  \cite{2001A&A...375..890N,2001A&A...368..939N,2006csxs.book..623T}. 

From the full \texttt{COSMIC} population, we identify AM CVn progenitors by selecting DWD binaries with RLOF initiated by the helium donor. These systems correspond to the onset of stable mass transfer between the two WDs. 
To obtain our fiducial population, we forward evolve the ZAMS grid and filter for He, C/O, or O/Ne primary WD stars and He secondary WD stars ($k_{\rm star 1} \in \{10, 11, 12\}$ and $k_{\rm star 2} \in \{10\}$ in COSMIC, respectively) undergoing RLOF. This selection produces $15{,}742$ AM CVn progenitor binaries and is consistent with our model for the continuum spectrum in Section~\ref{Sec:ForwardEM}, which only requires that the donor be a He WD. 

We explored variation in the common envelope efficiency over the range $0.2 < \alpha_{\rm CE} < 10$, and find that AM CVn progenitors are produced predominantly for $\alpha_{\rm CE} \gtrsim 6$, motivating our use of $\alpha_{\rm CE} = 8.6$ in constructing our ZAMS grid. 
We find that the evolutionary pathways of the population are sensitive to the bounds of $m_{2,\mathrm{ZAMS}}$ which govern the boundary in the parameter space between a population dominated by double‑core CEE (one recorded common‑envelope event that strips both stars) and one dominated by two separate CEE episodes. When $m_{2,\mathrm{ZAMS}} \lesssim 1.0\,M_\odot$, its main‑sequence lifetime exceeds $\sim10\,\mathrm{Gyr}$ and the first CEE event, triggered by the more massive primary, occurs before the secondary has left the main sequence.  The secondary is engulfed during the primary’s giant phase and stripped of its envelope in the same CEE episode, leaving two degenerate cores but only one recorded CEE event, and the pathway appears as a single, double‑core CEE.  
In contrast, for $m_{2,\mathrm{ZAMS}} \gtrsim 1.0\,M_\odot$ the secondary’s main‑sequence lifetime is short enough that it can become a giant, fill its Roche lobe, and initiate a second CEE episode, yielding the canonical pathway with two CEEs. In our fiducial grid in Eq.~(\ref{Eq:COSMICGrid}), we find that $\approx 80\%$ of the AM CVns evolve in the canonical pathway with two CEEs and the remaining $\approx 20\%$ evolve in the double-core CEE pathway. 
We emphasize that theoretical modeling of AM CVns remains uncertain \cite{2023A&A...678A..34B}. 

A notable feature of the AM CVn systems is that they have a narrow, non-Gaussian distribution of orbital periods at RLOF onset, typically with a peak of $P_{\rm orb} \approx 300$ s. This is due to the binary orbital configuration required for RLOF from a He WD donor which has a small range in mass, radius, and thus Roche radius. Consequently, the raw population does not span the full range of observed AM CVn orbital periods. To obtain systems at longer periods, it is necessary to evolve the binaries forward in time through the DWD mass-transfer phase.

\subsection{Forward Evolution Through the AM CVn Phase}
\label{SubSec:AMCVnEvolve}

We follow the same assumptions in Section~\ref{SubSec:GW_masstransfer}. 
Following RLOF onset, the DWD binaries evolve under the combined effects of GW emission and mass transfer. The orbital angular momentum loss due to gravitational radiation is given by Eq.~(\ref{Eq:Jdot}) for an accretor with mass $M_1$ and its less massive donor with mass $M_2$ and $a$ is the orbital separation \cite{1964PhRv..136.1224P}. 
For degenerate donors, mass loss causes the donor to expand, leading to orbital expansion after contact. The equilibrium mass-transfer rate $\dot{M}_2$, determined by requiring Roche-lobe contact, is given by Eq.~(\ref{Eq:Mdot}). 

We integrate the coupled evolution equations
\begin{equation}\label{Eq:AMCVnEvolve1}
\frac{dM_1}{dt} = -\frac{dM_2}{dt},
\end{equation}
\begin{equation}\label{Eq:AMCVnEvolve2}
\frac{dM_2}{dt} = \dot{M}_2(M_1,M_2)\,,
\end{equation}
while enforcing Roche-lobe contact at each timestep, and where $\dot{M}_2 < 0$ and given by Eq.~(\ref{Eq:Mdot}). 

The orbital period at each timestep is computed from the instantaneous binary configuration using Kepler's third law in Eq.~(\ref{Eq:Keplers3rd}), and the orbital separation is determined by the Roche-lobe constraint from Section~\ref{SubSec:GW_masstransfer},
\begin{align}
R_2(M_2) = R_{\rm L}(M_1,M_2,a)\,,
\end{align}
where $R_2$ is the donor radius in Eq.~(\ref{Eq:DonorRadius}) and $R_{\rm L}$ is the Roche-lobe radius in Eq.~(\ref{Eq:RocheRadius}). This procedure self-consistently couples mass transfer, orbital evolution, and GW emission, producing the characteristic widening of AM CVn systems following the onset of stable mass transfer. 

We scale the raw \texttt{COSMIC} output from the previous section to obtain a sizable number of AM CVn-evolved systems by drawing 7 uniformly random accretion times for each AM CVn binary in the raw output. This results in $110{,}194$ systems, which we evolve through the AM CVn phase and track the evolution of their masses and orbital period. 
Specifically, we assign evolution times with uniform draws between $0$ and $t_{\rm max}$, and integrate forward in time as described above. This procedure produces a broad orbital period distribution spanning $P_{\rm orb} \approx 5$ to $70$ min,
consistent with observed AM CVn systems \cite{2005ASPC..330...27N,2010PASP..122.1133S,2018MNRAS.476.3820R}. 
The final population is characterized by its masses $M_1$ and $M_2$, orbital period $P_{\rm orb}$, mass-transfer rate $\dot{M}$, and age $t_{\rm age}$. 

While we leave a systematic study of binary population uncertainties of AM CVns and their observational consequences to a future work, the population we obtain here provides an example for generating synthetic multi-messenger observations and performing inference on AM CVn binary parameters.

\section{Model for \textit{LISA} detections of AM CVn binaries}
\label{Sec:LISA}

Using the \textsc{legwork}\footnote{See the derivations in the \textsc{legwork} \href{https://legwork.readthedocs.io/en/latest/notebooks/Derivations.html}{documentation}.} package \cite{LEGWORK_apjs,LEGWORK_joss}, 
we compute GW detectability for all modeled AM CVn systems using instantaneous orbital parameters, treating each source as quasi--monochromatic over the \textit{LISA} mission lifetime. In this approximation, the GW frequency is assumed to remain constant during the observation, and we neglect mass-transfer--driven orbital evolution and any associated non-monotonic frequency behavior. While some AM CVn systems may formally satisfy $|\dot{f}|T^2 \gtrsim 1$, where $T_{\rm LISA}$ is the \textit{LISA} mission duration, the resulting error in the signal-to-noise ratio (SNR) $\rho_{\rm LISA}$ is typically of order unity and will not qualitatively affect our detectability estimates of an AM CVn population \cite{2004PhRvD..70h2002V,2009MNRAS.400L..24S}. 

For a circular compact binary with orbital frequency $f_{\rm orb}$, the GW frequency at leading order is twice the orbital frequency, 
$f_{\rm GW} = 2 f_{\rm orb}$, and the chirp mass is,
\begin{equation}\label{Eq:ChirpMass}
\mathcal{M}_c = \frac{(M_1 M_2)^{3/5}}{(M_1 + M_2)^{1/5}}\,,
\end{equation} 
for component masses $M_1$ and $M_2$. 
The strain amplitude $h_0$ depends on the chirp mass, orbital frequency, distance, eccentricity, and inclination, and includes the appropriate angular dependence of the GW polarizations. 
As AM CVn systems are generally circular, we compute $h_0$ with \textsc{legwork} and take the 2nd harmonic associated with GW emission from a circular binary. 
The characteristic strain $h_c$, which incorporates both the intrinsic strain amplitude and the coherent buildup of signal over the observation time, is \cite{2015CQGra..32a5014M},
\begin{equation}\label{Eq:CharacteristicStrain}
h_c(f_{\rm GW}) = h_0 \sqrt{N_{\rm cyc}},
\end{equation}
where the number of observed cycles is $N_{\rm cyc} = f_{\rm GW} T_{\rm LISA}$, 
which we convert into the amplitude spectral density, 
\begin{align}\label{Eq:AmplitudeSpectralDensity}
A(f_{\rm GW}) = h_c(f_{\rm GW})\,(f_{\rm GW})^{-1/2}\,, 
\end{align}
i.e., the square root of the power spectral density of the signal. 
We compute the SNR $\rho_{\rm LISA}$ of the source with \textsc{legwork}, which assumes a 6-link (3-arm) \textit{LISA} detector in the long wavelength limit given $A_n(f)$ the noise amplitude spectral density of \textit{LISA} which includes instrumental noise and the Galactic stochastic DWD foreground \cite{2019CQGra..36j5011R}. 
We fix the GW polarization to 10$^{\circ}$ for all binaries, and assume a 1 year LISA mission duration. 
We emphasize that our \textit{LISA} analysis provides an instantaneous snapshot of GW observability consistent with the EM forward model. A fully self-consistent treatment of coupled mass transfer, orbital evolution, and time-dependent GW phase evolution is deferred to future work.

\section{Model for the AM CVn binary Galactic location}
\label{Sec:GalacticLocation}

To model the spatial distribution of DWDs in the Milky Way, we adopt an analytic, axisymmetric disk model that captures the principal structural and dynamical properties of the Galactic stellar disk while remaining computationally efficient. The distribution of systems in Galactocentric cylindrical coordinates $(R,\phi,z)$ is constructed by sampling radial positions from an exponential disk profile, incorporating age‐dependent radial diffusion, and assigning vertical offsets that reflect progressive dynamical heating with stellar age. This approach is designed to reflect the dominant trends in stellar populations observed in the solar neighborhood and throughout the disk in a simplified treatment of the populations in the disk.

The radial distribution of binaries is drawn from an exponential profile,
\begin{equation}
    p(R) \propto \exp\!\left(-\frac{R}{R_d}\right),
\end{equation}
with disk scale length $R_d$ chosen to match observational constraints on the Milky Way’s old stellar disk. Observational and dynamical analyses of stellar surface density profiles find radial scale lengths in the range $2.0$--$3.0\,$kpc for the old disk and its chemically selected subcomponents \cite{2013ApJ...779..115B,2016ApJ...823...30B}. An exponential radial profile provides an approximation to the surface density of stars and binaries in the disk, consistent with extensive photometric and kinematic studies showing a roughly exponential decline in surface density with increasing Galactocentric radius \cite{2021Galax...9...29K}.

To capture secular changes in orbital radius driven by interactions with spiral arms, giant molecular clouds, and other perturbations, we model radial diffusion as a Gaussian broadening of the birth radius \cite{2018ApJ...865...96F},
\begin{equation}
    R = R_{\rm birth} + \Delta R, \quad
    \Delta R \sim \mathcal{N}(0,\sigma_R^2), \quad
    \sigma_R = \sigma_{R,0}\sqrt{t},
\end{equation}
where $t$ is the system age and $\sigma_{R,0}$ is a diffusion coefficient. This random‐walk approximation is widely used in analytic treatments of stellar migration in disk galaxies \cite{2012MNRAS.422.1363S}, and efficiently captures the age-radius correlation seen in detailed chemo-dynamical models without full orbit integrations.

The vertical distribution perpendicular to the plane is modeled with an age‐dependent exponential,
\begin{equation}\label{Eq:GalVertical}
    p(z) = \frac{1}{2z_d(t)}\exp\!\left(-\frac{|z|}{z_d(t)}\right),
\end{equation}
with scale height
\begin{equation}\label{Eq:GalScaleHeight}
    z_d(t) = z_{\min} + (z_{\max}-z_{\min})\left(\frac{t}{t_{\max}}\right)^\alpha.
\end{equation}
Here, $z_{\min} = 0.1$ kpc and $z_{\max} = 0.9$ kpc are the characteristic scale heights of the youngest and oldest disk populations \cite{2017MNRAS.465...76M}, respectively, and $\alpha$ parametrizes the age dependence. 

Once Galactocentric positions are sampled, we transform them to heliocentric Galactic longitude $l$, latitude $b$, and distance $d$ using an astrometric coordinate framework with Solar parameters consistent with the latest astrometric measurements, specifically a Solar Galactocentric radius $R_0 \approx 8.1\,$kpc \cite{2019A&A...625L..10G}. This transformation yields coordinates readily comparable to observational surveys and is essential for modeling both EM and gravitational wave detectability.

Interstellar absorption along each line of sight is computed using the three-dimensional dust reddening maps from Bayestar 2019 \cite{2019ApJ...887...93G}, which provide reddening $E(B-V)$ as a function of distance and sky position by combining Gaia parallaxes with Pan-STARRS1 and 2MASS photometry. 
We convert reddening to equivalent hydrogen column density $N_{\rm H}$ using the canonical relations in \ref{subSec:AbsorpExtinct}. 
This empirical scaling is widely used in Galactic absorption modeling due to its basis in measurements of supernova remnants and X-ray sources distributed across the disk \cite{GuverOzel2009}.

Because three-dimensional dust maps have incomplete sky coverage and limitations at high latitudes and large distances, we apply a fallback scheme for $N_{\rm H}$ when Bayestar returns undefined values. High‐latitude sightlines are assigned a low column density characteristic of dust‐poor halo environments, while distant, low-latitude sightlines are assigned a saturated disk column density consistent with integrated extinction in the Galactic midplane. This hybrid approach ensures robust, physically plausible absorption estimates across the full synthetic population and avoids pathological values that would otherwise bias EM detectability predictions.

In the multi-messenger pipeline, these derived extrinsic parameters — $l$, $b$, $d$, and $N_{\rm H}$ — are combined with intrinsic binary properties to compute ultraviolet and X-ray spectra. The resulting population synthesis accounts self-consistently for both spatial distribution and interstellar absorption effects, enabling accurate population forecasts for current and future observatories.

\section{Simulation-based inference with normalizing flows} 
\label{Sec:NormalFlows}

The forward multi-messenger model described in the previous section provides a deterministic mapping between intrinsic AM CVn binary parameters and observable EM and GW quantities. While this forward model is computationally efficient and physically interpretable, the inverse problem---inferring binary parameters from observed data---remains highly non-trivial. The mapping from intrinsic parameters to observables is strongly non-linear and degenerate, particularly when combining EM continuum emission with GW information. Moreover, the likelihood function corresponding to this forward model is analytically intractable, since the model incorporates multiple coupled physical processes including GW driven mass transfer, accretion disk emission, boundary layer physics, interstellar absorption, and instrumental response. To address this challenge, we adopt a simulation-based inference framework using conditional normalizing flows to learn the inverse mapping directly from simulated data.

We define the parameter vector of an AM CVn binary as, 
\begin{align}\label{Eq:ParameterVector}
\boldsymbol{\theta} =
\left(
M_1,\,
M_2,\,
P_{\rm orb},\,
i,\,
d,\,
N_{\rm H}
\right),
\end{align}
where $M_1$ and $M_2$ are the accretor and donor masses, $P_{\rm orb}$ is the orbital period, $i$ is the inclination angle, $d$ is the distance, and $N_{\rm H}$ is the hydrogen column density. We note that we do not use the Galactic location model of Section~\ref{Sec:GalacticLocation} in the training of the flow. 
The multi-messenger model, i.e., the combination of the EM continuum model presented in Section \ref{Sec:ForwardEM} and of the \textit{LISA} model in Section \ref{Sec:LISA}, produces a corresponding observation vector, 
\begin{align}\label{Eq:ObsVector}
\boldsymbol{x} = & \left( \log_{10} C_{\rm X},\, \log_{10} C_{\rm UV},\, \log_{10} (E F_E)_1,\, \dots \right. \nonumber\\
                 & \,\,\left. \log_{10} (E F_E)_{N_{\rm spec}},\, \log (1+\mathrm{SNR}_{\rm LISA}) \right)\,,
\end{align}
where $C_{\rm X}$ and $C_{\rm UV}$ are the count rates for an \textit{AXIS}-like detector and \textit{CASTOR}, respectively, $(E F_E)_k$ denotes the continuum spectral flux in the $k$-th energy bin, and $\rho_{\rm LISA}$ is the GW signal-to-noise ratio. We adopt $N_{\rm spec}=128$, yielding an observation vector of dimension 131. 

The inference problem is to determine the posterior probability distribution
\begin{equation}
p(\boldsymbol{\theta}|\boldsymbol{x}_{\rm obs}),
\end{equation}
for a given set of multi-messenger observations $\boldsymbol{x}$. Using Bayes' theorem, the posterior can be written as

\begin{equation}
p(\boldsymbol{\theta}|\boldsymbol{x})
=
\frac{
p(\boldsymbol{x}|\boldsymbol{\theta})
p(\boldsymbol{\theta})
}{
p(\boldsymbol{x})
}.
\end{equation}
However, the likelihood $p(\boldsymbol{x}|\boldsymbol{\theta})$ is not available in closed form, since the inverse model is not analytically tractable. We therefore employ simulation-based inference to approximate the posterior directly using neural density estimation \cite{2018arXiv180501683P,2019PNAS..116.22567P,2021JOSS....6.2505T}.

\subsection{Conditional normalizing flow architecture}

To approximate the posterior distribution, we adopt conditional normalizing flows implemented through the \texttt{sbi} framework \cite{2021JOSS....6.2505T}. Normalizing flows model complex probability distributions by transforming a simple base distribution through a sequence of invertible mappings \cite{2015arXiv150505770J}. Specifically, we introduce a latent variable $\boldsymbol{z}$ drawn from a standard multivariate Gaussian distribution,
\begin{equation}
\boldsymbol{z} \sim \mathcal{N}(0, I)\,,  \notag
\end{equation}
and define an invertible transformation
\begin{equation}
\boldsymbol{\theta}
=
T_\phi(\boldsymbol{z}; \boldsymbol{x})\,,  \notag
\end{equation}
where $T_\phi$ is parameterized by a neural net with parameters $\phi$. The resulting conditional density is obtained using the change-of-variables formula,
\begin{equation}
q_\phi(\boldsymbol{\theta}|\boldsymbol{x})
=
p(\boldsymbol{z})
\left|
\det
\frac{\partial T_\phi^{-1}}
{\partial \boldsymbol{\theta}}
\right|\,.  \notag
\end{equation}

We utilize the publicly available \texttt{sbi} module \cite{2021JOSS....6.2505T} to adopt neural spline flows (NSF), which use monotonic rational-quadratic splines to construct flexible invertible transformations \cite{2019NeurIPS..32.7503D}. These flows are well suited for modeling non-Gaussian and multi-modal posterior distributions commonly encountered in astrophysical inference problems. 

As the observation vector is high-dimensional and contains structured spectral information, we compress it using an embedding network based on a one-dimensional convolutional neural net (CNN),
\begin{equation}
\boldsymbol{h} = g_\psi(\boldsymbol{x}),
\end{equation}
where $g_\psi$ consists of a sequence of convolutional layers acting on the spectral component of $\boldsymbol{x}$, followed by fully connected layers that combine the spectral features with the scalar observables (count rates and \textit{LISA} SNR). The convolutional architecture enables the network to exploit local correlations and smooth structure in the continuum spectrum, which are not efficiently captured by a purely fully connected network. The resulting latent representation $\boldsymbol{h}$ provides a compact and informative summary of the multi-messenger data and is used as conditioning input to the normalizing flow. This architecture significantly improves the ability of the model to extract physically relevant features from the spectrum and reduces degeneracies in the inferred parameters. 
The embedding network consists of multiple one-dimensional convolutional layers with nonlinear activations, followed by fully connected layers that map to a latent feature vector of fixed dimension. The conditional normalizing flow itself is implemented using NSFs with multiple coupling layers and hidden layers in each transformation network, as provided by the \texttt{sbi} framework \cite{2021JOSS....6.2505T}. Hyperparameters such as the number of flow transformations, hidden units, and embedding dimensionality are chosen to balance flexibility and computational efficiency. 

The embedding network consists of two branches and proceeds in two stages. In the first stage, modality-specific feature extraction is performed. A scalar branch processes the three non-spectral observables $(\log_{10}C_{\rm X},\,\log_{10}C_{\rm UV},\,\log(1+\mathrm{SNR}_{\rm LISA}))$ using a two-layer multilayer perceptron (i.e., $3 \rightarrow 64 \rightarrow 32$) with Sigmoid Linear Unit (SiLU) activations. In parallel, the spectral branch processes the 128-bin continuum spectrum using a one-dimensional convolutional neural net with three convolutional layers (i.e., $1 \rightarrow 32 \rightarrow 64 \rightarrow 64$ channels) and kernel sizes of 9, 5, and 3, respectively, each followed by SiLU activations. A global average pooling layer then reduces the spectral features to a 64-dimensional representation. These intermediate representations form distinct latent spaces corresponding to spectral (64-dimensional) and scalar (32-dimensional) information.

\begin{figure*}
\centering
\includegraphics[width=\textwidth]{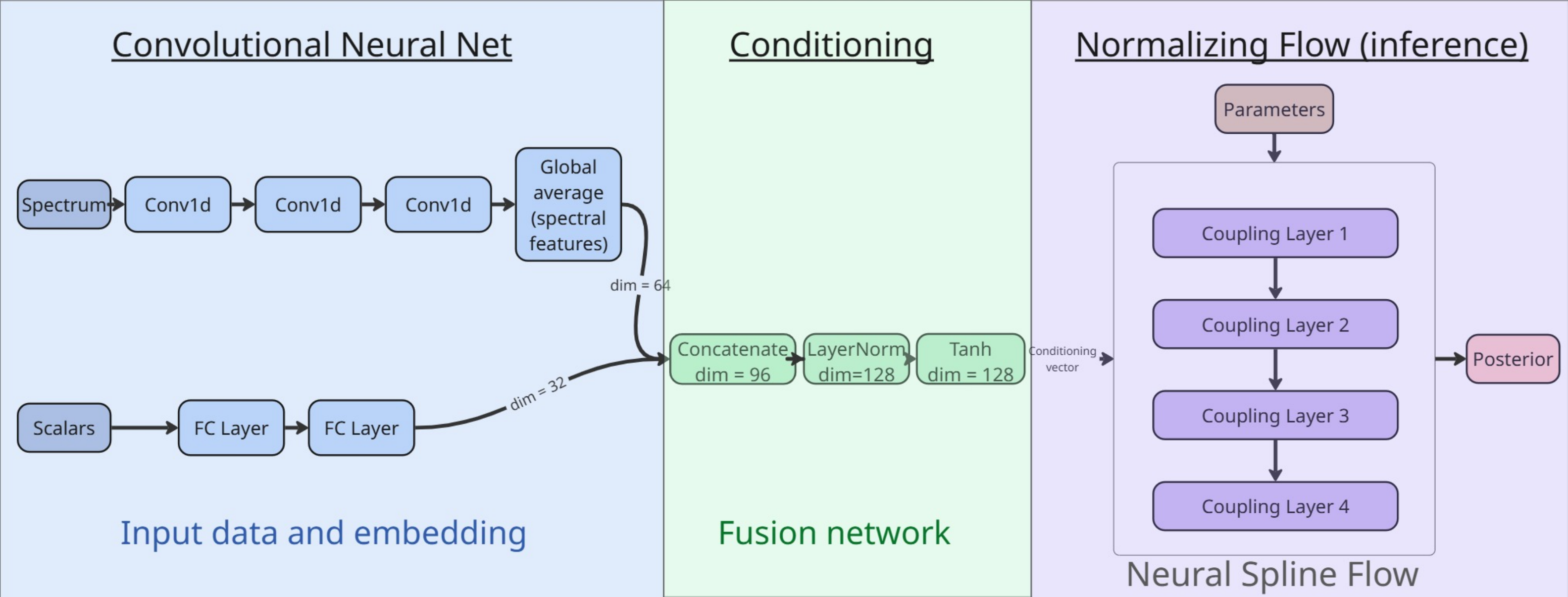}
\caption{Architecture of the conditional normalizing flow used for
simulation-based inference of AM CVn binary parameters
$\boldsymbol{\theta} = (M_1, M_2, P_{\rm orb}, i, d, N_{\rm H})$
from the multi-messenger observation vector $\boldsymbol{x}$
(Eqs.~\ref{Eq:ParameterVector}--\ref{Eq:ObsVector}).
The network proceeds in three stages.
\textit{Left (Convolutional Neural Net):} The 131-dimensional observation
vector is processed by a two-branch embedding network $g_\psi$.
The spectral branch ingests the 128-bin continuum spectrum
$\{\log_{10}(EF_E)_k\}$ through three successive one-dimensional
convolutional layers ($1\!\to\!32\!\to\!64\!\to\!64$ channels, kernel
sizes 9, 5, and 3) with SiLU activations, followed by global average
pooling to produce a 64-dimensional spectral feature vector.
In parallel, the scalar branch processes the three non-spectral
observables $(\log_{10}C_{\rm X},\,\log_{10}C_{\rm UV},\,
\log[1+\rho_{\rm LISA}])$ through a two-layer multilayer perceptron
($3\!\to\!64\!\to\!32$) with SiLU activations, producing a
32-dimensional scalar feature vector.
\textit{Center (Conditioning):} The two feature vectors are
concatenated into a 96-dimensional representation, which is passed
through two fully connected layers ($96\!\to\!128\!\to\!128$).
A layer normalization and SiLU activation follow the first layer,
and a $\tanh$ activation follows the second, yielding the final
128-dimensional conditioning vector
$\boldsymbol{h} = g_\psi(\boldsymbol{x}) \in \mathbb{R}^{128}$.
\textit{Right (Normalizing Flow (inference)):} The conditioning vector
$\boldsymbol{h}$ is passed to a neural spline flow (NSF) comprising
four sequential coupling layers implementing monotonic
rational-quadratic spline transformations \citep{2019NeurIPS..32.7503D},
as provided by the \texttt{sbi} framework. During training, the NSF transforms samples of $\boldsymbol{\theta}$
drawn from the prior into the latent Gaussian base distribution;
during inference, the inverse transformation maps latent samples to
posterior samples $q_\phi(\boldsymbol{\theta}|\boldsymbol{x})$
conditioned on the observed multi-messenger data.
} \label{F:FlowDiagram}
\end{figure*}

In the second stage, the scalar and spectral feature vectors are concatenated to form a 96-dimensional representation, which is passed through a fusion network consisting of two fully connected layers with hidden width 128 (96 $\rightarrow$ 128 $\rightarrow$ $d_{\rm embed}$), where we set $d_{\rm embed}=128$. The first layer is followed by a normalizing layer and a SiLU activation, and the final layer is followed by a $\tanh$ activation, producing the final conditioning vector $\boldsymbol{h} \in \mathbb{R}^{128}$. 
Thus, the dimensionality evolves as $128 \rightarrow 64$ for the spectral branch, $3 \rightarrow 32$ for the scalar branch, and $(64+32)=96 \rightarrow 128$ for the fused representation, which serves as the conditioning input to the normalizing flow. 
This architecture is summarized in the diagram of Figure~\ref{F:FlowDiagram}

During development, we explored multiple flow architectures and training configurations. Initial experiments with real-valued non-volume preserving flows were unable to adequately capture the complexity of the posterior distributions. We subsequently adopted neural spline flows, which provided improved flexibility but still exhibited deficiencies in calibration diagnostics such as probability--probability (PP) plots. These limitations motivated two key improvements: the introduction of a convolutional embedding network to better extract information from the spectral data, and the use of a joint prior over $(M_1, M_2)$ to more accurately reflect the physical parameter space. These modifications significantly improved the quality and calibration of the inferred posteriors. Future extensions of this framework will incorporate measurement uncertainties from \textit{LISA} parameter estimation and treat the EM spectrum and GW signal as time-dependent observables across the orbital evolution. This will provide additional constraints on the binary parameters, particularly in breaking degeneracies between $M_1$ and $M_2$, and will further improve the fidelity of the inferred posteriors.

\subsection{Training procedure}

The conditional density estimator is trained using neural posterior estimation (NPE) \cite{2018arXiv180501683P}. The objective function is the conditional log-likelihood,

\begin{equation}
\mathcal{L}(\phi)
=
\sum_{i=1}^{N}
\log
q_\phi
(
\boldsymbol{\theta}_i
|
\boldsymbol{x}_i
).
\end{equation}
where the parameter vector $\mathbf{\theta}$ is defined as in Eq.~(\ref{Eq:ParameterVector}) and the observation vector is defined as in Eq.~(\ref{Eq:ObsVector}). 
The forward model for multi-messenger observables defines the deterministic mapping
\begin{equation}
\boldsymbol{x} = f(\boldsymbol{\theta})\,,
\end{equation}
while the normalizing flow learns the inverse conditional distribution
\begin{equation}
q_\phi(\boldsymbol{\theta}|\boldsymbol{x}) \approx p(\boldsymbol{\theta}|\boldsymbol{x})\,.
\end{equation}

We obtain the training dataset by sampling binary parameters $\boldsymbol{\theta}$ from a prior distribution and generating corresponding observables using the multi-messenger model. 
The training dataset is constructed by sampling $\boldsymbol{\theta}$ from broad, uninformative priors covering the physically relevant AM CVn parameter space:
\begin{align}\label{Eq:Priors}
M_1 &\in [0.1, 1.0]\Msol, \nonumber\\
M_2 &\in [0.01, 0.2]\Msol, \nonumber\\
P_{\rm orb} &\in [200, 5000]~{\rm s}, \nonumber\\
i &\in [5^\circ, 85^\circ], \nonumber\\
d &\in [10^3, 1.2\times10^4]~{\rm pc}, \nonumber\\
N_{\rm H} &\in [10^{20}, 10^{22}]~{\rm cm}^{-2}.
\end{align}

The prior is defined as a uniform distribution in logarithmic parameter space,
\begin{equation}
\log \boldsymbol{\theta}
\sim
\mathcal{U}
(\log \boldsymbol{\theta}_{\rm min},
\log \boldsymbol{\theta}_{\rm max}),
\end{equation}
with bounds from Eq.~(\ref{Eq:Priors}).  
The inclination angle is sampled uniformly in $\cos i$ to ensure an isotropic orientation distribution. Rather than enforcing the constraint $M_2 < M_1$ through rejection sampling, we adopt a joint prior over $(M_1, M_2)$ that explicitly encodes this physical condition. Specifically, the donor mass is sampled conditionally given the accretor mass such that $M_2 < M_1$ is always satisfied. This joint conditional prior improves sampling efficiency and ensures that the training distribution more accurately reflects the physically allowed parameter space of AM CVn systems. 
This procedure produces a simulated dataset
$\mathcal{D} = \left\{ (\boldsymbol{\theta}_i,\boldsymbol{x}_i)
\right\}_{i=1}^{N}\,$, where we set the number of samples to be $N = 5 \times 10^5$, which is used for training the conditional density estimator. 
A detailed investigation of prior sensitivity and SBI calibration under alternative training distributions is deferred to future work.

Training is performed using stochastic gradient descent with mini-batches drawn from the simulated dataset, with early stopping based on validation loss to prevent overfitting. 
Once trained, posterior samples are generated directly by drawing latent variables $\boldsymbol{z}$ and applying the learned transformation. This direct sampling approach enables rapid inference across large populations of systems. 
We split the simulated dataset into training and validation subsets, and monitor the validation loss to determine convergence and implement early stopping. 

To assess posterior calibration in training, we perform simulation-based calibration (SBC) using posterior rank statistics computed from simulated test systems. For each test case $k$, we draw posterior samples
\begin{equation}
\left\{
\boldsymbol{\theta}^{(n)}_k
\right\}_{n=1}^{N_{\rm post}}
\sim
q_\phi(\boldsymbol{\theta}|\boldsymbol{x}_k),
\end{equation}
where $\boldsymbol{\theta}^{\rm true}_k$ denotes the true parameter vector used to generate the simulated observation $\boldsymbol{x}_k$. For each parameter $\theta_j$, we compute the posterior rank statistic, 
\begin{equation}
r_{k,j}
=
\frac{1}{N_{\rm post}}
\sum_{n=1}^{N_{\rm post}}
\mathbf{1}
\left[
\theta^{(n)}_{k,j}
<
\theta^{\rm true}_{k,j}
\right],
\end{equation}
where $\mathbf{1}[\cdot]$ is the indicator function. For a well-calibrated posterior estimator, the rank statistics should be uniformly distributed on the interval $[0,1]$. We therefore construct PP plots by comparing the empirical cumulative distribution function
\begin{equation}
\hat{F}_j(x)
=
\frac{1}{N_{\rm test}}
\sum_{k=1}^{N_{\rm test}}
\mathbf{1}
\left[
r_{k,j} \leq x
\right]
\end{equation}
to the ideal relation
\begin{equation}
\hat{F}_j(x)=x.
\end{equation}
Systematic deviations from the diagonal relation indicate posterior miscalibration. 
Throughout this work we set $N_{\rm tets} = N_{\rm post} = 500$ for each. In addition, we examine derived quantities such as the chirp mass to evaluate how well the learned posterior captures physically relevant parameter combinations. Finally, we compare the training prior to the effective learned distribution using samples from the trained flow to ensure that the model has not introduced unintended biases. These diagnostics provide a comprehensive validation of both the accuracy and calibration of the inference framework.

\subsection{Testing procedure}

To assess the astrophysical performance of the trained conditional normalizing flow, we evaluate the model on the population of AM CVn binaries generated using the \texttt{COSMIC} binary population synthesis framework presented in Section~\ref{Sec:Cosmic}. This provides an independent validation dataset drawn from a physically motivated generative model not used during training. 

\texttt{COSMIC} directly provides intrinsic parameters $(M_1, M_2, P_{\rm orb})$ for each system, and we assign the extrinsic parameters $(i, d, N_{\rm H})$ by sampling from the same prior distributions used during training. In particular, the inclination is drawn uniformly in $\cos i$, and the distance and column density are sampled log-uniformly within the bounds of Eq.~(\ref{Eq:Priors}). The intrinsic masses $(M_1, M_2)$ are taken directly from the \texttt{COSMIC} population, which is subsequently filtered to ensure consistency with the joint conditional mass prior used during training, i.e., enforcing $M_2 < M_1$ as part of the physically allowed parameter space. This produces the full parameter vector in Eq.~(\ref{Eq:ParameterVector}) and ensures consistency between training and testing distributions while preserving intrinsic correlations from \texttt{COSMIC}.

For each \texttt{COSMIC} system, we compute the corresponding multi-messenger observables using the forward model, i.e.,
$\boldsymbol{x}_{\rm COSMIC}
=
f(\boldsymbol{\theta}_{\rm COSMIC})\,$,
where $\boldsymbol{\theta}_{\rm COSMIC}$ denotes the full parameter vector including both COSMIC-derived intrinsic parameters and sampled extrinsic parameters. 
The observation vector $\boldsymbol{x}_{\rm COSMIC}$ is identical in structure as Eq.~(\ref{Eq:ObsVector}) and contains \textit{AXIS} and \textit{CASTOR} count rates, the intrinsic continuum spectrum, and the \textit{LISA} signal-to-noise ratio. This ensures that the trained normalizing flow is applied within its learned data representation. 
In particular, this representation matches the input structure expected by the convolutional embedding network described in Section~\ref{Sec:NormalFlows}, ensuring that spectral features are processed consistently between training and testing.

To ensure that inference is performed within the domain of the trained model, we restrict the \texttt{COSMIC} population to systems satisfying the training prior bounds,
\begin{equation}
\boldsymbol{\theta}_{\rm min}
\le
\boldsymbol{\theta}
\le
\boldsymbol{\theta}_{\rm max},
\end{equation}
together with the physically motivated constraint $M_2 < M_1$ implied by the joint conditional mass prior adopted during training. This filtering step ensures that all evaluated systems lie within the support of the learned posterior and avoids extrapolation into regions of parameter space not represented in the training data.

We consider two testing scenarios by varying the upper bound on the orbital period prior:
\begin{itemize}
\item Case 1 (full prior test): the same prior upper bound as used during training, i.e., 
\begin{equation}
P_{\rm orb} \in [200, 5000]~{\rm s}
\end{equation}
\item Case 2 (reduced prior test): the same trained flow is evaluated on a restricted subset of the \texttt{COSMIC} population with 
\begin{equation}
P_{\rm orb} \in [200, 2000]~{\rm s}
\end{equation}
\end{itemize}
All other parameter prior bounds remain unchanged. 

The two testing scenarios serve complementary purposes in evaluating the robustness and astrophysical applicability of the inference framework. The full prior test assesses the global performance of the trained normalizing flow across the entire parameter space on which it was trained, providing a baseline measure of accuracy and calibration when the training and testing domains are matched. In contrast, the reduced prior test isolates the short-period AM CVn population, which is of particular observational relevance for both EM surveys and \textit{LISA} detections. This regime is characterized by higher mass transfer rates, stronger GW signals, and more pronounced spectral features, and therefore provides a stringent test of whether the trained model can generalize to a physically distinct subpopulation without retraining. Together, these two cases quantify both interpolation performance within the training domain and the degree of robustness when restricting to an astrophysically motivated subset of the population. 

For each \texttt{COSMIC} system, we evaluate the learned posterior
\begin{equation}
q_\phi(\boldsymbol{\theta}|\boldsymbol{x}_{\rm COSMIC}),
\end{equation}
where $\boldsymbol{x}_{\rm COSMIC}$ contains the computed EM and GW observables. 
For $5{,}000$ test binaries from the \texttt{COSMIC} population, we draw $N_{\rm post} = 1{,}000$ posterior samples per system and compute summary statistics including posterior means, standard deviations, and credible regions $\boldsymbol{\theta}_{\rm lo}^{(\alpha)}, 
\boldsymbol{\theta}_{\rm hi}^{(\alpha)}\,$, 
with $\alpha \in \{0.68,0.90,0.95\}$. 

We assess the performance of the trained model using three complementary diagnostics. 
First, we evaluate parameter recovery on a system-by-system basis by comparing the posterior mean of each parameter to its true value from the \texttt{COSMIC} population. This quantifies both accuracy and dispersion, and we additionally compute summary statistics such as the Spearman rank correlation coefficient and mean absolute fractional error to characterize the inference across the population.

Second, we assess posterior calibration using empirical coverage fractions. For each parameter and each credible interval level $\alpha \in \{0.50, 0.68, 0.90, 0.95\}$, we compute
\begin{equation}
f_{\rm cov}(\alpha)
=
\frac{1}{N}
\sum_{i=1}^{N}
\mathbf{1}
\left(
\theta_{i,\rm true}
\in
[\theta_{i,\rm lo}^{(\alpha)},\theta_{i,\rm hi}^{(\alpha)}]
\right),
\end{equation}
and compare the empirical coverage $f_{\rm cov}$ to the nominal level $\alpha$. A well-calibrated posterior yields $f_{\rm cov} \approx \alpha$ across all parameters, while systematic deviations indicate over- or under-confidence in the inferred uncertainties. 

Finally, we evaluate population-level consistency by comparing the distribution of inferred posterior means to the true parameter distributions of the \texttt{COSMIC} sample. For each parameter, we construct normalized histograms of the true values and the inferred posterior means across all systems. This diagnostic tests whether the inference framework preserves the underlying population structure, and is sensitive to systematic biases or distortions introduced by the learned inverse mapping. 
Together, these diagnostics provide a comprehensive assessment of the inference performance, including accuracy, calibration, and the ability to recover population-level properties from multi-messenger observations.

\section{Results}
\label{Sec:Results}

\subsection{The continuum spectra of AM CVn binaries}

In this section we present the forward-model predictions for the coupled GW and EM observables of AM CVn systems. Each figure isolates a specific physical linkage in the model, and every trend shown below arises self-consistently from the GW--driven mass-transfer solution (Eqs.~\ref{Eq:Jdot}--\ref{Eq:Mdot}) together with the energetically normalized emission prescriptions described in Section~IIE. No phenomenological tuning is introduced at the level of observables.

\subsubsection{Mass transfer and spectral models}

\begin{figure*}
\centering
\includegraphics[width=\textwidth]{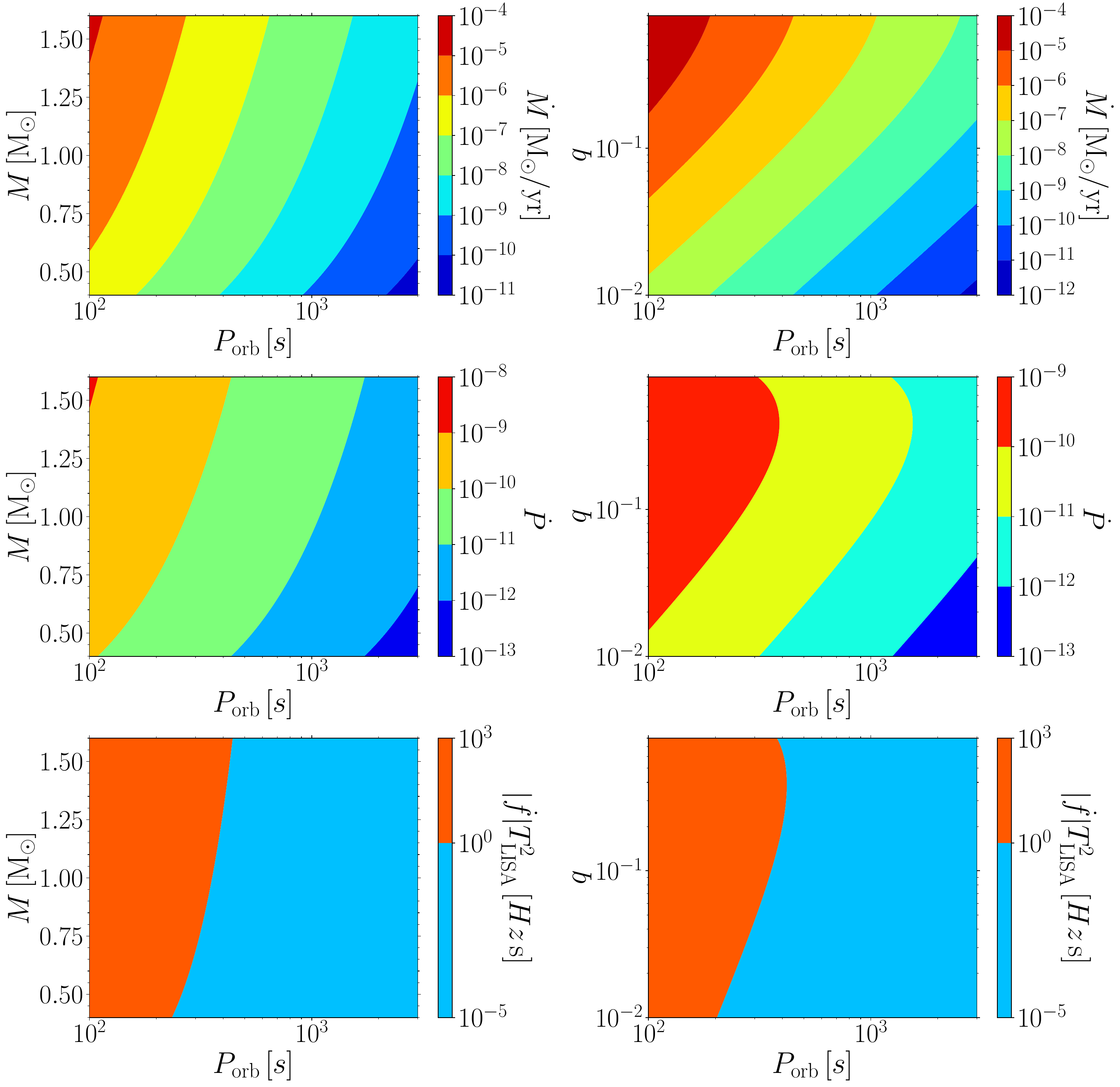}
\caption{Contours of the AM CVn mass transfer rate $\dot{M}$ (Eq.~(\ref{Eq:Mdot})), orbital period rate $\dot{P}$ (Eq.~(\ref{Eq:Pdot})), and monochromatic threshold $|\dot f|T_{\rm LISA}^2$ (Eq.~(\ref{Eq:Monochromaticity}) with $T_{\rm LISA} = 1$) as functions of the orbital period $P_{\rm orb}$ and of the total mass $M$ for the left-hand column (with $q = 0.1$) or of the mass ratio $q$ (with $M = 1\Msol$) for the right-column panels. 
} \label{F:MdotPdotFdot}
\end{figure*}

The forward model computes EM spectra over restricted ultraviolet and X-ray energy ranges chosen to be both observationally relevant and physically consistent with the adopted assumptions. In the X-ray band, spectra are evaluated over $0.1$--$10$~keV, matching the nominal sensitivity of \textit{AXIS} and encompassing the energies where optically thin boundary-layer emission and photoelectric absorption dominate. In the ultraviolet, spectra are computed over $2.0$--$8.27$~eV ($\sim620$--$150$~nm), corresponding to the \textit{CASTOR} bandpasses and capturing the long-wavelength tail of the disk and WD emission.

The model does not compute spectra in the extreme UV regime. Although AM CVn systems likely emit significant intrinsic power at these energies, interstellar absorption strongly suppresses the emergent flux, and the adopted continuum prescriptions are not physically valid in that regime. The UV/optical component is implemented as a bolometric redistribution of accretion power using optically thick continuum approximations rather than detailed atmosphere calculations. Consequently, helium edges, line blanketing, non-LTE effects, vertical disk structure, and reprocessing layers are not included. Evaluating the model beyond $\sim40$--$50$~eV would therefore violate its internal assumptions. Restricting the energy range ensures that all predicted observables remain physically consistent with the adopted emission physics. 

Figure~\ref{F:MdotPdotFdot} shows the basic output of the model for the GW-driven AM CVn mass transfer. The horizontal axis in all panels is $P_{\rm orb}$, while the vertical axes in the left and right hand columns of panels correspond to the total mass $M = M_1 + M_2$ with the mass ratio $q = M_2/M_1$ fixed to $q=0.1$, and to the mass ratio $q$ with the total mass fixed to $M = 1\Msol$, respectively. 
The three rows of panels are colored by (top to bottom) the mass-transfer rate $\dot{M}$, the orbital period derivative $\dot{P}_{\rm orb}$, and the monochromaticity criterion $|\dot{f}|T^2_{\rm LISA}$. 

For the panels in the top row of Fig.~\ref{F:MdotPdotFdot}, short-period systems sustain the highest mass transfer rates. The monotonic decline of $\dot{M}$ with increasing $P_{\rm orb}$ follows directly from GW angular-momentum losses (Eq.~(\ref{Eq:Jdot})) coupled to the Roche-lobe filling condition and the degenerate donor mass--radius relation through Eq.~(\ref{Eq:Mdot}). Since $\dot{J}_{\rm GW}/J \propto a^{-4}$ and $P_{\rm orb}\propto a^{3/2}$ from Kepler’s law (Eq.~(\ref{Eq:Keplers3rd})), the model predicts a steep decrease of $\dot{M}$ toward longer orbital periods, with the precise scaling set by the denominator $(\zeta_2-\zeta_L)$ in Eq.~(\ref{Eq:Mdot}). 
Systems with larger $M$ attain systematically higher mass-transfer rates at fixed $q$ because increasing $M$ raises both component masses while preserving the same mass ratio, which strengthens the GW torque in Eq.~(\ref{Eq:Jdot}). Since $a$ at fixed $P_{\rm orb}$ scales only weakly as $(M_1+M_2)^{1/3}$ from Eq.~(\ref{Eq:Keplers3rd}), the dominant effect is the increase of the mass term $M_1M_2(M_1+M_2)$ in Eq.~(\ref{Eq:Jdot}), giving larger $|\dot{J}_{\rm GW}/J|$ and therefore larger $\dot{M}$ through Eq.~(\ref{Eq:Mdot}). 
Similarly, at fixed total mass, increasing $q$ increases the donor mass $M_2$ and moves the system toward more equal masses, which increases the product $M_1M_2$ entering Eq.~(\ref{Eq:Jdot}) and also increases the normalization factor $M_2$ in Eq.~(\ref{Eq:Mdot}). Both effects act in the same direction, producing the monotonic rise of $\dot{M}$ with increasing $q$. Thus the upper-right panel shows the highest transfer rates in short-period, high-$q$ systems. 

The middle row shows the orbital period derivative computed from Eq.~(\ref{Eq:Pdot}). As in the top row, the strongest evolution occurs at short orbital periods where gravitational radiation is most efficient. For the detached inspiral term alone, $\dot{P}_{\rm orb}$ would be negative, but during stable mass transfer the second term in Eq.~(\ref{Eq:Pdot}),
\begin{equation}
3\dot M_2\left(\frac{1}{M_1}-\frac{1}{M_2}\right)\,,\notag
\end{equation}
acts in the opposite direction because $\dot M_2<0$ and $M_2<M_1$, driving orbital expansion and yielding the positive $\dot{P}_{\rm orb}$ values characteristic of AM CVn systems after contact. 
At fixed $q$, increasing the total mass increases both $|\dot{J}_{\rm GW}/J|$ and $|\dot M_2|$, so both terms in Eq.~(\ref{Eq:Pdot}) increase in magnitude producing the monotonic increase of $\dot{P}_{\rm orb}$ with $M$ seen in the left middle panel. 
At fixed total mass, the dependence on $q$ is complicated because the two terms in Eq.~(\ref{Eq:Pdot}) respond differently. Increasing $q$ strengthens GW losses through the larger $M_1M_2$ factor in Eq.~(\ref{Eq:Jdot}), but it also reduces the factor $(1/M_1-1/M_2)$ as the masses approach equality. Increasing through low and moderate values of $q$, the larger $\dot M_2$ dominates and $\dot{P}_{\rm orb}$ increases slightly, while at high $q$ the decreasing $(1/M_1-1/M_2)$ term partially offsets this effect, producing the weak non-monotonic flattening near the upper edge of the right-hand panel. 

The bottom row shows the monochromaticity criterion of Eq.~(\ref{Eq:Monochromaticity}), evaluated using Eq.~(\ref{Eq:fdot}) for a $T_{\rm LISA}=1\,{\rm yr}$ observation. Systems satisfying $|\dot f|T_{\rm LISA}^2<1$ are effectively monochromatic during the mission lifetime, while systems above this boundary exhibit measurable frequency evolution across multiple frequency bins. Eq.'s~(\ref{Eq:fdot}) imply that the strongest chirping occurs for short-period systems where both $f_{\rm GW}$ and $\dot P_{\rm orb}$ are largest. Consequently, the low-$P_{\rm orb}$ region of each panel lies above the monochromatic threshold, while long-period systems are predominantly monochromatic. 
The transition contour defined by $|\dot f|T_{\rm LISA}^2=1$ occurs near $P_{\rm orb}\sim300\,{\rm s}$ in both panels as the dominant scaling is set primarily by the strong dependence on orbital period rather than on the binary masses. Variations in total mass or mass ratio shift the monochromatic boundary weakly compared to the period dependence, implying that the boundary remains approximately vertical and centered near the same characteristic orbital period in both columns. This reflects the fact that AM CVn binaries with $P_{\rm orb}\lesssim300\,{\rm s}$ are generally evolving rapidly enough to be resolved as chirping \textit{LISA} sources. 

\begin{figure}
\centering
\includegraphics[width=0.49\textwidth]{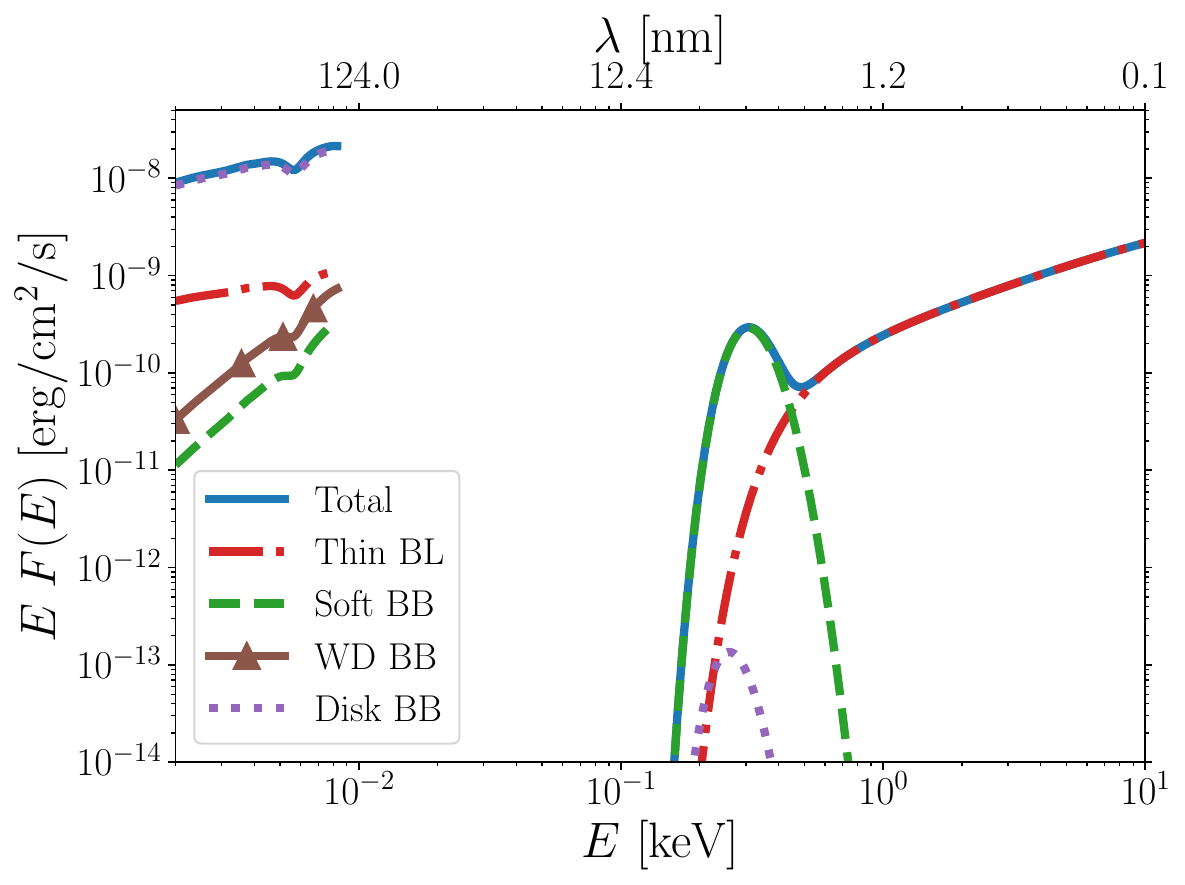}
\caption{The optical, UV and X-ray SEDs of the emission components in our model for the continuum spectrum of a single AM CVn binary with $M_1=0.8\Msol$, $M_2=0.1\Msol$, $P_{\rm orb}=500$ s, $d=100$ pc, and $N_{\rm H} = 10^{21}$ cm$^{-2}$ in the persistent-disk state. The total absorbed spectrum (blue solid line) is composed of thin boundary layer (red dash-dot line), soft blackbody (green dashed line), and disk blackbody (dotted purple line) emission. In the optical/UV, there is also a contribution from the WD blackbody component (brown line with triangles). } \label{F:SingleSpectrumComponents}
\end{figure}

Next, we show the continuum spectral energy distributions (SEDs) produced by the forward model, expressed in terms of the quantity $E\,F(E)$ as a function of photon energy $E$ in the ultraviolet (UV) and X-ray regimes. As described in Sec.~\ref{SubSec:SpectralModeling}, each spectral component is constructed as a normalized shape function and scaled such that its bolometric integral reproduces the luminosity assigned by the accretion energetics (Eq.~\ref{Eq:Lacc}), after which inclination and absorption effects are applied. The use of $E\,F(E)$ emphasizes the contribution to the total radiated power per logarithmic energy interval and provides a direct visualization of the dominant emission components across the spectrum.

Figure~\ref{F:SingleSpectrumComponents} shows the decomposition of the continuum spectrum for a representative system into its constituent emission components, separately for the X-ray (AXIS) and ultraviolet/optical (CASTOR) bands, along with the combined total spectrum. In optical/UV, the spectrum is dominated by the multi-temperature disk component described by Eq.~(\ref{Eq:MultiDisk3}), which exhibits the characteristic $E^{1/3}$ scaling at low energies and peaks at energies set by the inner disk temperature (Eq.~\ref{Eq:MultiDisk2}). The heated WD surface contributes a secondary blackbody component at lower temperatures, and the optically thin BL (bremsstrahlung) and optically thick BL (Soft BB in this accretion state) components are subdominant. 

In contrast, in the X-ray band the hierarchy of components is reversed. The dominant contribution arises from the optically thick boundary-layer emission, modeled as a blackbody with temperature set by the Stefan--Boltzmann relation and the covering factor $f_{\rm BL}$. This component produces a prominent soft X-ray peak that is sensitive to the orbital period. At higher energies, the optically thin boundary-layer emission, described by a bremsstrahlung spectrum with characteristic temperature near the virial value, contributes a hard X-ray tail with an exponential cutoff. The disk component contributes only weakly in the X-ray band, appearing as a low-level extension of its high-energy tail, consistent with the analytic form of Eq.~(\ref{Eq:MultiDisk3}). This illustrates explicitly the band-dependent hierarchy of emission components: disk-dominated emission in the optical/UV and boundary-layer-dominated emission in the X-ray regime, consistent with the physical partitioning of accretion power described in Sec.~\ref{subsubSec:DiskComponents}.

\begin{figure}
\centering
\includegraphics[width=0.49\textwidth]{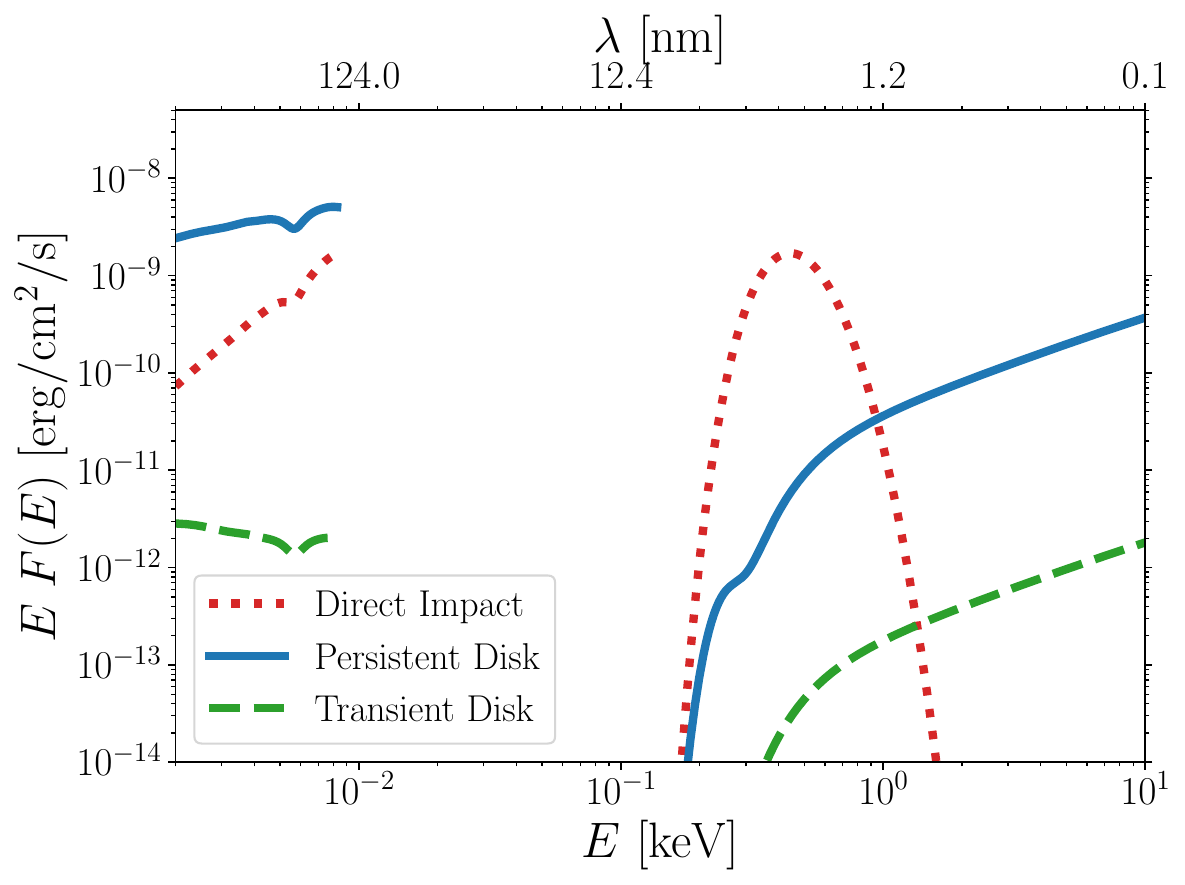}
\caption{Three representative systems in the observational states possible in our AM CVn continuum spectrum model. The red dotted line (Direct Impact) assumes $M_1=0.5\Msol$, $M_2=0.25\Msol$, $P_{\rm orb}=400$ s, the blue solid line (Persistent Disk) assumes $M_1=0.8\Msol$, $M_2=0.1\Msol$, $P_{\rm orb}=1000$ s, the green dashed line (Transient Disk) assumes $M_1=0.8\Msol$, $M_2=0.01\Msol$, $P_{\rm orb}=4750$ s, and all three systems have $d=100$ pc. 
} \label{F:SpectralClasses}
\end{figure}

Figure~\ref{F:SpectralClasses} shows the continuum spectra for three systems that differ in orbital period and mass, and consequently in mass-transfer rate $\dot{M}$, illustrating the transition between the three accretion regimes: direct-impact, persistent disk, and transient disk. 
At the shortest orbital period, corresponding to the highest $\dot{M}$ and smallest binary separation, the system satisfies the direct-impact condition (i.e., the circularization radius lies within the WD radius), and no disk forms. In this regime, the spectrum is dominated by the localized hotspot emission described in Sec.~\ref{subsubSec:DiskComponents}, with luminosity $L_{\rm soft} = \epsilon_{\rm rad} L_{\rm acc}$. The small emitting area $A_{\rm DI} \propto f_{\rm DI}$ leads to a high effective temperature, producing a soft X-ray--peaked spectrum. The strong X-ray emission shown in this case in Fig.~\ref{F:SpectralClasses} (red dotted line) is due to the large accretor mass of $M_1 = 0.8\Msol$. 

At intermediate orbital periods, the system enters the persistent disk regime, where $\dot{M} > \dot{M}_{\rm crit}$. In this case, the full accretion luminosity is partitioned equally between the disk and boundary layer. The boundary layer is predominantly optically thick, so that $L_{\rm soft} \approx L_{\rm BL}$, producing a soft X-ray component, while the disk contributes a strong optical/UV continuum. The resulting spectrum thus exhibits two prominent features: an optical/UV peak from the disk and a soft or hard X-ray peak (depending on the accretor mass) from the boundary layer. 

At the longest orbital periods, the system transitions to the transient disk regime with $\dot{M} < \dot{M}_{\rm crit}$. In this regime, the boundary layer becomes optically thin and a significant fraction $f_{\rm hard}$ of $L_{\rm BL}$ is emitted as hard X-rays. Consequently, the soft blackbody component is suppressed, and the spectrum is dominated by bremsstrahlung emission extending to higher energies. The disk, now cooler due to the lower $\dot{M}$ (Eq.~\ref{Eq:MultiDisk2}), shifts its emission to lower energies, reducing the optical/UV flux.  
The systematic evolution of spectral shape across these three cases therefore reflects the underlying dependence of the spectral model on $\dot{M}$ (and hence the masses and orbital period).

\begin{figure}
\centering
\includegraphics[width=0.49\textwidth]{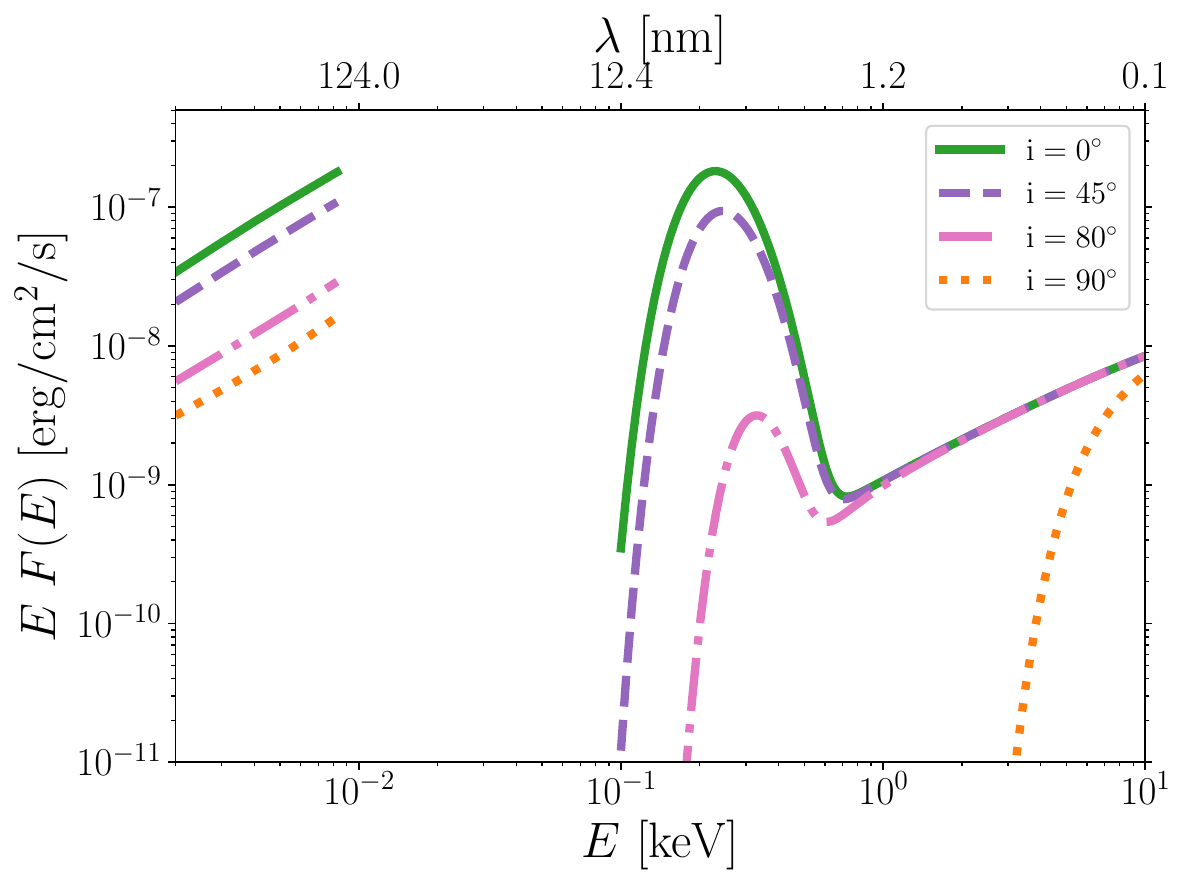}
\caption{The total absorbed SED of an AM CVn system with $M_1=0.8\Msol$, $M_2=0.1\Msol$, $P_{\rm orb}=300$ s, $d=100$ pc, and $N_H = \num{1e19}$ cm$^{-2}$ and for several inclinations: $i=0^{\rm o}$ (green solid line), $i=45^{\rm o}$ (purple dashed line), $i=80^{\rm o}$ (pink dash-dot line), and $i=90^{\rm o}$ (orange dotted line). 
} \label{F:SingleSpectrumInclination}
\end{figure}

\begin{figure}
\centering
\includegraphics[width=0.49\textwidth]{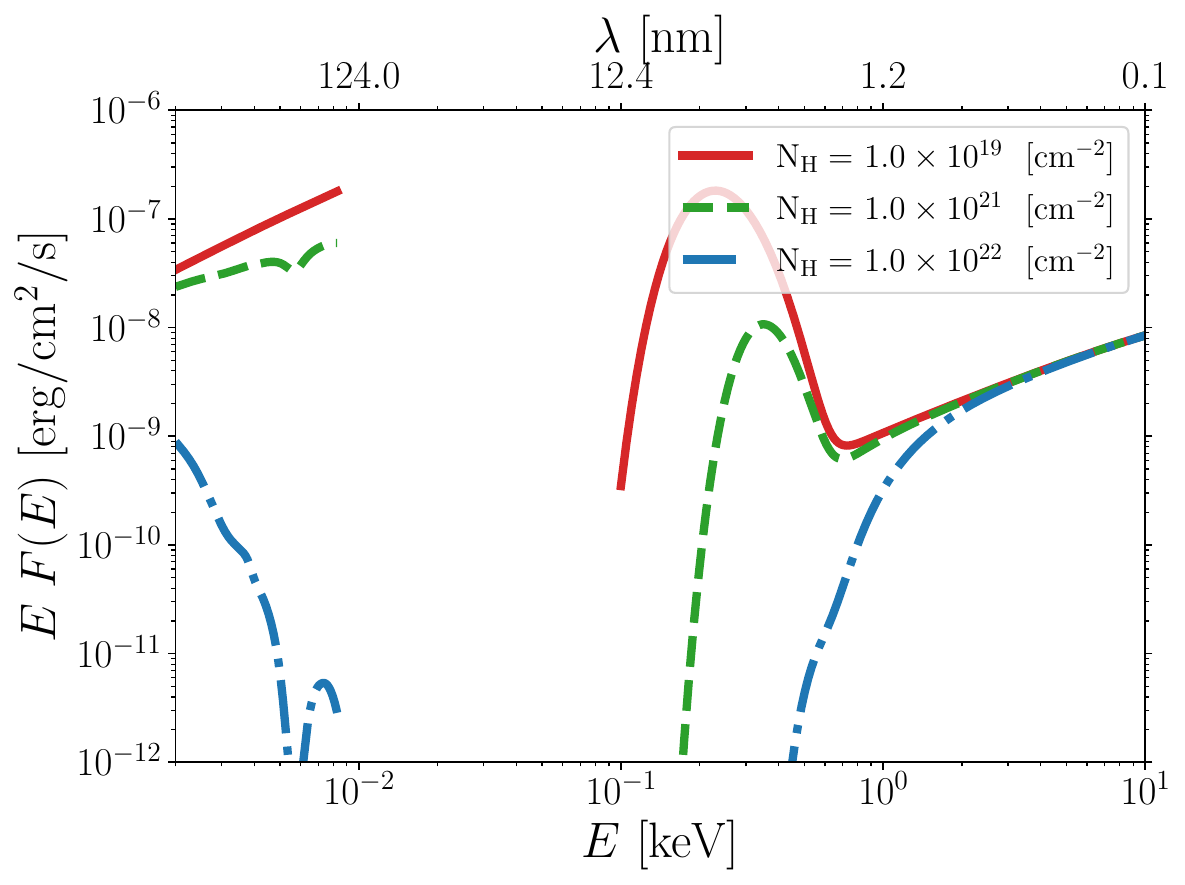}
\caption{The total absorbed SED of an AM CVn system with $M_1=0.8\Msol$, $M_2=0.1\Msol$, $P_{\rm orb}=300$ s, $d=100$ pc, and $i = 0^{\rm o}$ and for several column densities: $N_H= \num{1e19}$ (red solid line), $N_H= \num{1e21}$ (green dashed line), and $N_H= \num{1e22}$ (blue dash-dot line). 
} \label{F:SingleSpectrumNH}
\end{figure}

\begin{figure}
\centering
\includegraphics[width=0.49\textwidth]{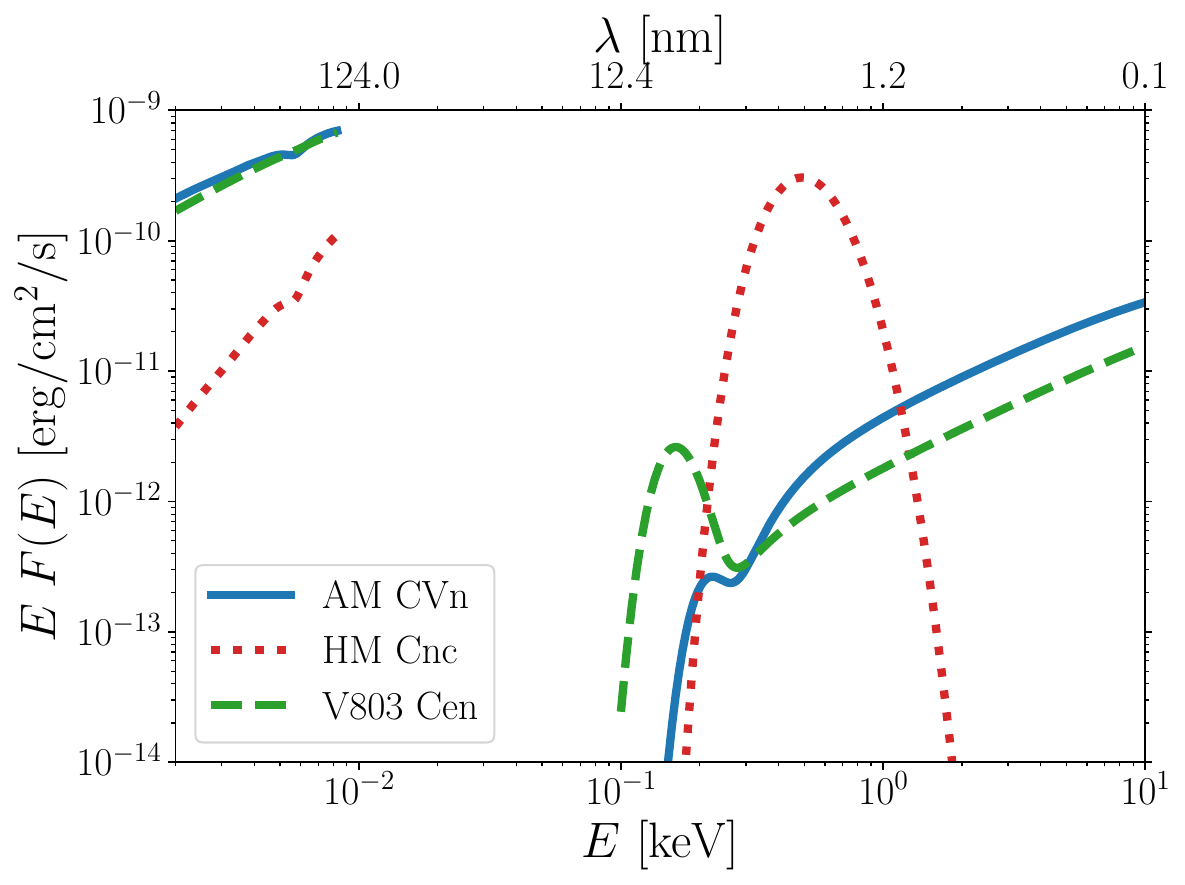}
\caption{The total absorbed SEDs of three systems, the AM CVn prototype binary, HM Cnc, and V803 Cen, using their observed parameters given in Table~\ref{Tab:ObservedBinaries}. 
} \label{F:ObservedBinaries}
\end{figure}

\begin{table*}[th!]
\centering
\renewcommand{\arraystretch}{1.8} 
\setlength{\tabcolsep}{15pt} 
\caption{Representative parameters for three illustrative example AM CVn targeted as VBs for LISA. The values are taken from Table 1 of Finch {\it et al} \cite{2023MNRAS.522.5358F}. The orbital period $P_{\rm orb}$ is in seconds, inclination $i$ and sky location in degrees, distance $d$ in parsecs, and masses $M_1$ and $M_2$ in solar mass. Right ascension and declination are given in the ICRS frame.
}\label{Tab:ObservedBinaries}
\begin{tabular}{lccccccc}
\hline
Source & $P_{\rm orb}$ [s] & $M_1$ [$M_\odot$] & $M_2$ [$M_\odot$] & $d$ [pc] & $i$ [deg] & RA [deg] & Dec [deg] \\
\hline
AM CVn & 1029.0 & 0.68 & 0.125 & 302 & 43.0 & 129 & 49 \\
HM Cnc & 321.5 & 0.55 & 0.27 & 500 & 38.0 & 8 & 58 \\
V803 Cen & 1596.4 & 0.97 & 0.085 & 287 & 13.5 & 76 & -15 \\    
\hline
\end{tabular}
\end{table*}

Figure~\ref{F:SingleSpectrumInclination} shows the dependence of the continuum spectrum on inclination angle $i$. As described in Sec.~\ref{SubSec:SpectralModeling}, inclination effects enter purely through geometric projection factors applied after bolometric normalization.
The spectrum is monotonically suppressed with increasing inclination, reflecting the $\cos i$ scaling of optically thick components. In particular, the disk emission follows Eq.~(\ref{Eq:DiskInclination}), while the optically thick boundary-layer emission is additionally modulated by the occultation factor $f_{\rm occ}(i)$, leading to a stronger suppression at high inclinations.

In contrast, the optically thin bremsstrahlung component remains independent of inclination. As a result, at high inclinations the relative contribution of the hard X-ray tail increases compared to the suppressed soft X-ray and optical/UV components. This produces a mild hardening of the observed spectrum with increasing $i$, even though the intrinsic emission remains unchanged.
The monotonic behavior seen in the figure therefore follows directly from the geometric projection factors applied to each component, without invoking any changes in intrinsic luminosity or spectral shape.

Figure~\ref{F:SingleSpectrumNH} shows the effect of varying the hydrogen column density $N_{\rm H}$ on the observed spectrum. As described in Sec.~\ref{subSec:AbsorpExtinct}, absorption is applied via an exponential attenuation in the X-ray band and a wavelength-dependent extinction curve in the UV.
The spectrum is monotonically suppressed with increasing $N_{\rm H}$ across all energies. However, the suppression is significantly stronger in the optical/UV band than in the X-ray band. This behavior arises because the optical/UV extinction $A(\lambda)$ increases steeply toward shorter wavelengths, leading to strong attenuation via Eq.~(\ref{Eq:UVExtinction}) while the X-ray attenuation from Eq.~(\ref{Eq:XrayAbsorption}) decreases with increasing photon energy due to the approximate scaling $\sigma(E) \propto E^{-2.6}$. 
As a result, lower energy photons are preferentially absorbed, while high-energy X-rays are comparatively less affected. This produces a progressive hardening of the observed spectrum with increasing column density. 
This differential attenuation between optical/UV and X-ray bands is a direct consequence of the distinct physical absorption mechanisms operating in these regimes and is consistent with the adopted extinction and absorption prescriptions.

\begin{figure*}
\centering
\includegraphics[width=\textwidth]{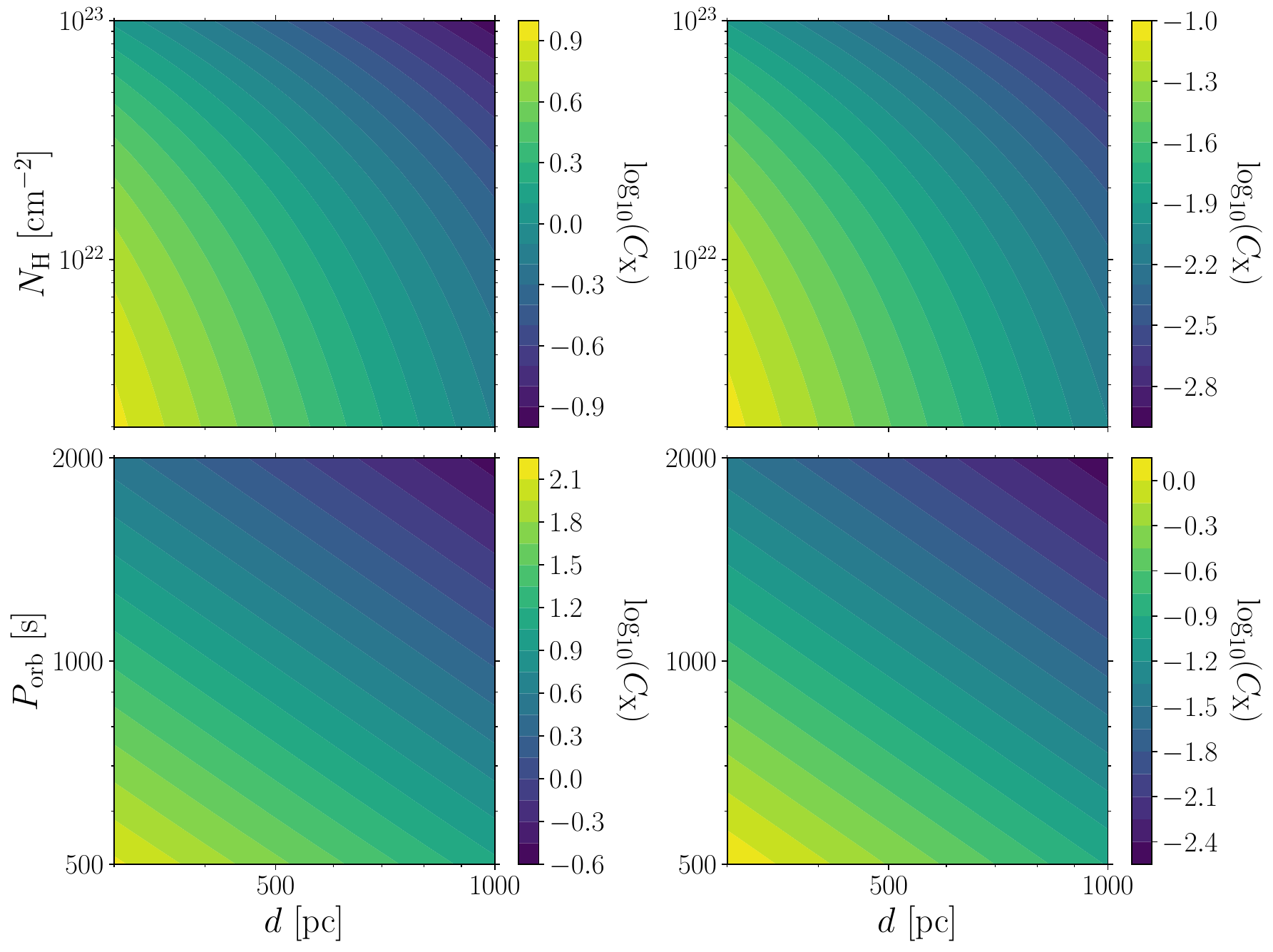}
\caption{The X-ray count rate for an \textit{AXIS}-like detector computed with the method in Section~\ref{Subsubsec:XrayCounts} as a function of distance and column density with fixed orbital period $P_{\rm orb} = 1500$ s (top row panels) or as a function of distance and orbital period with fixed column density $N_H = \num{5e21}$ cm$^{-2}$ (bottom row panels) assuming $i=0^{\rm o}$. Binaries in the left-hand column panels assume $M_1 = 1\Msol$ and $M_2=0.1\Msol$ and binaries in the right-hand column panels assume $M_1 = 1\Msol$ and $M_2=0.01\Msol$. 
} \label{F:XrayCounts}
\end{figure*}

\begin{figure*}
\centering
\includegraphics[width=\textwidth]{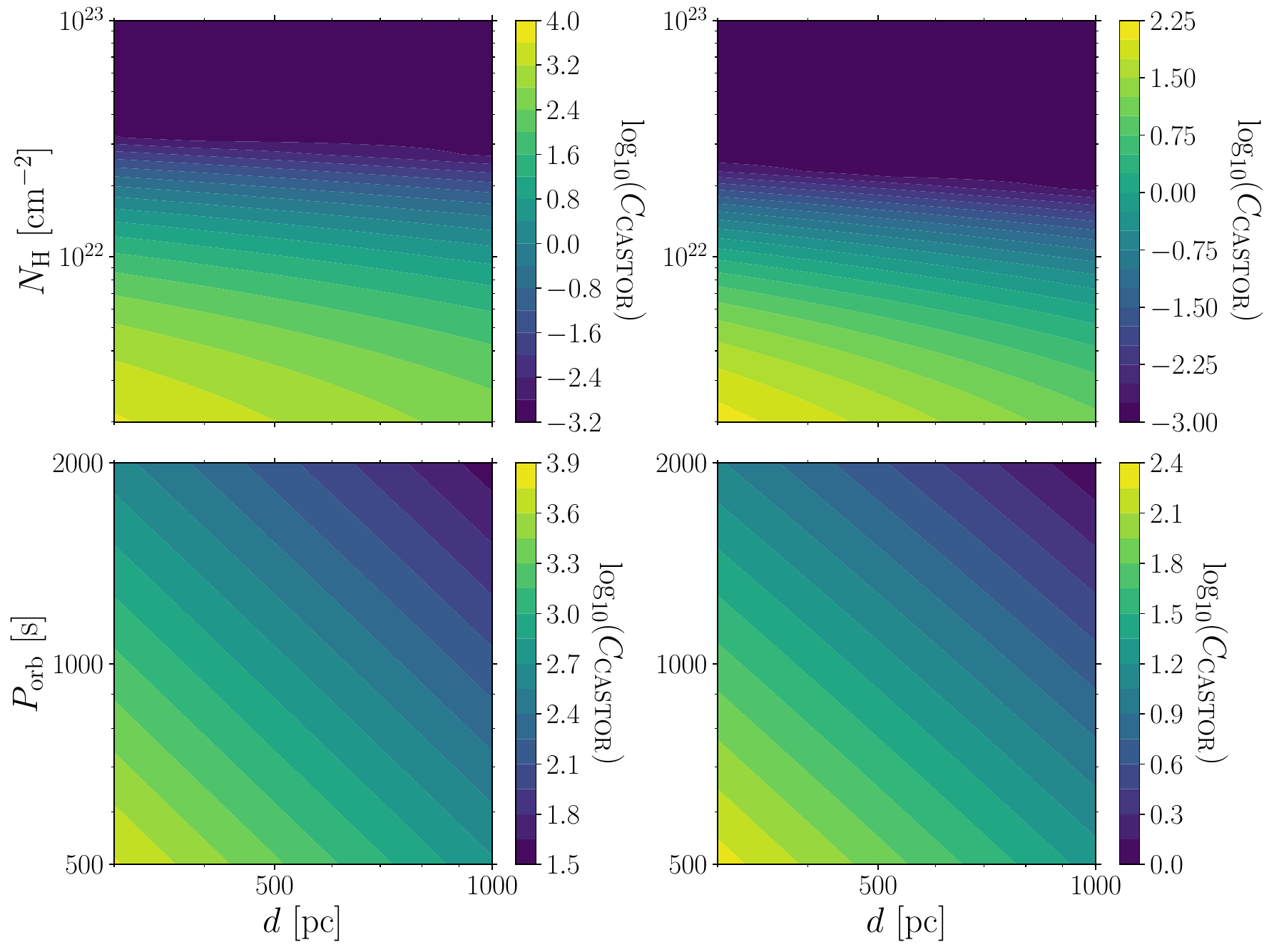}
\caption{The count rate for \textit{CASTOR} computed with the method in Section~\ref{Subsubsec:CASTORCounts} and the same settings as in Figure~\ref{F:XrayCounts}. 
} \label{F:CASTORCounts}
\end{figure*}

Figure~\ref{F:ObservedBinaries} shows the model spectra for three representative AM CVn binaries with observed parameters given in Table~\ref{Tab:ObservedBinaries}. The differences in their spectral shapes arise directly from the mapping between their binary parameters and the corresponding accretion regimes. 
We compute the column density $N_H$ of each system using the Bayestar 2019 \cite{2019ApJ...887...93G} data with the \textsc{dustmaps} module \cite{2018JOSS....3..695G}, which depends on the system's distance and sky location. 
We emphasize that this mapping is highly uncertain, and our model is missing many important processes needed to produce realistic spectra for observed binaries. Nevertheless, we cautiously apply our model to these three observed systems to establish a simple understanding of how the model's predictions relate to observed systems. 

HM~Cnc, with the shortest orbital period and highest mass ratio, resides firmly in the direct-impact regime with a high mass transfer rate of $\dot{M} \approx \num{1e-6}\Msol$/yr. Its spectrum is dominated by a compact, high-temperature hotspot, producing a soft X-ray--dominated SED with reduced optical/UV emission, consistent with the small radiative efficiency $\epsilon_{\rm rad}$ and localized emitting area. 
The column density of $N_H \approx \num{1e21}$ cm$^{-2}$ suppresses the total SED by $\approx$1 order of magnitude in both optical/UV and X-ray. 
AM CVn, with an intermediate orbital period, lies in the persistent disk regime with a moderate mass transfer rate of $\dot{M} \approx \num{1e-8}\Msol$/yr. 
The inclination of AM CVn, the largest of the three systems, and its moderate column density of $N_H \approx \num{6e20}$ cm$^{-2}$ suppress the soft X-rays of the thick boundary layer component. 
V803~Cen has the highest accretor mass $M_1$ of all three systems, and falls in the persistent disk regime despite its larger orbital period and lower $\dot{M} \approx \num{2e-9}\Msol$/yr than AM CVn. 
Its spectrum shows both a strong optical/UV disk component and a prominent optically thick soft X-ray BL component. 
The column density of $N_H \approx \num{2e20}$ cm$^{-2}$ and its inclination are smaller than compared to those of AM CVn, resulting in less suppression of the soft X-rays but its larger period compared to AM CVn produces relatively dimmer hard X-rays from the thin BL. 

This shows that the observed diversity of AM CVn spectra can arise as a consequence of the complicated interplay between the orbital period, component masses, and extrinsic parameters without requiring variation in the underlying emissivity parameters for a given observational state/regime. However, further work will be required to improve our binary mass transfer and continuum spectrum models for a detailed analysis of observed systems. 
For examples, including further terms in Eq.~(\ref{Eq:Jdot}) corresponding to  winds, or including effects of metallicity and disk truncation in the spectral model which would primarily modify the disk temperature structure, opacity, emitting area, and boundary-layer energetics with additional physical dependencies in both the spectral shapes and the relative hierarchy of emission components.

\subsubsection{Count rates for \textit{AXIS} and CASTOR}

Figure~\ref{F:XrayCounts} shows the predicted \textit{AXIS} count rates, $C_{\rm X}$, computed using the forward-folding procedure described by Equation~\ref{Eq:XrayCountRate}. In all panels, the intrinsic spectrum $F_E(E)$ is first converted to a photon flux via $\Phi(E)=F_E(E)/E$, and subsequently folded through the instrument effective area $A_{\rm eff}^{\rm AXIS}(E)$ derived directly from the \textit{AXIS} ARF, such that the resulting count rate reflects the full energy-dependent response of the detector. 
Each column corresponds to a different binary mass configuration, with $(M_1,M_2)=(1.0,0.1)$ in the left panels and $(1.0,0.01)$ in the right panels, thereby isolating the dependence of the observable signal on the donor mass and the associated mass-transfer rate.

In the top row, the count rate is shown as a function of source distance $d$ and hydrogen column density $N_{\rm H}$. The dominant trends in these panels arise from two physically distinct effects. First, the overall normalization of $C_{\rm X}$ decreases monotonically with increasing distance, reflecting the $d^{-2}$ scaling of the observed energy flux entering $F_E(E)$ prior to application of Equation~\ref{Eq:XrayCountRate}. Second, increasing $N_{\rm H}$ produces a strong suppression of the count rate due to photoelectric absorption, which preferentially attenuates the soft X-ray portion of the spectrum where a substantial fraction of the emission resides. Because the integrand in Equation~\ref{Eq:XrayCountRate} weights the photon flux by $A_{\rm eff}^{\rm AXIS}(E)$, the impact of absorption is amplified in energy ranges where the effective area is large, leading to the pronounced vertical gradients observed across the panels. 
The dependence on binary mass is evident when comparing the left and right columns. Systems with $M_2=0.1$ exhibit systematically higher count rates, reflecting the higher GW driven mass transfer rate and correspondingly higher luminosity entering $F_E(E)$. In contrast, the $M_2=0.01$ systems produce lower flux levels, shifting the contours of constant $C_{\rm X}$ toward smaller distances and lower column densities.

In the bottom row, the count rate is shown as a function of orbital period $P_{\rm orb}$ and distance, with the column density fixed. The primary driver of the structure in these panels is the strong dependence of the accretion rate on orbital period in the GW driven regime. At shorter periods, the higher mass-transfer rates yield hotter disks and boundary layers, enhancing the X-ray emission and increasing $\Phi(E)$ across the \textit{AXIS} bandpass. This results in higher values of $C_{\rm X}$ at small $P_{\rm orb}$. As the orbital period increases, the mass-transfer rate declines, leading to cooler emission components and a rapid decrease in X-ray flux, which is reflected in the downward trend of $C_{\rm X}$ with increasing $P_{\rm orb}$.

The interplay between this intrinsic luminosity evolution and the geometric dilution with distance produces the characteristic diagonal structure of the contours. For both mass configurations, the highest count rates occur at short periods and small distances, while systems at longer periods become increasingly difficult to detect except at the nearest distances. The contrast between the two columns again reflects the underlying scaling of the accretion luminosity with donor mass, with the higher-mass systems remaining detectable over a broader region of parameter space.

Figure~\ref{F:CASTORCounts} presents the corresponding \textit{CASTOR} count rates, $C_{\rm UV}$, computed using the band-integrated formalism described by Equations~\ref{Eq:UVAverageFlux} and \ref{Eq:UVMag}. In this case, the absorption-modified spectrum is converted to $F_\nu(\nu)$ and averaged over each passband using the transmission curves $T_b(\nu)$, yielding $\langle F_\nu \rangle_b$ that is subsequently mapped to AB magnitude and then to detector count rate via the instrument zeropoints. 
As in the X-ray case, the top row shows the dependence on distance and column density. The overall decrease of $C_{\rm UV}$ with increasing distance again reflects the geometric dilution of the flux entering $F_E(E)$ prior to the transformations leading to Equation~\ref{Eq:UVAverageFlux}. However, in contrast to the X-ray panels, the dependence on $N_{\rm H}$ is significantly weaker. This is because ultraviolet and optical extinction, while present, does not suppress the flux as strongly across the \textit{CASTOR} bandpasses as photoelectric absorption does in the soft X-ray regime. Consequently, the contours exhibit a more gradual variation with $N_{\rm H}$, with the dominant gradient running horizontally with distance rather than vertically with column density. 
The dependence on binary mass follows the same qualitative behavior as in the X-ray case. Systems with higher donor mass produce larger $\langle F_\nu \rangle_b$ and therefore brighter AB magnitudes (Equation~\ref{Eq:UVMag}), which translate into higher count rates. The lower-mass systems are correspondingly fainter, shifting the detectable region toward smaller distances.

\begin{figure*}
\centering
\includegraphics[width=\textwidth]{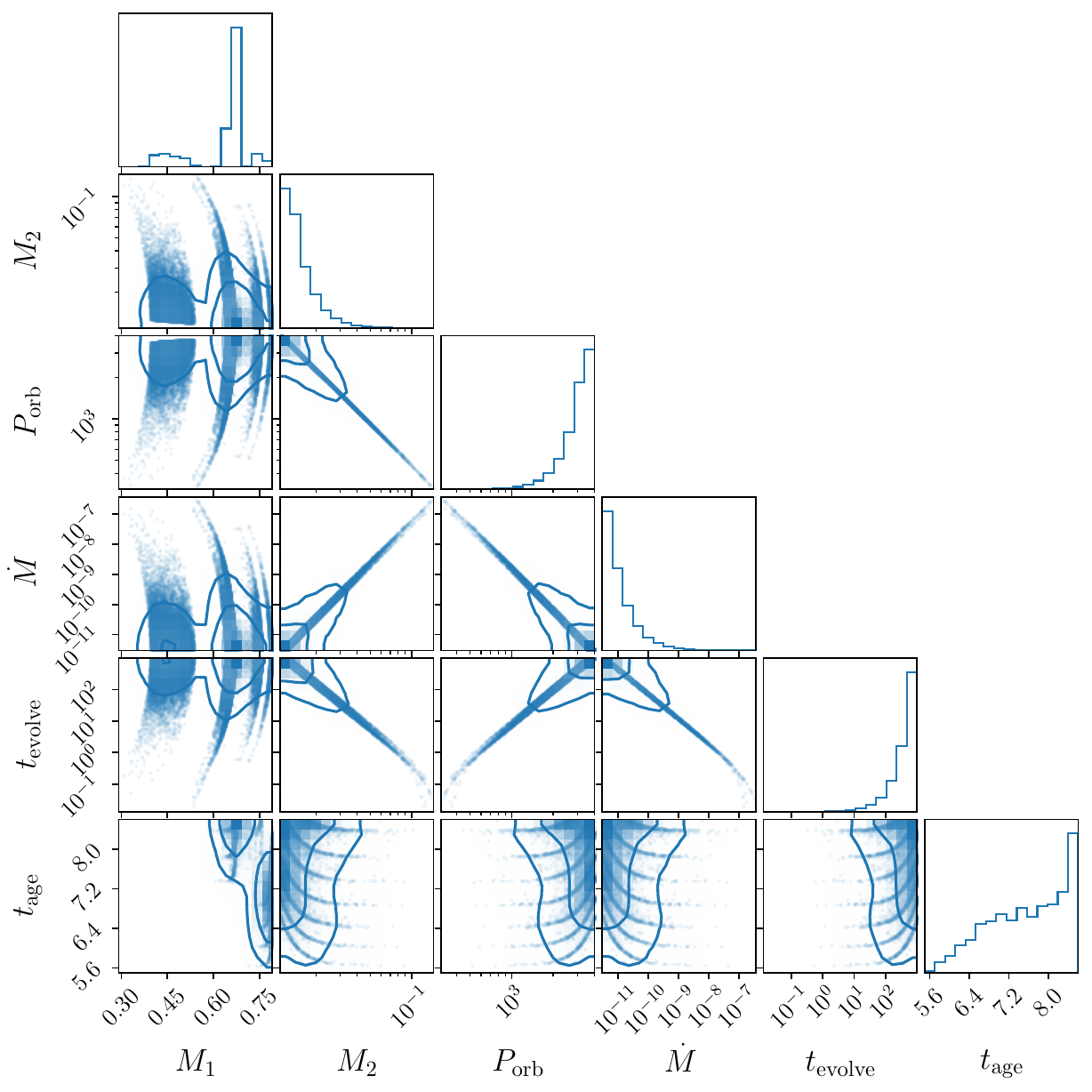}
\caption{The parameter distributions of the AM CVn binaries from \texttt{COSMIC} that have been evolved from RLOF onset through the AM CVn phase, as described in Section~\ref{SubSec:AMCVnEvolve}. The WD accretor mass $M_1$ ($\Msol$), WD donor mass $M_2$ ($\Msol$), orbital period $P_{\rm orb}$ (s),  and DWD formation time $t_{\rm form}$ (Myr) are outputs of COSMIC, while the mass transfer rate $\dot{M}$ and AM CVn evolution time $t_{\rm evolve}$ (Myr) are from our model for AM CVn evolution from Section~\ref{SubSec:GW_masstransfer}. The age of the AM CVns is $t_{\rm age} = t_{\rm form} + t_{\rm evolve}$ (Gyr). The two contour lines in the 2D histograms correspond to the 68\% and 95\% credible regions. 
} \label{F:COSMICPopParameters}
\end{figure*}

\begin{figure*}
\centering
\includegraphics[width=\textwidth]{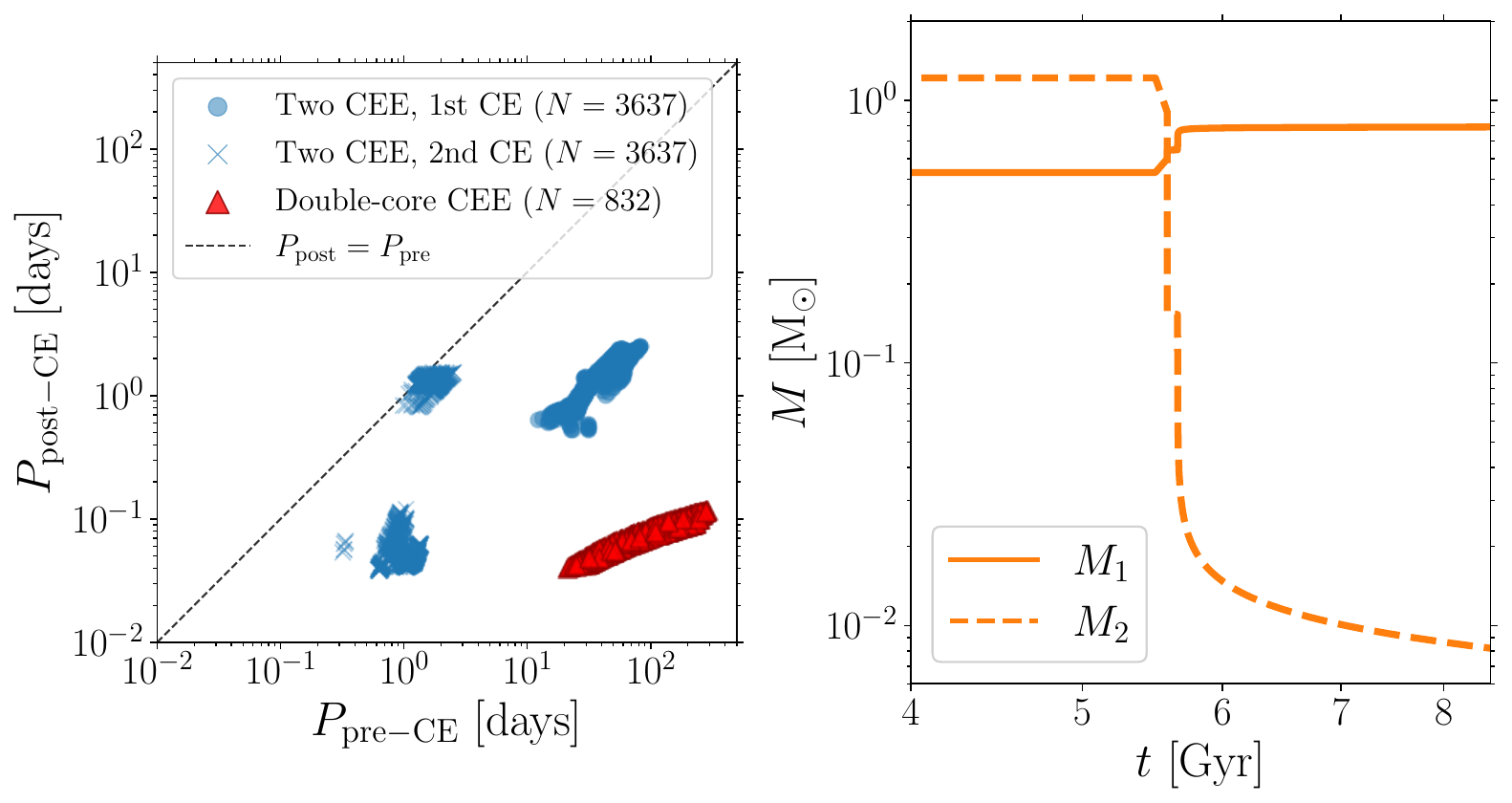}
\caption{Details of the evolutionary history of a subset of the AM CVn population from Figure~\ref{F:COSMICPopParameters}. The left panel shows the change in $P_{\rm orb}$ from the CEE episodes for each evolutionary pathway (blue dots and crossed correspond to the first and second CEE episodes in the pathway with two CEES, and the red triangles correspond to the minority of binaries in the pathway with one double-core CEE episode). 
The right panel shows the mass evolution of a binary from the pathway with two CEE episodes after the primary has already formed into a WD where the secondary initiates the second CEE, forms into a WD, and then donates mass through the AM CVn evolution phase beginning at $\approx 5.5$ Gyr. 
} \label{F:COSMICPopHistory}
\end{figure*}

\begin{figure*}
\centering
\includegraphics[width=\textwidth]{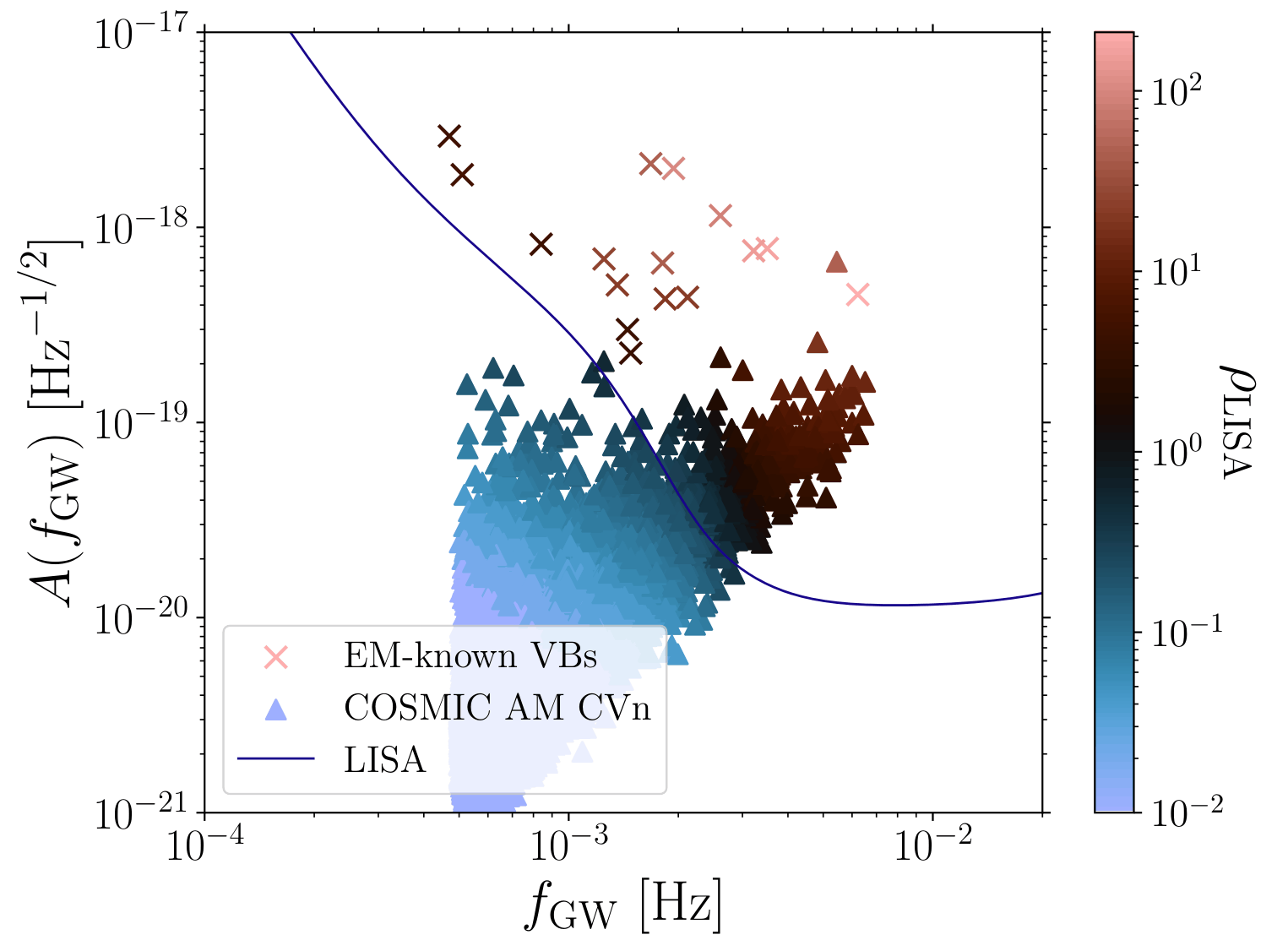}
\caption{The amplitude spectral density of the GW strain of the \texttt{COSMIC} AM CVn population from Figure~\ref{F:COSMICPopParameters} (triangles) and of a few EM-known VBs (x's), which are taken from the \textsc{legwork} package \cite{LEGWORK_apjs,LEGWORK_joss}. Both sets of binaries are colored according to their SNR by \textit{LISA} $\rho_{\rm LISA}$. The amplitude spectral density of the analytic \textit{LISA} sensitivity curve (black solid line) contains instrumental noise and the DWD stochastic foreground (note that the foreground is not computed from our \texttt{COSMIC} AM CVn population). 
} \label{F:COSMICPopStrainSNR}
\end{figure*}

\begin{figure*}
\centering
\includegraphics[width=\textwidth]{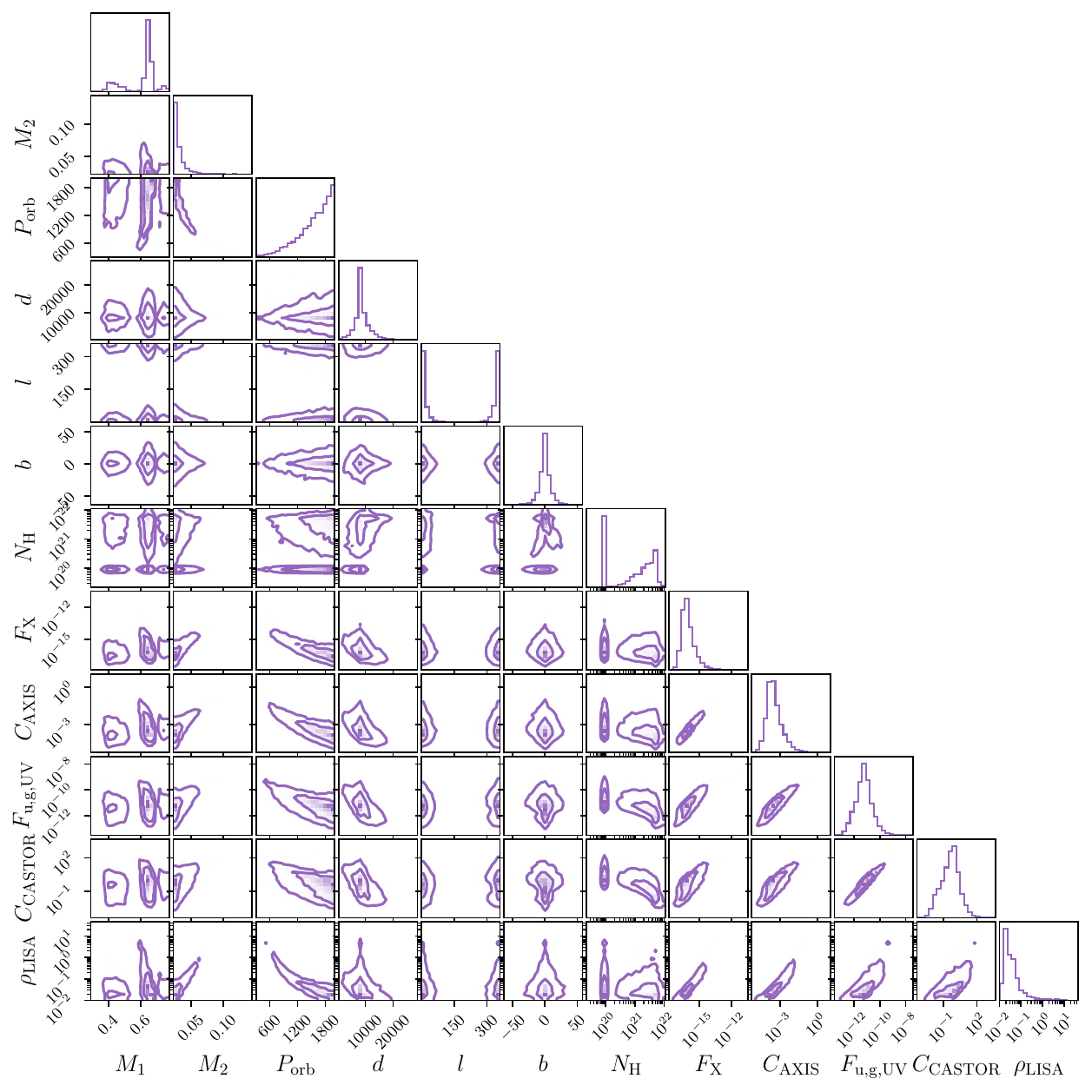}
\caption{The complete output of the multi-messenger pipeline of the \texttt{COSMIC} AM CVn population from Figure~\ref{F:COSMICPopParameters}, but sampled to only include systems with $P_{\rm orb} \leq 2000$ s. The masses $M_1$ and $M_2$ and the orbital period $P_{\rm orb}$ are subsets of the data in Figure~\ref{F:COSMICPopParameters}. The distance $d$ (pc) and Galactic longitude $l$ and latitude $b$ are drawn from the Galatic location model (Section~\ref{Sec:GalacticLocation}) and used to compute the column density $N_H$ from Bayestar 2019 data where systems with sky locations unavailable from the Bayestar data are assumed to have $N_H = \num{1e20}$. The multi-messenger observables are the maximum X-ray flux density $F_X$ (erg/cm$^2$/s), the \textit{AXIS} count rate $C_{\rm AXIS}$, the maximum optical/UV flux density $F_{\rm u,g,UV}$ (erg/cm$^2$/s), the \textit{CASTOR} count rate $C_{\rm CASTOR}$, and the SNR of \textit{LISA} $\rho_{\rm LISA}$. 
} \label{F:MultiMessPop}
\end{figure*}

In the bottom row, the variation of $C_{\rm UV}$ with orbital period and distance reflects the spectral redistribution of the emission as the accretion rate changes. At shorter orbital periods, the higher accretion rates produce hotter emission that peaks at higher energies, with a smaller fraction of the luminosity falling within the ultraviolet bandpasses. As a result, the \textit{CASTOR} count rates do not increase as steeply toward short periods as in the X-ray case. Instead, the \textit{CASTOR} sensitivity is maximized at intermediate periods where the disk temperature is such that a significant portion of the emission lies within the optical/UV bands, leading to the broad regions of enhanced $C_{\rm UV}$ seen in the contours. 
At longer orbital periods, the declining mass-transfer rate reduces the overall luminosity, causing a decrease in $\langle F_\nu \rangle_b$ and hence in $C_{\rm UV}$. The combination of these effects produces a comparatively flatter dependence on $P_{\rm orb}$ than in the X-ray panels, with detectability primarily limited by distance rather than by a sharp transition in intrinsic luminosity.

Overall, Figures~\ref{F:XrayCounts} and \ref{F:CASTORCounts} together illustrate how the mapping from intrinsic spectral energy distribution to observable count rates, as governed by Equations~\ref{Eq:XrayCountRate}--\ref{Eq:UVMag}, leads to qualitatively different sensitivities in the X-ray and ultraviolet bands, reflecting both the energy dependence of the emission processes and the instrument-specific response functions.

\subsection{A Galactic population of AM CVns}

Figure~\ref{F:COSMICPopParameters} shows the joint distribution of the intrinsic binary parameters and evolutionary timescales for the AM CVn population produced by \texttt{COSMIC} and the subsequent AM CVn evolution. The diagonal panels display the one-dimensional marginalized distributions, while the off-diagonal panels show the pairwise correlations. 
The accretor mass $M_1$ is distributed around $\sim 0.5\,\Msol$, reflecting the requirement that the accretor be a helium WD consistent with our selection criteria. In contrast, the donor mass $M_2$ spans several orders of magnitude, extending to very low values due to the continued mass loss governed by Eq.~(\ref{Eq:Mdot}). This produces a strong anti-correlation between $M_2$ and $t_{\rm evolve}$, as systems that have evolved longer through the AM CVn phase have undergone more extensive mass stripping.

The orbital period $P_{\rm orb}$ exhibits a broad logarithmic distribution extending well beyond the narrow $\sim 300~{\rm s}$ value at RLOF onset, showing the effect of forward evolution under GW-driven mass transfer. This widening follows directly from the coupled evolution equations and the Roche-lobe constraint,
$R_2(M_2) = R_{\rm L}(M_1,M_2,a),$, 
together with the donor expansion implied by Eq.~(\ref{Eq:DonorRadius}), which forces the orbit to expand as $M_2$ decreases. Consequently, $P_{\rm orb}$ is strongly anti-correlated with $\dot{M}$, since the mass-transfer rate decreases as the system evolves to longer periods according to Eq.~(\ref{Eq:Mdot}). 

The physical formation time $t_{\rm form}$ spans several Gyr and reflects the timescale for stellar evolution from ZAMS to DWD formation and RLOF onset. In contrast, $t_{\rm evolve}$ is typically much shorter, indicating that the AM CVn phase occupies only a small fraction of the total system lifetime. The total age $t_{\rm age} = t_{\rm form} + t_{\rm evolve}$ is therefore dominated by $t_{\rm form}$, producing a strong correlation between these two quantities. The absence of a strong correlation between $t_{\rm evolve}$ and $t_{\rm form}$ reflects the random assignment of evolution times, which decouples the post-RLOF evolution from the prior stellar evolution history, and the weak metallicity-age-radius correlation in our model. 

Figure~\ref{F:COSMICPopHistory} summarizes the evolutionary pathways leading to the AM CVn population. 
The left panel shows the orbital period before and after each CEE event. Deviation from the $P_{\rm post}=P_{\rm pre}$ line shows the dramatic orbital shrinkage predicted by the energy formalism in Eq.~(\ref{Eq:CommonEnvelope}). 
The points are marked according to their evolutionary pathway, with blue circles (x's) corresponding to the CEE initiated by the primary (secondary) in the dominant pathway with two CEE episodes, and the red triangles correspond to the pathway with one CEE that removes the envelopes of both progenitor stars. The formation of these tight-period, mass-transferring DWDs necessitates evolutionary pathways with efficient CEE episodes. 
In Eq.~(\ref{Eq:CommonEnvelope}),
a fraction $\alpha_{\rm CE}$ of the orbital energy released during the inspiral 
$\Delta E_{\rm orb} = G M_{1} M_{2}/(2a_{i}) - G M_{\rm core,1} M_{2}/(2a_{f})$ is available to unbind the envelope. We use a high efficiency $\alpha_{\rm CE}=8.6$, which favors CE survival. 

In the pathway with two CEEs, the first episode involves a relatively massive primary ($3\lesssim M_{1,\rm ZAMS}\lesssim 6\,M_{\odot}$) and a moderately wide orbit ($P_{\rm pre}\sim10$--$100\,\mathrm{days}$. The donor's envelope has a large binding energy, so even with the high $\alpha_{\rm CE}$ the released orbital energy can only eject the primary’s own hydrogen envelope; the orbit shrinks by about one order of magnitude (to $P_{\rm post}\approx 2\,\mathrm{days}$). A second CE episode, initiated by the secondary, is required to further shrink the orbit to ${\sim}0.1\,\mathrm{days}$. 

The orbital period after the second CEE episode depends on the stellar type of the donor secondary. This causes the population to split into two clusters (both marked by blue x's) corresponding to different donor types, where the cluster centered on the point (2, 1) days contains binaries with a secondary on the main sequence and the cluster centered on (1, 0.07) days contains secondaries that reach the Hertzsprung Gap and First Giant Branch, respectively. 
In the double-core CEE pathway, the primary is less massive ($1 \lesssim M_{1,\rm ZAMS}\lesssim 1.2\,M_{\odot}$) and the initial orbit can be slightly wider ($P_{\rm pre}\sim20$ to $130\,\mathrm{days}$).
The combination of a large orbital energy reservoir, a lower envelope binding energy, and the high $\alpha_{\rm CE}$ allows a single CE event to eject both stellar envelopes simultaneously and shrink the orbit by nearly three orders of magnitude (to $P_{\rm post}\approx 0.1\,\mathrm{days}$), directly producing a compact binary that can begin AM CVn mass transfer without a second CE phase. 

The right panel of Fig.~\ref{F:COSMICPopHistory} illustrates the mass evolution of one representative system in the dominant formation pathway. The initial stellar masses evolve through envelope loss and compact object formation, followed by the AM CVn phase in which $M_2$ decreases and $M_1$ increases according to conservative mass transfer via Eq.~(\ref{Eq:AMCVnEvolve1}). The subsequent decline in $\dot{M}$ and expansion of the orbit are consistent with the GW-driven evolution described in Section~\ref{SubSec:AMCVnEvolve}. 

We also compared $(P_{\rm orb}$ and $\dot{M})$ of the \texttt{COSMIC} binaries with their EM accretion regime, and find that the overall trend of decreasing $\dot{M}$ with increasing $P_{\rm orb}$ directly reflects the solution to Eq.~(\ref{Eq:Mdot}) under GW angular momentum loss (Eq.~(\ref{Eq:Jdot})). 
At short orbital periods and high mass-transfer rates, systems are in the direct-impact regime, where the accretion stream directly strikes the accretor due to the small orbital separation. As the system evolves to slightly longer periods, a persistent accretion disk forms, corresponding to intermediate $\dot{M}$. At even longer periods and lower $\dot{M}$, the disk becomes thermally unstable, producing the transient regime. The clear stratification of these regimes in the $(P_{\rm orb}, \dot{M})$ plane shows that the EM state is fundamentally governed by the mass-transfer evolution.

Figure~\ref{F:COSMICPopStrainSNR} presents the GW amplitude spectral density $A(f_{\rm GW})$ as a function of frequency. The population follows the expected scaling with chirp mass and frequency through the characteristic strain, Eq.~(\ref{Eq:CharacteristicStrain}), 
and its conversion to the amplitude spectral density,
Eq.~(\ref{Eq:AmplitudeSpectralDensity}). 
The \texttt{COSMIC} population populates a broad band in frequency, with higher-frequency systems corresponding to shorter orbital periods. At fixed frequency, the spread in $A(f_{\rm GW})$ is driven primarily by variations in chirp mass and distance. Systems with higher $\mathcal{M}_c$ produce stronger GW signals, while distant systems are suppressed by the $1/d$ scaling of the strain. 
The color coding by $\rho_{\rm LISA}$ shows that detectability is highest for systems at high frequency and large strain, as expected from the comparison with the \textit{LISA} sensitivity curve. Systems approaching or exceeding the sensitivity curve achieve $\rho_{\rm LISA} \gtrsim 1$, while the bulk of the population lies below the detection threshold. The EM-known VBs occupy the high-SNR region, signifying the peak of the loudest systems. As the \texttt{COSMIC} population has $P_{\rm orb} \gtrsim 10$ min, the sources are essentially monochromatic in our framework. 

Figure~\ref{F:MultiMessPop} shows the full multi-messenger parameter space for systems with $P_{\rm orb} < 2000~{\rm s}$. The intrinsic parameters $(M_1, M_2, P_{\rm orb})$ retain the correlations described above, with $M_2$ decreasing and $P_{\rm orb}$ increasing along the evolutionary sequence. 
The distance distribution peaks at several kpc, consistent with the exponential disk model in Eq.~(\ref{Eq:GalVertical}) and the transformation to heliocentric coordinates. The Galactic longitude $l_{\rm Gal}$ is broadly distributed, while the latitude $b_{\rm Gal}$ is concentrated near the plane, reflecting the thin-disk structure and the age-dependent scale height in Eq.~(\ref{Eq:GalScaleHeight}). 
The hydrogen column density $N_{\rm H}$ correlates strongly with both distance and Galactic latitude, increasing for distant systems in the plane due to larger integrated extinction along the line of sight. This directly impacts the observed EM fluxes through absorption. 

The X-ray and optical/UV fluxes and count rates exhibit strong correlations with $\dot{M}$ and distance. Systems with higher $\dot{M}$ produce larger intrinsic luminosities, while the observed flux scales as $1/d^2$ and is attenuated by $N_{\rm H}$. Consequently, high-$\dot{M}$, nearby systems dominate the observable EM population. The count rates further reflect instrumental responses but preserve these underlying physical trends. 
Finally, the \textit{LISA} SNR correlates with both orbital period and distance, with short-period, nearby systems producing the highest $\rho_{\rm LISA}$. Together, this shows that EM and GW observables probe complementary regions of parameter space: GW detectability is primarily sensitive to compact, high-frequency systems, while EM detectability depends strongly on $\dot{M}$ and line-of-sight absorption. This confirms that the multi-messenger pipeline consistently links intrinsic binary evolution, Galactic structure, and observational selection effects. 

\renewcommand{\arraystretch}{1.6} 
\setlength{\tabcolsep}{3pt}     
\begin{table*}[th!]
\centering
\footnotesize
\caption{Combined multi-messenger detection fractions and subpopulation
properties for the synthetic AM CVn population at fiducial detection
thresholds: \textit{LISA} SNR $\rho_{\mathrm{LISA}}>7$; \textit{AXIS} source counts $\geq10$ per 10~ks observation (count rate $\geq0.00100$~cps); \textit{CASTOR} source counts $\geq25$ per 10~ks observation (count rate
$\geq0.00250$~cps). The total synthetic population contains $N_{\mathrm{tot}}=110{,}194$ binaries. The column titled 'Fraction' specifies the weighted detected fraction for each Detection Combination, computed as the sum of astrophysical prior weights over all detected systems normalized to the total population weight. $Y_{\rm det}$ is the corresponding weighted detection yield; weights are assigned under a Kroupa IMF, uniform mass-ratio distribution, log-uniform period distribution, and uniform eccentricity distribution, and vary by two orders of magnitude across the grid.
$\tilde{\rho}_{\mathrm{LISA}}$ is the median \textit{LISA} signal-to-noise ratio;
$\tilde{C}_{\mathrm{AXIS}}$ and $\tilde{C}_{\mathrm{CASTOR}}$ are
the median \textit{AXIS} and \textit{CASTOR} source count rates (counts/s);
$\tilde{d}$ is the median heliocentric distance;
$\tilde{P}_{\mathrm{orb}}$ is the median orbital period in minutes with
the [min,\,max] range given in parentheses;
$\tilde{M}_1$ and $\tilde{M}_2$ are the median accretor and donor masses
(${\rm M}_\odot$);
$\tilde{N}_H$ is the median line-of-sight hydrogen column
density (cm$^{-2}$) derived from Bayestar 2019 dust maps
\citep{Green2019Bayestar}. Note that these detection yields arise from our grid-based synthetic population and
are not normalized to an absolute Galactic AM CVn number density in this table.}
\begin{tabular}{l||c|c||c|c|c|c||c|c|c|c}
Detection Combination
  & $Y_{\rm det}$ & Fraction 
  & $\tilde{\rho}_{\mathrm{LISA}}$
  & $\tilde{C}_{\mathrm{AXIS}}$ 
  & $\tilde{C}_{\mathrm{CASTOR}}$ 
  & $\tilde{d}$ [pc]
  & $\tilde{P}_{\mathrm{orb}}$ [min]
  & $\tilde{M}_{1}\ [{\rm M}_\odot]$
  & $\tilde{M}_{2}\ [{\rm M}_\odot]$
  & $\tilde{N}_H$ [cm$^{-2}$] \\
\hline
\hline
\textit{LISA} only
  & $1.5\times10^{-4}$ & 0.0001
  & 8     & 0.17 & 46.9 & 7679
  & $6.1$ (5.1--8.7)    & 0.54 & 0.12 & $1\times10^{20}$ \\
\hline
\textit{AXIS} only
  & $5.2\times10^{-2}$ & 0.052
  & 0     & 0.0017 & 0.27  & 4922
  & $46.8$ (5.1--66.4)  & 0.47 & 0.016 & $1\times10^{20}$ \\
\hline
\textit{CASTOR} only
  & $8.5\times10^{-1}$ & 0.85
  & 0     & 0.0002 & 0.031  & 8196
  & $49.4$ (5.1--66.8)  & 0.44 & 0.015 & $9.3\times10^{20}$ \\
\hline
\hline
\textit{LISA} $+$ \textit{AXIS}
  & $1.5\times10^{-4}$ & 0.0001
  & 8     & 0.17 & 46.9 & 7679
  & $6.1$ (5.1--8.7)    & 0.54 & 0.12 & $1\times10^{20}$ \\
\hline
\textit{LISA} $+$ \textit{CASTOR}
  & $1.5\times10^{-4}$ & 0.0001
  & 8     & 0.17 & 46.9 & 7679
  & $6.1$ (5.1--8.7)    & 0.54 & 0.12 & $1\times10^{20}$ \\
\hline
\textit{AXIS} $+$ \textit{CASTOR}
  & $5.2\times10^{-2}$ & 0.052
  & 0     & 0.0017 & 0.27  & 4923
  & $46.8$ (5.1--66.4)  & 0.47 & 0.016 & $1\times10^{20}$ \\
\hline
\textit{LISA}$+$\textit{AXIS}$+$\textit{CASTOR}
  & $1.5\times10^{-4}$ & 0.0001
  & 8     & 0.17 & 46.9 & 7679
  & $6.1$ (5.1--8.7)    & 0.54 & 0.12 & $1\times10^{20}$ \\
\hline
\end{tabular}
\label{Tab:CombinedDetection}
\end{table*}

\begin{table*}[th!]
\centering
\caption{Sensitivity of multi-messenger detection yields to the adopted \textit{LISA} SNR threshold $\rho_{\mathrm{thr}}$. EM thresholds are fixed at the fiducial values: \textit{AXIS} counts $\geq 10$ per 10~ks; \textit{CASTOR} counts $\geq 25$  per 10~ks. $Y_{\mathrm{LISA}}$ and $f_{\mathrm{LISA}}$ are the weighted detection yield and weighted fraction for systems detected by LISA alone; $Y_{\mathrm{L+A}}$, $Y_{\mathrm{L+C}}$, and $Y_{\mathrm{all}}$ give the corresponding weighted yields for systems detected by \textit{LISA} $+$ \textit{AXIS}, \textit{LISA} $+$ \textit{CASTOR}, and all three instruments simultaneously. The fiducial threshold $\rho_{\mathrm{thr}} = 7$ is highlighted by the central row.}
\begin{tabular}{r||r|r|r|r|r}
$\rho_{\mathrm{thr}}$ & $Y_{\mathrm{LISA}}$ & $f_{\mathrm{LISA}}$ & $Y_{\mathrm{L+A}}$ & $Y_{\mathrm{L+C}}$ & $Y_{\mathrm{all}}$ \\
\hline
\hline
$5$ & $6.7\times10^{-4}$ & 0.0007 & $6.1\times10^{-4}$ & $6.7\times10^{-4}$ & $6.1\times10^{-4}$ \\
\hline
$7$$^{\dagger}$ & $1.5\times10^{-4}$ & 0.0001 & $1.5\times10^{-4}$ & $1.5\times10^{-4}$ & $1.5\times10^{-4}$ \\
\hline
$10$ & $3.7\times10^{-5}$ & 0.0000 & $3.7\times10^{-5}$ & $3.7\times10^{-5}$ & $3.7\times10^{-5}$ \\
\hline
\multicolumn{6}{l}{\small $^\dagger$ Fiducial \textit{LISA} SNR threshold.} \\
\end{tabular}
\label{Tab:LISASNRSens}
\end{table*}

\begin{table*}[th!]
\centering
\small
\caption{Sensitivity of EM and multi-messenger detection yields to the minimum source count threshold $N_{\mathrm{min}}$ and observation duration $T_{\mathrm{obs}}$, with the \textit{LISA} SNR threshold fixed at the fiducial value $\rho_{\mathrm{LISA}} > 7$. The count-rate threshold for each instrument is $N_{\mathrm{min}}/T_{\mathrm{obs}}$. $N_{\mathrm{min,AXIS}}$ and $N_{\mathrm{min,CASTOR}}$ are the minimum source counts required for a detection with \textit{AXIS} and \textit{CASTOR} respectively. Columns give the number of systems detected by \textit{AXIS} alone, \textit{CASTOR} alone, \textit{LISA} $+$ \textit{AXIS}, \textit{LISA} $+$ \textit{CASTOR}, and all three instruments simultaneously. A horizontal rule separates groups of rows sharing the same $T_{\mathrm{obs}}$. The fiducial row ($T_{\mathrm{obs}} = 10$ ks, $N_{\mathrm{min,AXIS}} = 10$, $N_{\mathrm{min,CASTOR}} = 25$) is marked with a dagger.}
\begin{tabular}{r|r|r||r|r|r|r|r}
$T_{\mathrm{obs}}$ [ks] & $N_{\mathrm{min,AXIS}}$ & $N_{\mathrm{min,CASTOR}}$ & $N_{\mathrm{AXIS}}$ & $N_{\mathrm{CASTOR}}$ & $N_{\mathrm{L+A}}$ & $N_{\mathrm{L+C}}$ & $N_{\mathrm{all}}$ \\
\hline
\hline
$1$ & $10$ & $25$ & 575 & 62,042 & 32 & 32 & 32 \\
\hline
$1$ & $10$ & $10$ & 575 & 79,546 & 32 & 32 & 32 \\
\hline
$1$ & $25$ & $50$ & 236 & 43,442 & 32 & 32 & 32 \\
\hline
\hline
$10$$^{\dagger}$ & $10$ & $25$ & 8,505 & 97,507 & 32 & 32 & 32 \\
\hline
$10$ & $10$ & $10$ & 8,505 & 105,670 & 32 & 32 & 32 \\
\hline
$10$ & $25$ & $50$ & 2,512 & 89,711 & 32 & 32 & 32 \\
\hline
\hline
$100$ & $10$ & $25$ & 83,510 & 109,726 & 32 & 32 & 32 \\
\hline
$100$ & $10$ & $10$ & 83,510 & 110,036 & 32 & 32 & 32 \\
\hline
$100$ & $25$ & $50$ & 52,723 & 108,765 & 32 & 32 & 32 \\
\hline
\multicolumn{8}{l}{\small $^\dagger$ Fiducial EM thresholds.} \\
\end{tabular}
\label{Tab:EMSens}
\end{table*}

\subsection{Multi-messenger detection yields}

We compute multi-messenger detection yields directly from the synthetic AM CVn population generated by the hybrid \texttt{COSMIC} + ODE framework, applying instrument-specific detectability criteria to each system. The total synthetic population contains $N_{\mathrm{tot}}=110{,}194$ binaries after application of the Galactic column-density mask. Unlike our previous unweighted analysis, the present calculation incorporates astrophysical prior weights derived from the ZAMS progenitor distributions used in the \texttt{COSMIC} population synthesis. Since the underlying \texttt{COSMIC} population is constructed as a deterministic grid rather than a Monte Carlo realization, raw occupancy counts do not directly correspond to astrophysical occurrence rates. We therefore assign each surviving AM CVn system a statistical weight
\begin{equation}
w_i \propto p(M_{1,\mathrm{ZAMS}})\,
p(q_{\mathrm{ZAMS}})\,
p(P_{\mathrm{orb,ZAMS}})\,
p(e_{\mathrm{ZAMS}}),
\end{equation}
where the priors follow a Kroupa IMF for the primary mass, a uniform mass-ratio distribution, a log-uniform orbital-period distribution, and a uniform eccentricity distribution: 
\begin{align}
p(M_1) &\propto M_1^{-\alpha},\\
p(q) &= \mathrm{const},\\
p(P_{\rm orb}) &\propto P_{\rm orb}^{-1},\\
p(e) &= \mathrm{const},
\end{align}
with the Kroupa IMF slope $\alpha$ defined piecewise following \cite{2001MNRAS.322..231K}. The resulting weights span approximately two orders of magnitude across the synthetic grid. Detection yields are then computed as weighted sums over detectable systems,
\begin{equation}
Y_{\rm det} = \sum_{i\in {\rm det}} w_i,
\end{equation}
while the weighted detection fraction is
\begin{equation}
F_{\rm det} =
\frac{\sum_{i\in {\rm det}} w_i}
{\sum_i w_i}.
\end{equation}
Throughout this section, quoted detection fractions correspond to these weighted astrophysical fractions rather than to raw synthetic occupancy counts.

Detection in each channel is defined at the level of individual systems using threshold criteria that map directly onto instrumental sensitivity. For \textit{LISA}, a signal-to-noise ratio threshold $\rho_{\mathrm{LISA}}>7$ is adopted as the fiducial detection criterion. In the EM bands, detectability is determined by requiring a minimum number of source counts over a fixed exposure time, such that
\begin{equation}
N_{\rm src} = \dot{C}\,T_{\rm obs} \geq N_{\rm min},
\end{equation}
where $\dot{C}$ is the source count rate. The fiducial configuration corresponds to $T_{\rm obs}=10$~ks, $N_{\rm min,AXIS}=10$ for \textit{AXIS}, and $N_{\rm min,CASTOR}=25$ for \textit{CASTOR}, yielding count-rate thresholds of $10^{-3}$~cps and $2.5\times10^{-3}$~cps respectively. Note that within the scope of this work, we ignore contributions to the UV background from Earthshine, zodiacal light, and geocoronal emission lines, as well as instrumental effects, which would somewhat reduce the signal-to-noise of our model CASTOR observations, reserving such considerations for after CASTOR's upcoming Phase A.

Detection thresholds in all three observational channels are intended to provide a physically motivated and simple estimate of detectability. For \textit{LISA}, the \textsc{legwork} calculations include the instrumental sensitivity curve and the Galactic DWD confusion foreground, but assume quasi-monochromatic emission over the mission lifetime and neglect the effects of mass-transfer--driven frequency evolution and orbital-frequency derivatives. The resulting SNRs should be interpreted as estimates of instantaneous GW observability. 

For the EM channels, we do not explicitly model source detection pipelines, background estimation, source confusion, or survey selection effects. For \textit{AXIS}, count rates are obtained by directly folding the absorbed source spectrum through the published ARF but assumes that energy-resolution losses are unimportant for integrated count-rates. Likewise, for \textit{CASTOR}, count rates are derived from passband-integrated fluxes using the published filter transmission curves and photometric zero-point calibrations. 

In the source-dominated Poisson limit the fiducial thresholds ($N_{\rm min,AXIS}=10$ counts for \textit{AXIS} and $N_{\rm min,CASTOR}=25$ counts for CASTOR) correspond to effective signal-to-noise ratios of approximately $\sqrt{10}\simeq3.2$ for \textit{AXIS} and $\sqrt{25}=5$ for CASTOR. The more conservative \textit{CASTOR} threshold partially compensates for the fact that UV and optical backgrounds and detector-noise contributions are not explicitly included in the present model. 
We explore a range of exposure times and minimum-count thresholds to quantify the sensitivity of the detection yields to these assumptions. 

The resulting weighted detection fractions, summarized in Table~\ref{Tab:CombinedDetection}, exhibit a highly asymmetric distribution across the three observational channels. The \textit{CASTOR} detection fraction dominates at $F_{\rm CASTOR}\simeq0.85$, corresponding to a weighted detection yield $Y_{\rm det}=8.5\times10^{-1}$, while \textit{AXIS} detects a substantially smaller subset with $F_{\rm AXIS}\simeq0.052$ and $Y_{\rm det}=5.2\times10^{-2}$. In contrast, the \textit{LISA}-detectable population remains extremely sparse, with a weighted detection fraction of only $F_{\rm LISA}\simeq1.5\times10^{-4}$ and weighted yield $Y_{\rm det}=1.5\times10^{-4}$.

This hierarchy reflects the different characteristics of the three instruments. \textit{CASTOR} probes the optical/UV emission from the accretion disk and donor-heated components, which remain detectable over a broad range of orbital periods and mass-transfer rates. Consequently, it recovers nearly the entire long-period AM CVn population, characterized by low-mass donors ($\tilde{M}_2\simeq0.015\,M_\odot$) and orbital periods $\tilde{P}_{\rm orb}\approx49$~min. \textit{AXIS}, while still sensitive to accretion-powered emission, preferentially selects systems with higher mass-transfer rates and thus shorter periods, yielding a somewhat reduced but still substantial detection fraction with $\tilde{P}_{\rm orb}\approx47$~min.

By contrast, \textit{LISA} selects only the most compact systems, where the GW strain amplitude is sufficiently large. The detected systems are confined to $\tilde{P}_{\rm orb}\simeq6$~min with relatively massive donors ($\tilde{M}_2\simeq0.12\,M_\odot$), consistent with early-stage mass transfer. This strong bias toward ultra-compact configurations explains both the extremely low weighted detection fraction and the near-complete overlap between the \textit{LISA}-detected sample and the subsets detected by \textit{AXIS} and \textit{CASTOR}. Indeed, the weighted yields for the LISA-only, LISA+\textit{AXIS}, \textit{LISA}+\textit{CASTOR}, and three-instrument combinations are effectively identical in Table~\ref{Tab:CombinedDetection}. This is not coincidental but instead reflects the fact that systems luminous enough in gravitational waves to exceed $\rho_{\mathrm{LISA}}>7$ are also sufficiently bright in both X-ray and optical/UV bands under the adopted accretion model.

The subpopulation properties listed in Table~\ref{Tab:CombinedDetection} reinforce this interpretation. The \textit{LISA}-detectable systems occupy a distinct region of parameter space, with high count rates ($\tilde{C}_{\mathrm{AXIS}}\simeq0.17$~cps, $\tilde{C}_{\mathrm{CASTOR}}\simeq46.9$~cps) despite their large median distance ($\tilde{d}\approx7.7$~kpc), indicating intrinsically luminous sources. In contrast, the \textit{AXIS}- and \textit{CASTOR}-detected populations extend to longer periods and lower accretion luminosities, with correspondingly smaller count rates and a broader distribution in distance.

The sensitivity of these results to the adopted \textit{LISA} detection threshold is quantified in Table~\ref{Tab:LISASNRSens}. Lowering the threshold from $\rho_{\mathrm{thr}}=7$ to $5$ increases the weighted \textit{LISA} detection yield from $1.5\times10^{-4}$ to $6.7\times10^{-4}$, corresponding to an increase in the weighted detectable fraction from $\sim10^{-4}$ to $\sim7\times10^{-4}$. Importantly, this increase propagates almost linearly into the multi-messenger yields, with the triple-coincidence yield increasing from $1.5\times10^{-4}$ to $6.099\times10^{-4}$. This reflects the steepness of the SNR distribution near the detection boundary: a modest relaxation of the threshold admits a substantial number of marginal systems. Conversely, increasing the threshold to $\rho_{\mathrm{thr}}=10$ reduces the weighted detectable population to only $3.654\times10^{-5}$, implying that the \textit{LISA}-detectable population is dominated by sources near the fiducial sensitivity limit.

The dependence on EM observing strategy is explored in Table~\ref{Tab:EMSens}. Increasing the exposure time from $1$~ks to $100$~ks dramatically enhances the \textit{AXIS} detection yield, reflecting the linear scaling of the count threshold with $T_{\rm obs}$. \textit{CASTOR} exhibits a similar but less pronounced trend due to its already high rate at shorter exposures. In contrast, the multi-messenger yields involving \textit{LISA} are essentially invariant across all EM configurations. Even under the conservative EM thresholds considered here, all systems satisfying the fiducial \textit{LISA} criterion remain detectable by both EM detectors. This insensitivity arises because the \textit{LISA}-detectable systems are intrinsically bright in both X-ray and optical/UV emission and exceed the adopted EM count thresholds. As a result, EM depth does not limit the joint-detection sample; instead, the bottleneck in our model is set by the GW detectability. 

The weighted yields incorporate the astrophysical prior distributions of the progenitor population and therefore provide a self-consistent relative estimate of the detectable AM CVn subpopulations within the synthetic framework itself. The dominant remaining uncertainty is associated not with instrumental sensitivity, but with the underlying binary-evolution physics and the still poorly constrained absolute Galactic AM CVn formation rate.

We conclude that joint \textit{LISA}+\textit{AXIS}+\textit{CASTOR} detections correspond to an astrophysically distinct subset of AM CVn binaries, some fraction of which will serve as VBs for \textit{LISA}. Multi-messenger observability is fundamentally limited by the small fraction of systems that attain sufficiently high GW amplitudes to be individually resolved by \textit{LISA}, whereas EM detectability remains comparatively efficient across much of the AM CVn parameter space. 
Under fiducial detection thresholds, our model predicts that only $\sim1.5\times10^{-4}$ of Galactic AM CVn systems are jointly detectable by \textit{LISA}, \textit{AXIS}, and \textit{CASTOR}, corresponding to approximately one multi-messenger detection per $\sim7\times10^{3}$ AM CVn binaries.

\subsection{Normalizing flows for multi-messenger inference}

\begin{figure*}
\centering
\includegraphics[width=\textwidth]{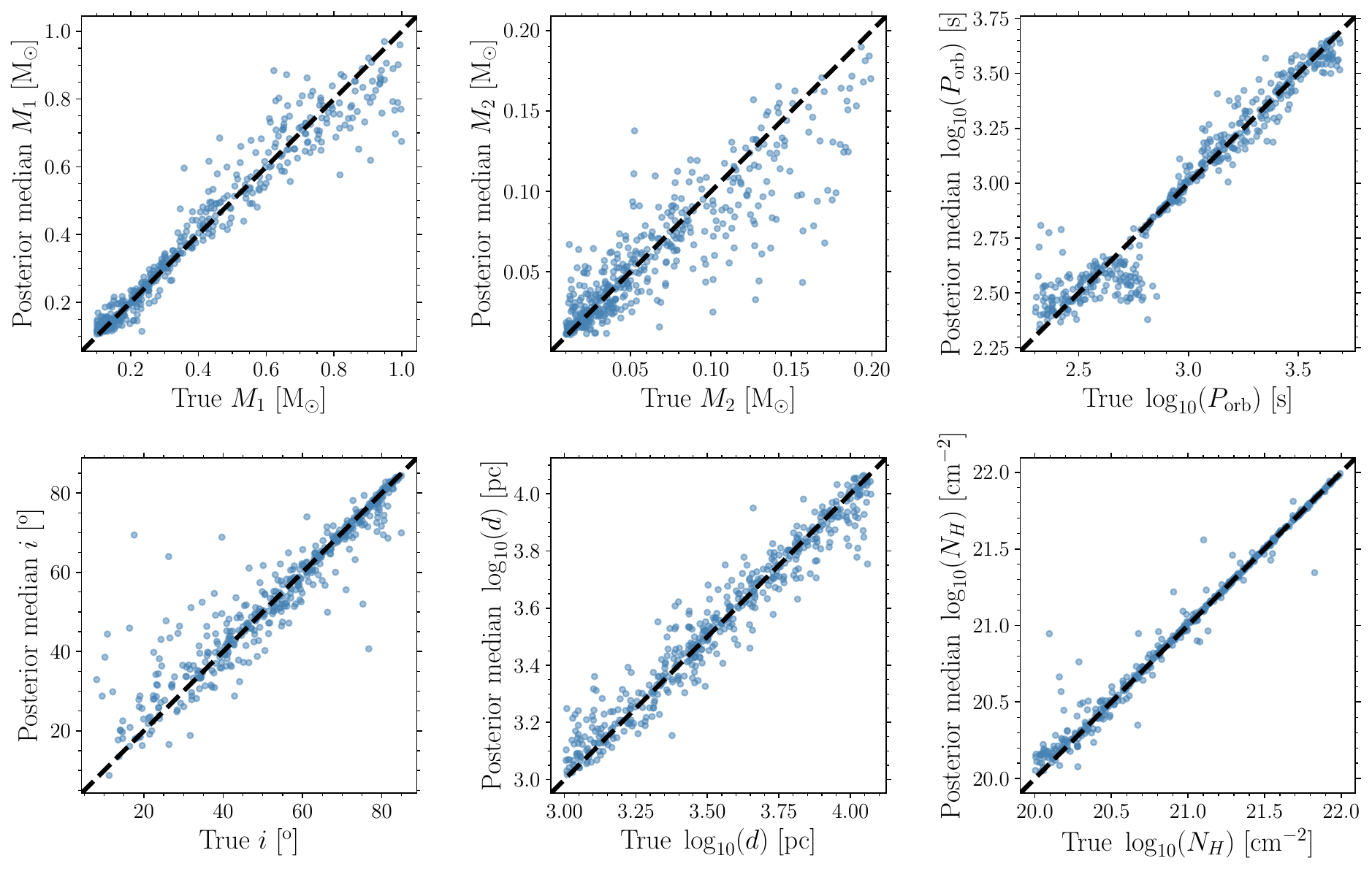}
\caption{Recovery performance of the trained conditional normalizing flow on held-out validation samples drawn from the same prior distribution as used during training. Each panel compares the posterior median inferred from the flow to the true parameter value for the accretor mass $M_1$ ($\Msol$), donor mass $M_2$ ($\Msol$), orbital period $P_{\rm orb}$ (s), inclination $i$ ($^\circ$), distance $d$ (pc), and hydrogen column density $N_H$ (${\rm cm}^{-2}$). The dashed black line denotes the one-to-one relation corresponding to perfect recovery. Parameters sampled logarithmically during training are plotted in $\log_{10}$ space. 
} \label{F:TrainingRecovery}
\end{figure*}

\begin{figure*}
\centering
\includegraphics[width=\textwidth]{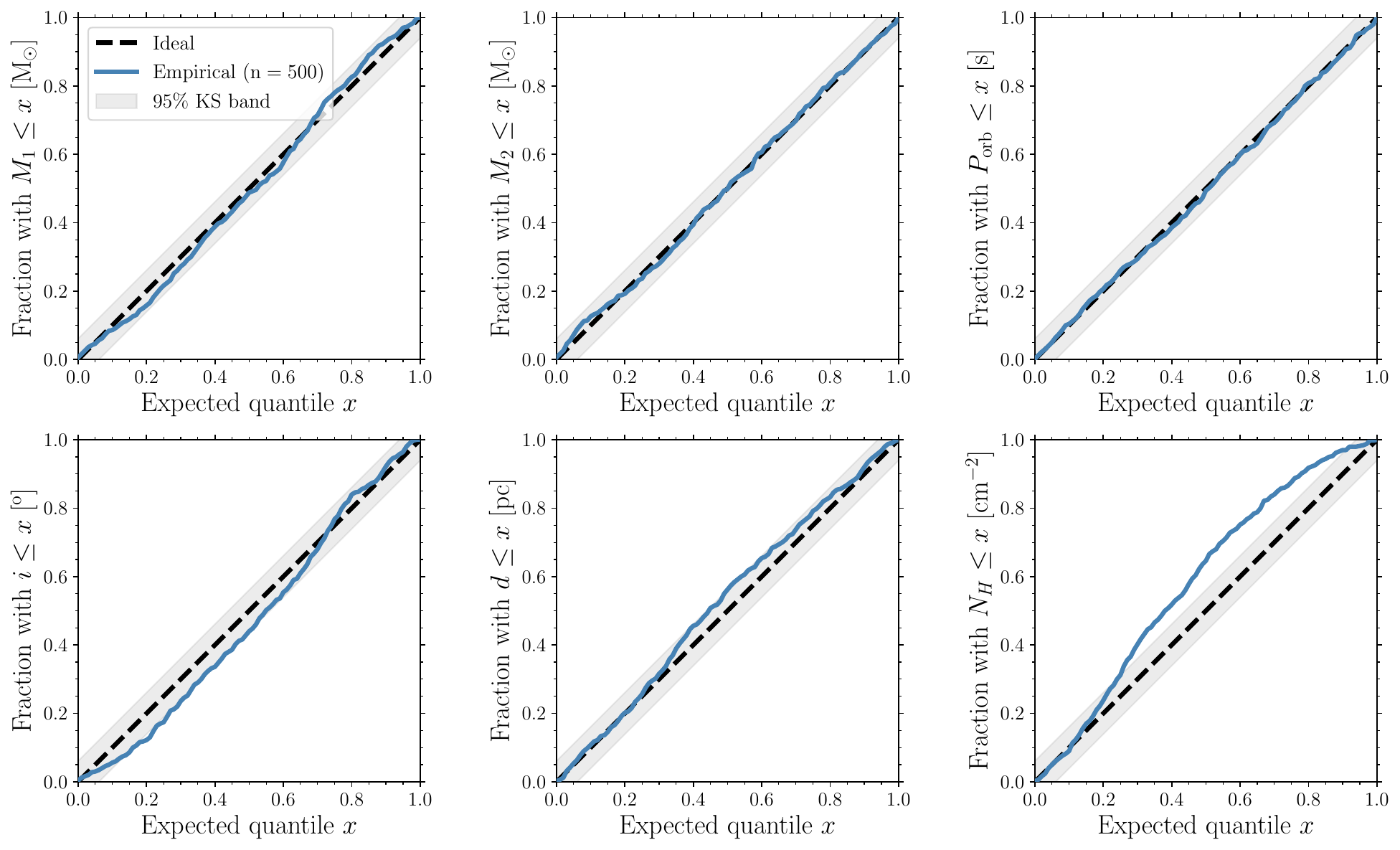}
\caption{
Calibration diagnostics for the trained conditional normalizing flow evaluated on held-out validation samples. Each panel shows a PP plot, i.e., the empirical cumulative distribution of posterior ranks for the inferred parameters $M_1$, $M_2$, $P_{\rm orb}$, $i$, $d$, and $N_H$, compared to the ideal diagonal relation expected for a perfectly calibrated posterior. The grey shaded region corresponds to the 95\% Kolmogorov--Smirnov confidence band. 
} \label{F:TrainingPP}
\end{figure*}

\begin{figure*}
\centering
\includegraphics[width=0.9\textwidth]{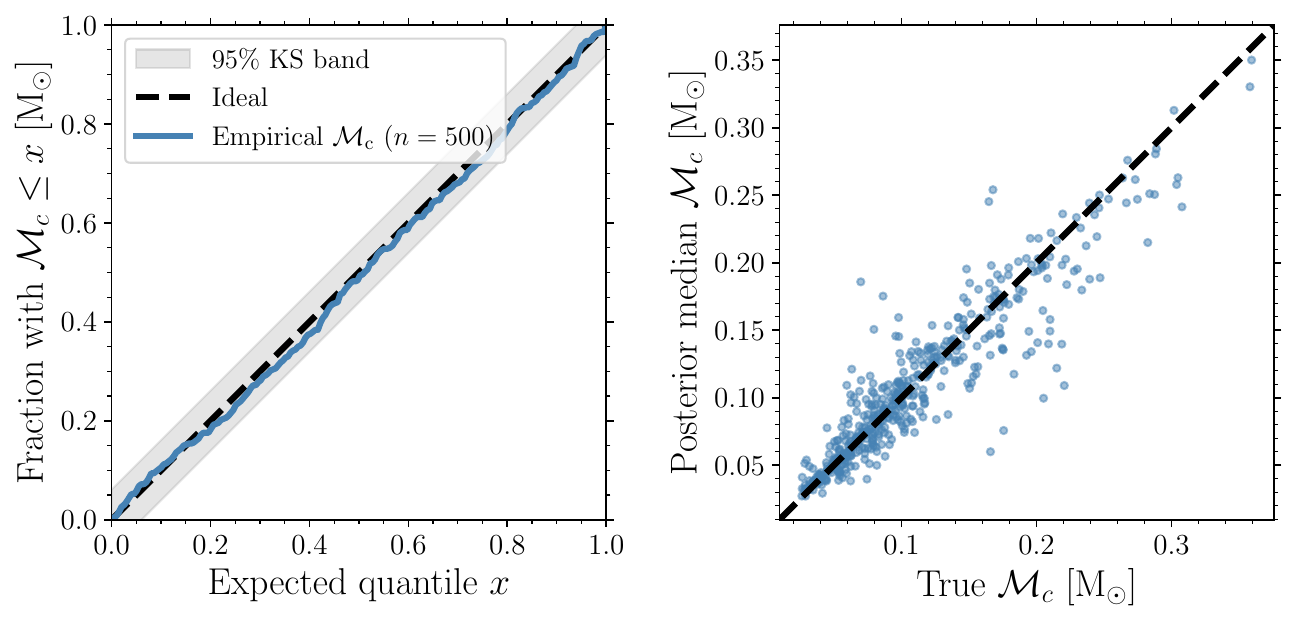}
\caption{Chirp-mass recovery diagnostics for the trained conditional normalizing flow on held-out validation samples. The left panel shows the calibration diagnostic for the inferred chirp mass, while the right panel compares the posterior median chirp mass to the true chirp mass of each system. The chirp mass is computed as in Eq.~(\ref{Eq:ChirpMass}). The dashed black line in the right panel denotes the one-to-one relation corresponding to perfect recovery. 
} \label{F:TrainingChirpMass}
\end{figure*}

\begin{figure*}
\centering
\includegraphics[width=\textwidth]{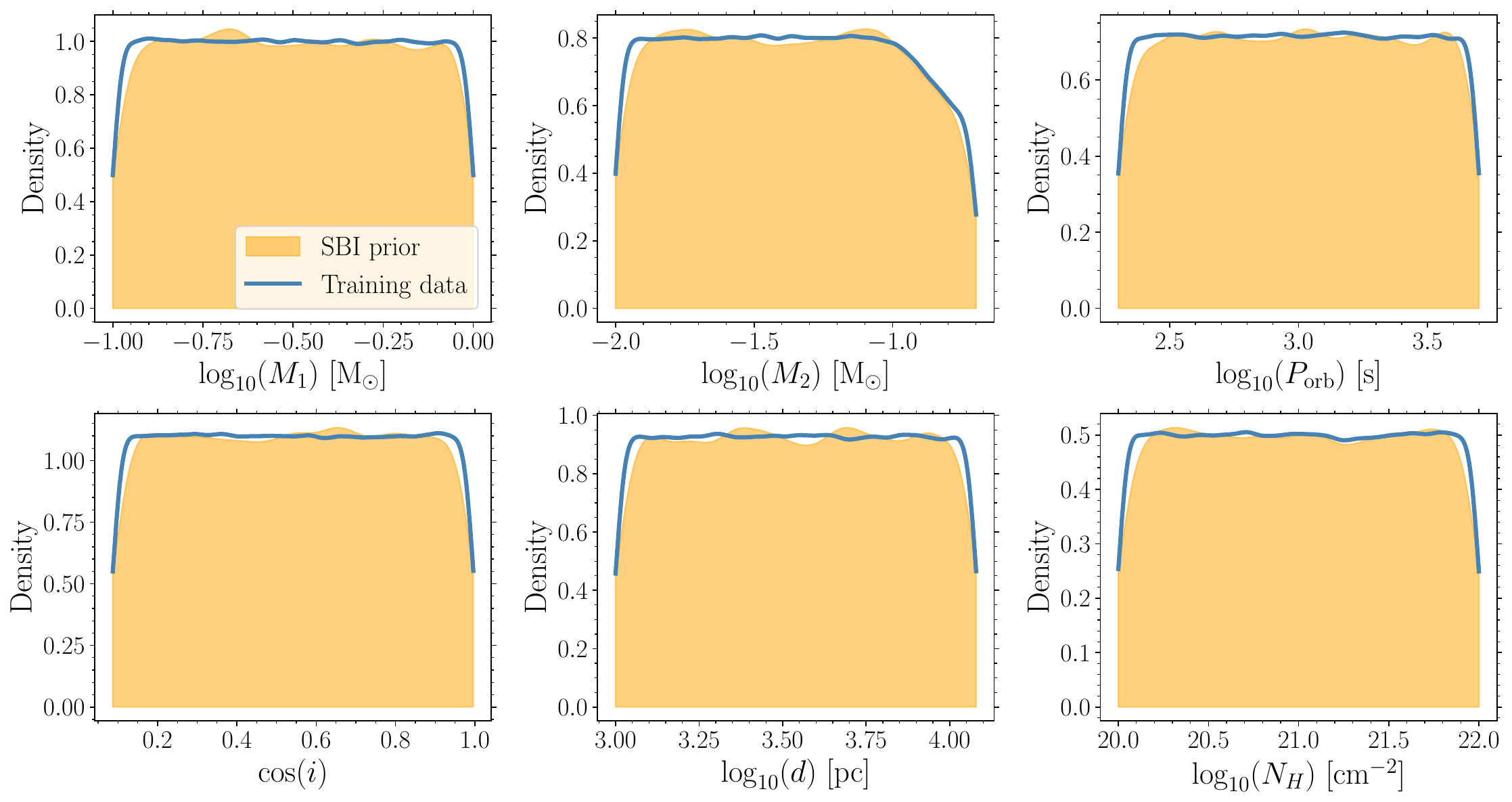}
\caption{Comparison between the training prior distribution and samples reconstructed from the trained conditional normalizing flow. The panels show the parameter distributions for the accretor mass $M_1$ ($\Msol$), donor mass $M_2$ ($\Msol$), orbital period $P_{\rm orb}$ (s), inclination $i$ ($^\circ$), distance $d$ (pc), and hydrogen column density $N_H$ (${\rm cm}^{-2}$). The comparison shows the consistency between the effective learned distribution and the parameter space sampled during training. 
} \label{F:TrainingVsPrior}
\end{figure*}

\begin{figure*}
\centering
\includegraphics[width=0.85\textwidth]{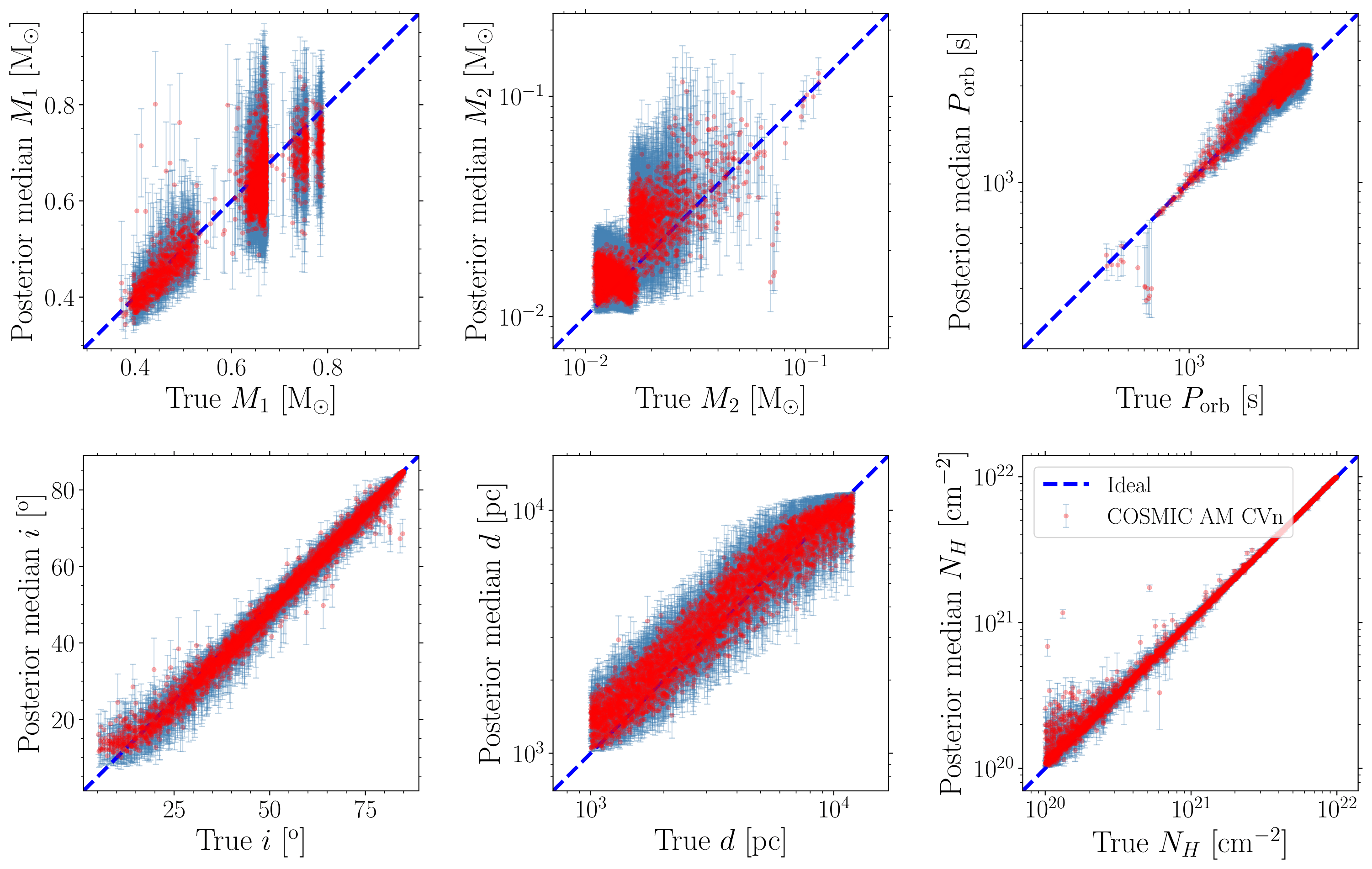}
\caption{Recovery performance of the trained conditional normalizing flow evaluated on the independent AM CVn population generated with \texttt{COSMIC} for the Case~1 test with $P_{\rm orb} \in [200,5000]~{\rm s}$. Each panel compares the posterior median inferred from the flow to the true parameter value for $M_1$, $M_2$, $P_{\rm orb}$, $i$, $d$, and $N_H$. The dashed black line denotes the one-to-one relation corresponding to perfect recovery. 
The error bars are the asymmetric 68\% central credible interval. 
} \label{F:TestingRecovery}
\end{figure*}

\begin{figure*}
\centering
\includegraphics[width=\textwidth]{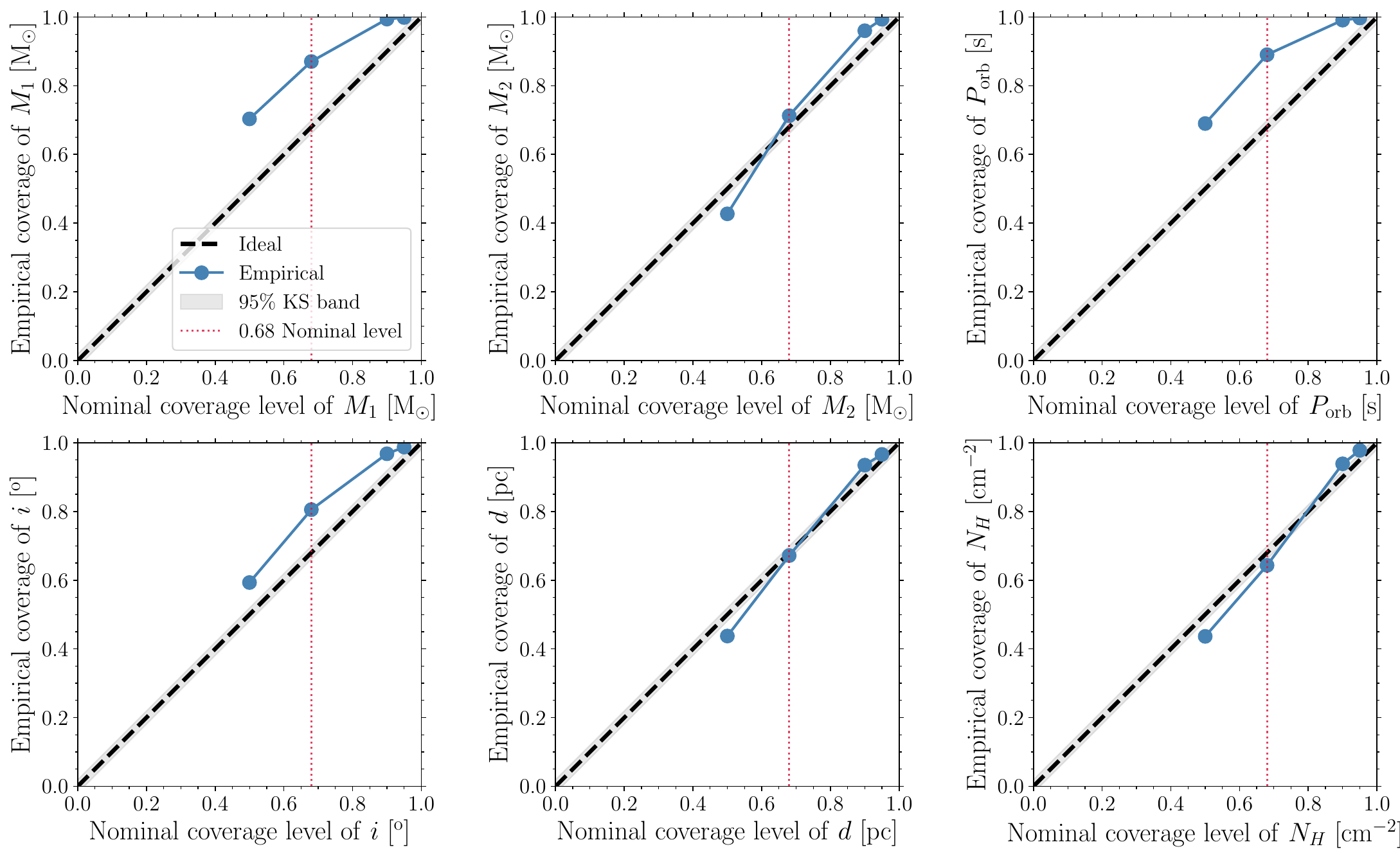}
\caption{Posterior calibration diagnostics for the independent \texttt{COSMIC} population in the Case~1 test with $P_{\rm orb} \in [200,5000]~{\rm s}$. Each panel shows the empirical coverage fraction as a function of the nominal credible interval level for the inferred parameters $M_1$, $M_2$, $P_{\rm orb}$, $i$, $d$, and $N_H$. The dashed black line corresponds to perfect calibration where the empirical coverage equals the nominal posterior credible interval. 
} \label{F:TestingCalibration}
\end{figure*}

\begin{figure*}
\centering
\includegraphics[width=0.85\textwidth]{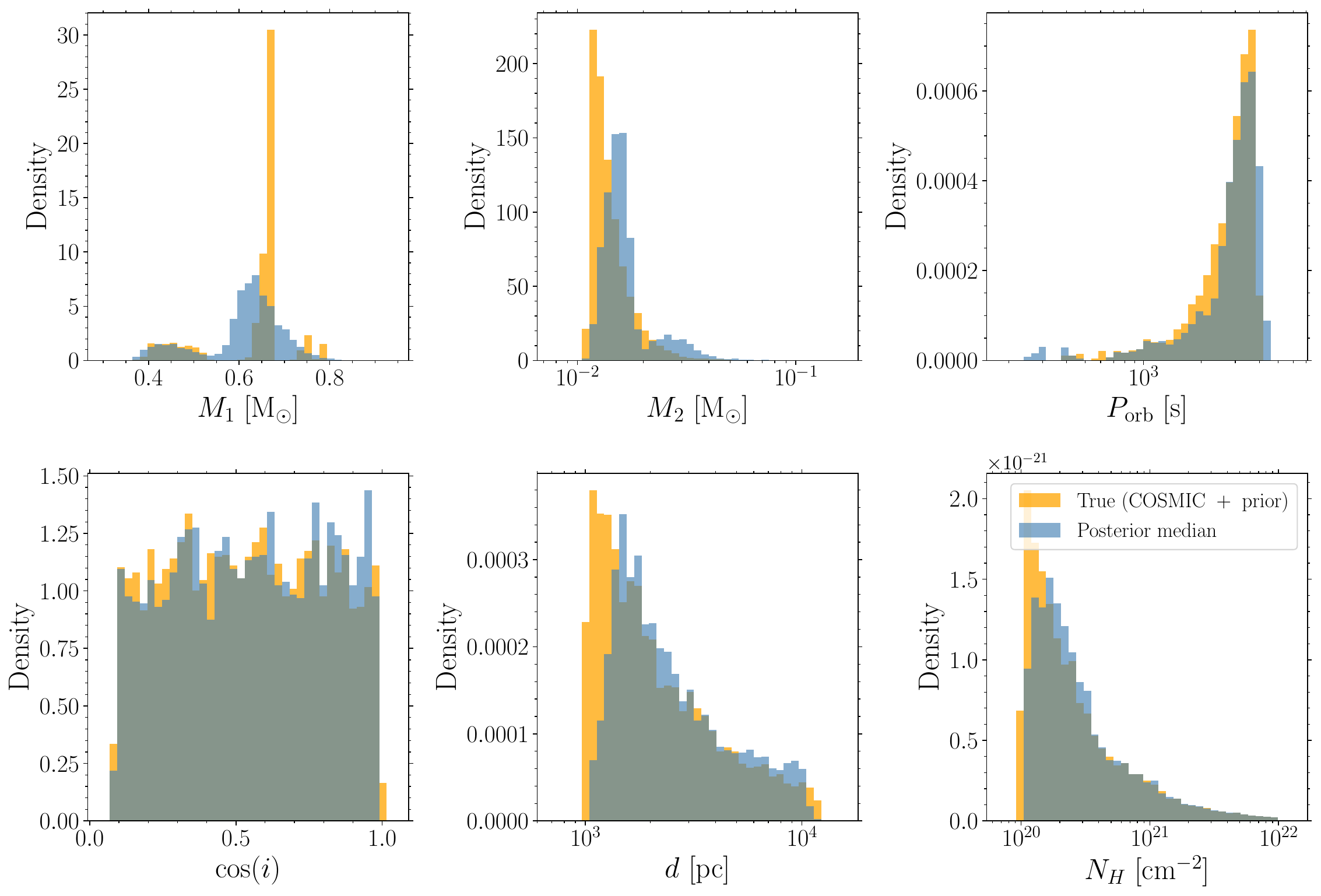}
\caption{Comparison between the true parameter distributions of the \texttt{COSMIC} AM CVn population and the distributions reconstructed from the posterior medians inferred by the trained conditional normalizing flow for the Case~1 test with $P_{\rm orb} \in [200,5000]~{\rm s}$ prior. The panels show the distributions for the accretor mass $M_1$ ($\Msol$), donor mass $M_2$ ($\Msol$), orbital period $P_{\rm orb}$ (s), inclination $i$ ($^\circ$), distance $d$ (pc), and hydrogen column density $N_H$ (${\rm cm}^{-2}$). 
} \label{F:TestingPosteriors}
\end{figure*}

\begin{figure*}
\centering
\includegraphics[width=0.85\textwidth]{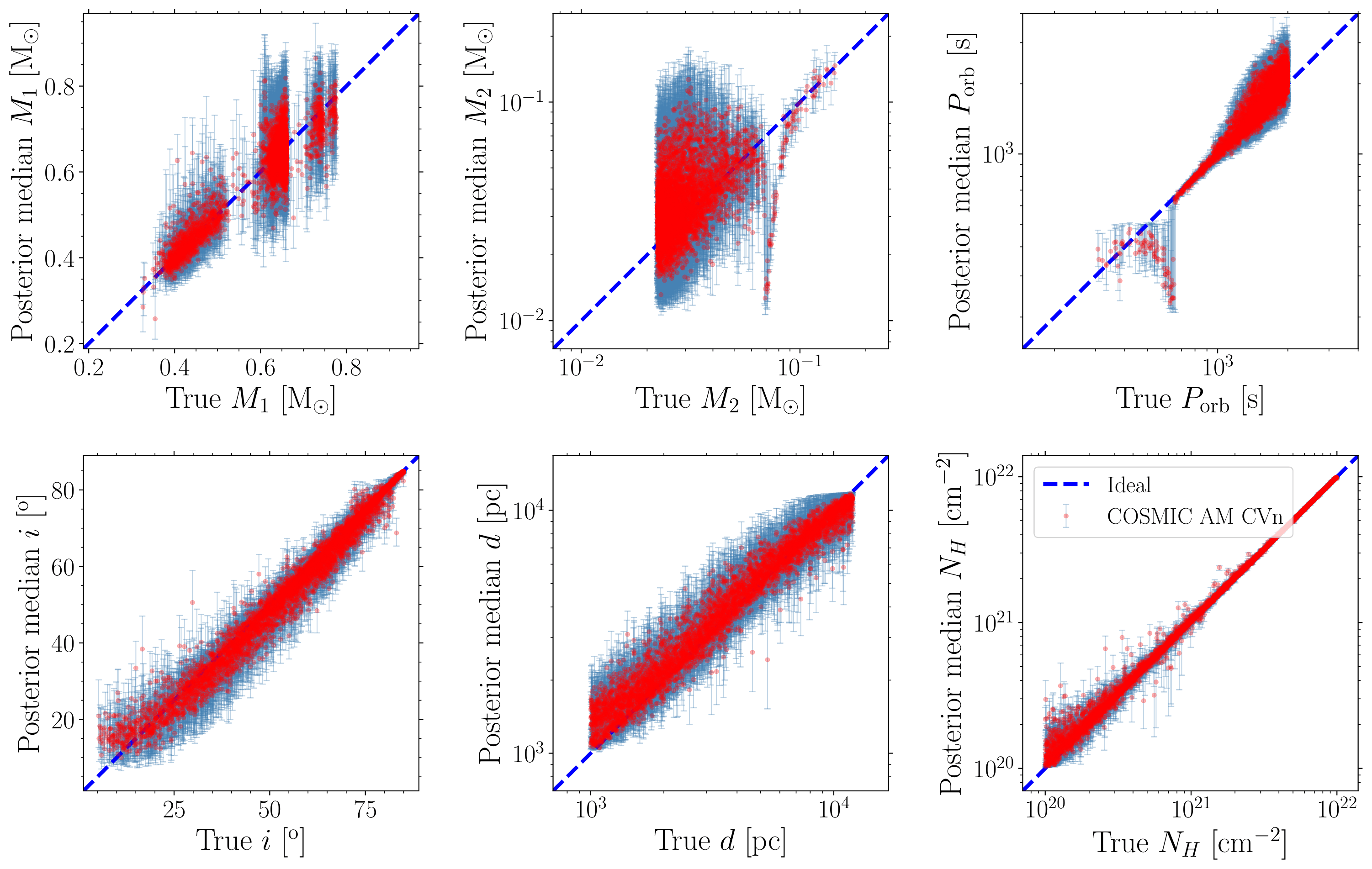}
\caption{Recovery performance of the trained conditional normalizing flow evaluated on the restricted \texttt{COSMIC} AM CVn population for the Case~2 test with $P_{\rm orb} \in [200,2000]~{\rm s}$ prior. Each panel compares the posterior median inferred from the flow to the true parameter value for $M_1$, $M_2$, $P_{\rm orb}$, $i$, $d$, and $N_H$. The dashed black line denotes the one-to-one relation corresponding to perfect recovery. The error bars are the asymmetric 68\% central credible interval. 
} \label{F:TestingRecoveryNarrow}
\end{figure*}

\begin{figure*}
\centering
\includegraphics[width=0.85\textwidth]{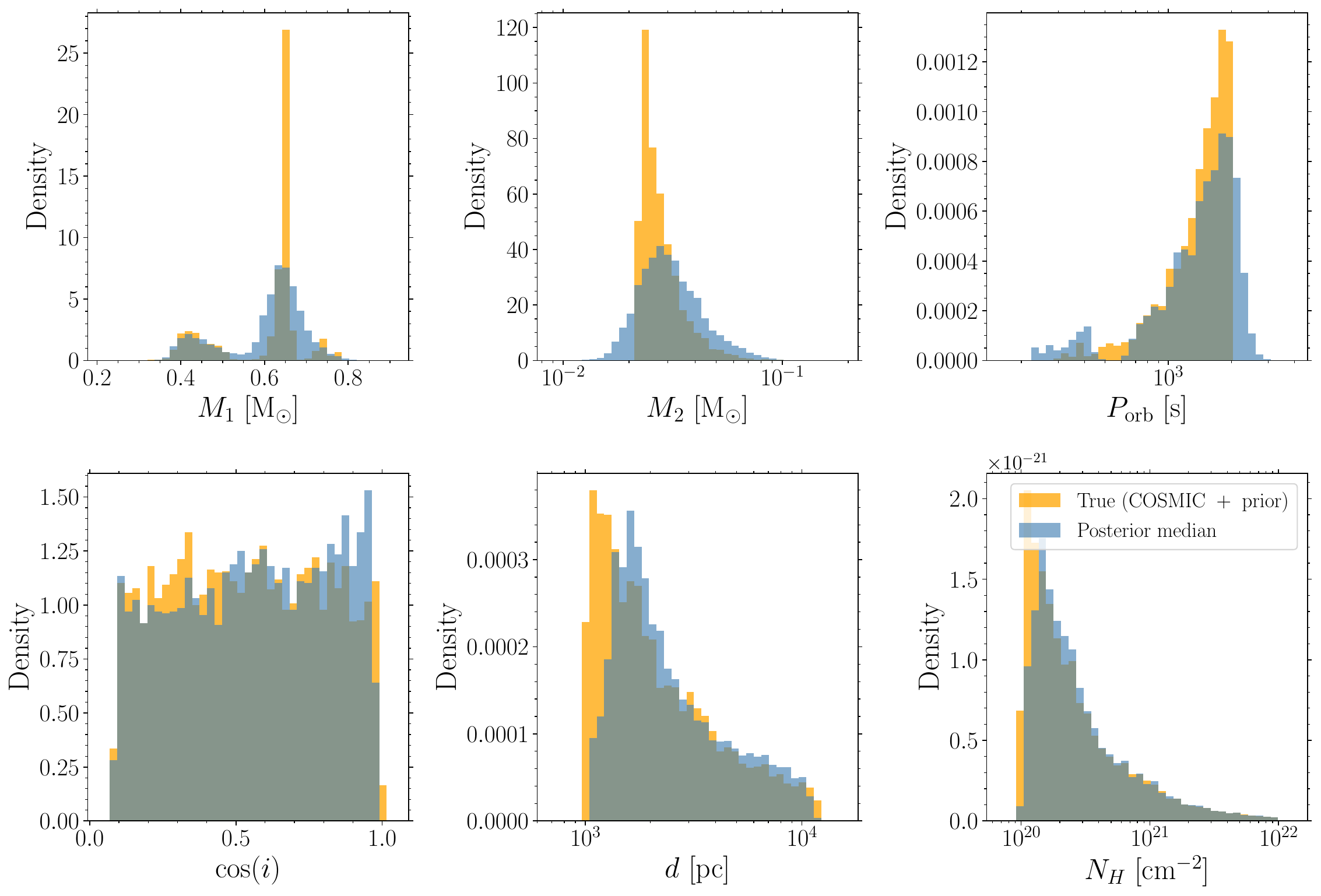}
\caption{Comparison between the true parameter distributions of the restricted \texttt{COSMIC} AM CVn population and the distributions reconstructed from the posterior medians inferred by the trained conditional normalizing flow for the Case~2 test with $P_{\rm orb} \in [200,2000]~{\rm s}$ prior. The panels show the distributions for the accretor mass $M_1$ ($\Msol$), donor mass $M_2$ ($\Msol$), orbital period $P_{\rm orb}$ (s), inclination $i$ ($^\circ$), distance $d$ (pc), and hydrogen column density $N_H$ (${\rm cm}^{-2}$). 
} \label{F:TestingPosteriorsNarrow}
\end{figure*}

We now assess the performance of the multi-messenger simulation-based inference framework described in Section~\ref{Sec:NormalFlows}. The trained conditional normalizing flow is evaluated both on held-out samples drawn directly from the training prior and on an independent synthetic AM CVn population generated using \texttt{COSMIC}. These tests quantify the ability of the learned inverse model to recover intrinsic and extrinsic binary parameters from combined EM and GW observables, and provide insight into the physical degeneracies governing the inference problem.

\subsubsection{Training-set recovery and calibration}

Figure~\ref{F:TrainingRecovery} shows the parameter recovery performance of the trained normalizing flow on held-out validation samples drawn from the same prior distribution as the training set. Each panel compares the posterior median inferred from the flow to the true parameter value used to generate the multi-messenger observables. Overall, the flow successfully reconstructs the global structure of the inverse mapping across the six-dimensional parameter space, showing that the learned posterior captures the dominant physical dependencies encoded in the forward model.

The recovery performance for the accretor mass $M_1$ exhibits a clear mass dependence, with systematically tighter recovery at lower masses and increased dispersion toward larger $M_1$. This behavior arises directly from the structure of the accretion physics encoded in the forward EM model of Section~\ref{Sec:ForwardEM}. 
The boundary layer luminosity scales approximately as $L_{\rm BL} \propto (M_1/R_1) \dot{M}$ (Eq.~(\ref{Eq:Lacc})) and the WD mass--radius $R_1 \propto M_1^{-1/3}$ (\ref{Eq:DonorRadius}))
lead to the approximate scaling $L_{\rm BL} \propto M_1^{4/3}\dot{M}$. 
At relatively low accretor masses ($M_1 \lesssim 0.5\,\Msol$), the white dwarf radius is larger and the boundary layer temperature remains well matched to the soft X-ray sensitivity window of the \textit{AXIS}-like detector. Variations in $M_1$ therefore produce strong and localized changes in the continuum spectral shape and peak energy distribution, allowing the convolutional embedding network to efficiently isolate physically informative spectral features. In this regime, the flow receives strong constraints from both the detailed spectrum and the X-ray count rate, yielding relatively tight posterior recovery.

At larger accretor masses, however, the recovery degrades substantially. This occurs for two related physical reasons. First, the white dwarf mass--radius relation steepens as the accretor approaches the relativistic regime, producing stronger nonlinear coupling between $M_1$ and the boundary layer temperature. Multiple combinations of $(M_1,\dot{M})$ can therefore generate similar continuum spectra by trading a smaller, hotter boundary layer against a larger, cooler one while preserving the integrated flux. Second, the increasing boundary layer temperature progressively shifts the spectral peak toward or beyond the upper edge of the \textit{AXIS} bandpass. Once the dominant emission moves outside the observed energy range, the inference becomes increasingly dependent on integrated count rates and on the \textit{LISA} signal-to-noise ratio rather than on detailed spectral morphology. These observables exhibit stronger degeneracies with $M_2$, $d$, and $N_H$, naturally broadening the posterior distribution at high $M_1$.

The donor mass $M_2$ exhibits qualitatively different behavior. In contrast to $M_1$, the scatter about the one-to-one relation remains relatively uniform across the parameter range. This reflects the fact that the donor mass enters the forward model only indirectly through the GW-driven mass transfer rate and the chirp mass (\ref{Eq:ChirpMass}). 
Across the prior range adopted in Eq.~(\ref{Eq:Priors}), the dependence of the continuum spectrum and \textit{LISA} signal on $M_2$ is comparatively weak and does not introduce a strongly localized spectral feature analogous to the boundary layer temperature dependence on $M_1$. The helium donor mass--radius relation (\ref{Eq:DonorRadius})   controls the Roche geometry and therefore the mass transfer rate relatively smoothly across the parameter space. As a consequence, the inferred posterior uncertainty on $M_2$ remains broadly distributed and approximately symmetric across the prior range.

The orbital period recovery reveals three distinct regimes that directly reflect the underlying accretion physics and GW detectability. The first regime corresponds to systems below approximately $P_{\rm orb} \sim 300\,\rm s$, where a distinct clump of outliers is visible in Fig.~\ref{F:TrainingRecovery}. These systems correspond primarily to direct-impact accretors, which are included in the forward model through the \texttt{use\_direct\_impact=True} prescription. In this regime, the ballistic accretion stream impacts the accretor surface directly rather than circularizing into a Keplerian accretion disk. The resulting continuum morphology differs substantially from the standard disk-plus-boundary-layer configuration that dominates the remainder of the training set. The flow therefore encounters a partially bimodal mapping between spectral morphology and orbital period, leading to a distinct population of recovery failures at very short periods.

The second regime occurs near $P_{\rm orb} \sim 800$--$1000\,\rm s$, where the recovery is exceptionally tight. This region lies close to the AM CVn orbital period minimum and simultaneously maximizes the constraining power of both the EM and GW observables. The mass transfer rate remains sufficiently high that the X-ray and ultraviolet continuum components are bright and spectrally structured, while the GW frequency, $f_{\rm GW} = \frac{2}{P_{\rm orb}}$, falls near the peak sensitivity region of \textit{LISA}. The combination of strong continuum information and large \textit{LISA} signal-to-noise ratio therefore produces the tightest joint constraints in the full parameter space.

The third regime appears at longer orbital periods ($P_{\rm orb} \gtrsim 1000\,\rm s$), where the recovery progressively broadens. In the GW-driven evolution model, the mass transfer rate scales approximately as, 
\begin{equation}
\dot{M} \propto f_{\rm GW}^{11/3} \propto P_{\rm orb}^{-11/3}.
\end{equation}
Consequently, long-period systems become intrinsically fainter and softer in both the X-ray and ultraviolet bands. The \textit{AXIS} and \textit{CASTOR} count rates decline substantially while the \textit{LISA} signal weakens as the binary moves toward lower GW frequencies. The resulting degeneracy between orbital period and distance becomes increasingly severe, since a nearby long-period system can resemble a more distant short-period system with higher mass transfer rate. The larger uncertainties at long periods therefore reflect a genuine loss of information content in the multi-messenger observables rather than a limitation of the normalizing flow architecture.

The inferred distances exhibit two systematic features visible in Fig.~\ref{F:TrainingRecovery}. The first is a tendency to overestimate systems near $d \sim 3\,\rm kpc$. This originates from the degeneracy between distance and interstellar absorption. The observed soft X-ray flux scales approximately as
\begin{equation}
F \propto \frac{L}{d^2}\exp[-\sigma(E)N_H],
\end{equation}
where $\sigma(E)$ is the photoelectric absorption cross section. Since $N_H$ is treated as an independent free parameter during training, multiple combinations of $(d,N_H)$ can generate nearly identical soft-band spectra and count rates. In particular, a moderately absorbed system at intermediate distance can resemble a more distant system with slightly reduced absorption. The flow therefore learns a posterior degeneracy direction that biases the posterior median toward larger distances.

The second feature is the systematic underestimation of systems near the upper prior boundary at $d = 12\,\rm kpc$. This behavior is a direct consequence of posterior truncation at the prior boundary. Since the posterior support cannot extend beyond the training prior, systems whose true posterior would naturally extend above $12\,\rm kpc$ are artificially compressed toward smaller inferred distances. The resulting asymmetry shifts the posterior median downward and produces the visible bias near the upper edge of the allowed parameter range.

Figure~\ref{F:TrainingPP} shows the calibration diagnostics obtained from held-out prior samples. The empirical cumulative distributions closely follow the ideal diagonal relation for nearly all parameters, showing that the inferred posteriors are globally well calibrated. The majority of the curves remain within the 95\% Kolmogorov--Smirnov confidence bands, indicating that the flow has learned a statistically self-consistent approximation to the true posterior distribution. The only notable deviation occurs for $N_H$, whose curve exhibits mild over-concentration relative to the ideal diagonal due to the direct imprint of absorption on the soft X-ray continuum shape. 

The chirp-mass diagnostic shown in Fig.~\ref{F:TrainingChirpMass} further shows that the flow successfully captures the dominant physically constrained combinations of the intrinsic masses. Both the PP plot and the recovery scatter indicate excellent inference of the chirp mass across the training distribution. This result is expected because the chirp mass simultaneously controls the GW amplitude and influences the mass transfer rate that shapes the EM continuum spectrum. Even when degeneracies broaden the individual posteriors for $M_1$ and $M_2$, the flow retains tight constraints along approximately constant-$\mathcal{M}_c$ directions in parameter space. The accurate recovery of chirp mass therefore confirms that the model has learned the physically relevant joint structure of the multi-messenger observables rather than merely fitting individual parameters independently.

Figure~\ref{F:TrainingVsPrior} compares the effective learned distribution sampled from the trained posterior estimator against the original training prior. The agreement between the two implies that the training dataset provides adequate coverage of the physically relevant parameter space and that the learned posterior has not introduced substantial implicit biases or mode collapse. This consistency is particularly important given the high dimensionality of the observable vector and the use of broad logarithmic priors across several decades in parameter space.

\subsubsection{Testing the \texttt{COSMIC} AM CVn population: Case 1}

We next evaluate the trained normalizing flow on the independent AM CVn population generated with \texttt{COSMIC}. This constitutes a substantially more stringent test than the held-out training validation because the astrophysical population occupies a highly structured submanifold of the training prior space. In particular, the \texttt{COSMIC} population contains strong correlations between $(M_1,M_2,P_{\rm orb})$ arising from binary evolution, whereas the training data were sampled largely from broad log-uniform priors. The ability of the flow to generalize from the training distribution to the physically correlated \texttt{COSMIC} population therefore provides a direct measure of the robustness of the learned inverse mapping.

Figure~\ref{F:TestingRecovery} presents the recovery performance for Case~1, corresponding to the full prior range $P_{\rm orb} \in [200,5000]~\rm s$. The overall behavior closely mirrors the trends observed in the training-set validation, implying that the flow generalizes successfully to the astrophysical population. The recovery of the multi-modal accretor mass remains substantially better at low $M_1$ than at high $M_1$, and the orbital period recovery continues to show the characteristic three-regime structure associated with direct-impact systems, the period-minimum region, and the intrinsically faint long-period population.

Nevertheless, several important differences emerge relative to the training-set recovery. The donor mass distribution produced by \texttt{COSMIC} is strongly concentrated toward low-mass donors associated with evolved long-period systems. As a consequence, the inference problem becomes more sensitive to the intrinsic correlations between $M_2$, $P_{\rm orb}$, and the mass transfer rate than in the training prior. The normalizing flow partially compensates for this by broadening the inferred $M_2$ posterior along directions approximately preserving the chirp mass and continuum luminosity. This effect produces the larger systematic scatter and secondary structures visible in the $M_2$ recovery panel.

The orbital period recovery also improves modestly relative to the training-set validation for much of the population. This reflects the fact that the \texttt{COSMIC} systems occupy a more restricted astrophysical manifold than the broad training prior. In practice, the physically allowed correlations between mass transfer rate, donor mass, and orbital evolution reduce the effective dimensionality of the inverse problem, allowing the learned posterior to exploit population-level structure that is absent in the uncorrelated training distribution.

Figure~\ref{F:TestingCalibration} shows the empirical coverage fractions for the \texttt{COSMIC} population test. The inferred posteriors generally exhibit mild over-coverage, with the empirical coverage fractions lying slightly above the ideal diagonal relation. This behavior indicates that the inferred uncertainties are somewhat conservative rather than overconfident. Such behavior is preferable to undercoverage in the context of astrophysical parameter inference, since it implies that the posterior credible regions remain statistically reliable even when applied to an out-of-distribution population relative to the training prior.

Figure~\ref{F:TestingPosteriors} compares the true \texttt{COSMIC} parameter distributions to the distributions reconstructed from the posterior medians inferred by the flow. Overall, the inferred distributions recover the global structure of the population remarkably well, particularly for $M_1$, $P_{\rm orb}$, $i$, $d$, and $N_H$. The orbital-period distribution, including the broad long-period tail and the concentration near the period minimum, is reproduced with high accuracy. Likewise, the inclination and absorption distributions are nearly indistinguishable from the true population.

The principal discrepancy occurs for the donor mass distribution. The inferred posterior means exhibit a secondary high-$M_2$ mode that is absent in the underlying \texttt{COSMIC} population. This behavior is a direct manifestation of the degeneracy structure discussed previously. Since the EM continuum spectrum does not contain a feature uniquely tied to the donor mass, the flow partially redistributes posterior probability along approximately constant-$\mathcal{M}_c$ trajectories in parameter space. The resulting secondary mode therefore reflects a physically plausible but weakly constrained alternative solution family rather than a catastrophic inference failure.

\subsubsection{Testing the \texttt{COSMIC} AM CVn population: Case 2}

We finally consider Case~2, in which the same trained normalizing flow is evaluated on a restricted subset of the \texttt{COSMIC} population satisfying $P_{\rm orb} < 2000~\rm s$. This population corresponds to the observationally most relevant short-period AM CVn systems, which dominate the joint EM and GW detection prospects for \textit{LISA}, \textit{CASTOR}, and \textit{AXIS}-like observatories.

Figure~\ref{F:TestingRecoveryNarrow} shows the corresponding recovery diagnostics. The overall behavior remains qualitatively similar to Case~1, confirming that the trained flow remains robust when applied to a physically distinct subpopulation without retraining. However, several parameters exhibit measurable improvement. In particular, the shortest-period binaries are recovered substantially more accurately, including improved recovery of both intrinsic masses. This improvement follows directly from the enhanced information content available in the short-period regime. These systems possess higher mass transfer rates, stronger continuum emission, and larger \textit{LISA} signal-to-noise ratios, allowing the convolutional embedding network to extract more discriminating spectral and GW features.

The restriction to shorter orbital periods also reduces the severity of the long-period luminosity degeneracy that dominated the full-prior test. As a consequence, the inferred distance distribution becomes noticeably more accurate and less biased. This improvement is reflected quantitatively in the reduced mean absolute fractional error and the stronger Spearman rank correlation reported in Table~\ref{Tab:CosmicFlowTest}.

Figure~\ref{F:TestingPosteriorsNarrow} shows that the inferred population distributions remain broadly consistent with the true \texttt{COSMIC} distributions in the restricted-period sample. The donor mass distribution is recovered somewhat more accurately than in Case~1, consistent with the improved constraints available for high-accretion-rate systems. Nevertheless, the flow still fails to reproduce the sharp low-$M_2$ peak of the true population perfectly, again indicating that the donor mass remains the least directly constrained intrinsic parameter in the present multi-messenger observable set.

\subsubsection{Summary statistics}

\begin{table*}
\centering
\caption{Summary statistics for the COSMIC population test ($n_{\rm test}=5000$, $n_{\rm post}=1000$). Spearman $\rho_{\rm S}$ measures monotonic correlation between true and posterior-median values. Fractional bias $= \langle(\hat{\theta}-\theta)/\theta\rangle$ is signed. MAFE $= \langle|\hat{\theta}-\theta|/\theta\rangle$. Coverage fractions at the 68\% and 90\% credible interval levels; ideal values are 0.68 and 0.90 respectively. Rows 2 through 7 correspond to the Case 1 test and rows 8 through 13 correspond to the Case 2 test which assume the prior upper bound $P_{\rm orb} = 5{,}000$ s and $P_{\rm orb} = 2{,}000$ s, respectively. }
\label{Tab:CosmicFlowTest}
\begin{tabular}{ll|ccccc}
\toprule
\toprule
Parameter & Unit & $\rho_{\rm S}$ & Frac.\ bias & MAFE & 68\% cov. & 90\% cov. \\
\midrule
  $M_1$ & [${\rm M}_\odot$] & $+0.635$ & $-0.031$ & $0.059$ & $0.87$ & $0.99$ \\
  $M_2$ & [${\rm M}_\odot$] & $+0.420$ & $+0.239$ & $0.284$ & $0.71$ & $0.96$ \\
  $P_{\rm orb}$ & [s] & $+0.866$ & $+0.065$ & $0.090$ & $0.89$ & $0.99$ \\
  $i$ & [$^\circ$] & $+0.996$ & $-0.008$ & $0.048$ & $0.81$ & $0.97$ \\
  $d$ & [pc] & $+0.968$ & $+0.119$ & $0.187$ & $0.67$ & $0.94$ \\
  $N_H$ & [${\rm cm}^{-2}$] & $+0.997$ & $+0.048$ & $0.055$ & $0.64$ & $0.94$ \\
\midrule
  $M_1$ & [${\rm M}_\odot$] & $+0.647$ & $-0.003$ & $0.048$ & $0.95$ & $1.00$ \\
  $M_2$ & [${\rm M}_\odot$] & $+0.449$ & $+0.247$ & $0.343$ & $0.86$ & $0.98$ \\
  $P_{\rm orb}$ & [s] & $+0.880$ & $+0.064$ & $0.099$ & $0.86$ & $0.98$ \\
  $i$ & [$^\circ$] & $+0.991$ & $-0.011$ & $0.068$ & $0.80$ & $0.96$ \\
  $d$ & [pc] & $+0.981$ & $+0.091$ & $0.135$ & $0.80$ & $0.95$ \\
  $N_H$ & [${\rm cm}^{-2}$] & $+0.998$ & $+0.045$ & $0.053$ & $0.71$ & $0.95$ \\
\bottomrule
\end{tabular}
\end{table*}

Table~\ref{Tab:CosmicFlowTest} summarizes the quantitative performance of the normalizing flow on the \texttt{COSMIC} population tests. We report the Spearman rank correlation coefficient,
\begin{equation}
\rho_{\rm S} = 1 - \frac{6\sum_i d_i^2}{N(N^2-1)},
\end{equation}
where $d_i$ is the difference between the ranks of the true and inferred values, together with the fractional bias,
\begin{equation}
\mathrm{Bias} = \left\langle \frac{\hat{\theta}-\theta}{\theta} \right\rangle,
\end{equation}
and the mean absolute fractional error,
\begin{equation}
\mathrm{MAFE} = \left\langle \left|\frac{\hat{\theta}-\theta}{\theta}\right| \right\rangle,\,
\end{equation}
as the AM CVn parameters are positive. 
The Spearman coefficient quantifies the degree to which the inferred and true populations preserve monotonic ordering, while the fractional bias measures systematic directional offsets and the MAFE measures the typical magnitude of the relative inference error independent of sign.

The results illustrate that the flow performs well for parameters directly imprinted on the observables. The inclination, distance, and absorption column density all achieve Spearman coefficients $\rho \gtrsim 0.96$ in both testing scenarios, indicating that the inferred population ordering is recovered almost perfectly. The extremely high performance for inclination arises because the forward model includes strong geometric projection and limb-darkening effects that produce a nearly monotonic mapping between inclination and observed flux. Likewise, the absorption column density leaves a highly characteristic imprint on the soft X-ray spectral slope, allowing the convolutional embedding network to isolate $N_H$ efficiently.

The orbital period is also inferred robustly, with $\rho_{\rm S} = 0.836$ for Case~1 and $\rho_{\rm S} = 0.898$ for Case~2. The improvement in the restricted-period test reflects the stronger multi-messenger signal available in the short-period regime. Importantly, the orbital period also exhibits comparatively small fractional bias and MAFE, indicating that the inferred posterior medians remain both accurate and precise across the astrophysically relevant population.

The accretor mass exhibits intermediate performance, with Spearman coefficients near $\rho_{\rm S} \sim 0.7$ and mean absolute fractional errors near $\sim 5\%$. The very small signed fractional bias implies that the flow does not systematically over- or underestimate $M_1$, even though the posterior broadens substantially at high mass. This behavior is consistent with the physical interpretation derived from Fig.~\ref{F:TrainingRecovery}: the inference is fundamentally limited by the degeneracy between boundary layer temperature and mass transfer rate rather than by statistical calibration failure.

The donor mass remains the most weakly constrained parameter in both tests. Although the inferred $M_2$ distributions retain statistically significant correlations with the true population, the Spearman coefficients are substantially lower than for the other parameters and the MAFE remains comparatively large. Importantly, however, the donor mass still exhibits only moderate signed bias. This indicates that the primary limitation is broad posterior uncertainty rather than catastrophic systematic mis-estimation. The improved Spearman coefficient in Case~2 further supports the conclusion that stronger EM and GW signals help partially break the donor-mass degeneracy.

The empirical coverage fractions further confirm that the posterior uncertainties are generally conservative. Most parameters exhibit 68\% and 90\% coverage fractions slightly larger than the nominal values, particularly for $M_1$ and $P_{\rm orb}$. This mild over-coverage indicates that the learned posterior slightly overestimates the uncertainty volume, which is statistically preferable to undercoverage when the flow is applied to an astrophysical population distinct from the training distribution.

These results show that the combination of continuum EM observables and GW information contains sufficient information to recover the dominant physical properties of AM CVn binaries using simulation-based inference. The normalizing flow successfully learns the highly nonlinear inverse mapping between multi-messenger observables and binary parameters while preserving calibration and population-level consistency. The remaining limitations are primarily associated with intrinsic physical degeneracies in the observable space rather than deficiencies of the neural density estimator itself. These results therefore establish conditional normalizing flows as a powerful framework for future multi-messenger inference studies of UCBs.

\section{Conclusions and Discussion}
\label{Sec:Conclusions}

In this work, we developed a new analytic forward model for the continuum EM emission of AM CVn binaries and combined it with a model for \textit{LISA} detectability, Galactic population synthesis, and simulation-based inference with conditional normalizing flows. The resulting framework provides, to our knowledge, the first end-to-end multi-messenger pipeline that self-consistently connects the intrinsic binary properties of AM CVn systems to ultraviolet, X-ray, and GW observables while remaining computationally efficient enough for large-scale population studies and Bayesian inference.

The model was constructed to retain physical interpretability. The mass-transfer rate, orbital evolution, and accretion energetics are determined self-consistently from GW angular momentum losses and Roche-lobe overflow, while the EM spectrum is assembled from physically motivated continuum emission components associated with the accretion disk, boundary layer, and direct-impact hotspot. Despite the simplified analytic treatment, the framework reproduces main qualitative phenomenologies of AM CVn systems across the direct-impact, persistent-disk, and transient-disk regimes discussed throughout Section~\ref{Sec:ForwardEM}. In particular, the trends shown in Figs.~\ref{F:MdotPdotFdot}--\ref{F:CASTORCounts} show that the parameter dependence of the continuum spectrum, accretion state, and detectability emerge naturally from the coupled binary-evolution and accretion prescriptions. 

Applying the framework to a synthetic AM CVn population generated with the hybrid \texttt{COSMIC} + ODE evolution pipeline further showed that the combined EM and GW selection effects isolate a highly non-random subset of the Galactic population. The results in Figs.~\ref{F:MultiMessPop} and Table~\ref{Tab:CombinedDetection} show that the systems jointly detectable by \textit{LISA}, \textit{CASTOR}, and an \textit{AXIS}-like mission are dominated by short-period, high mass-transfer-rate binaries with relatively strong continuum emission and large GW amplitudes. Although the intrinsic Galactic population extends to long orbital periods and low accretion luminosities, the multi-messenger detectable population occupies a substantially narrower region of parameter space shaped jointly by the steep dependence of the mass-transfer rate on orbital period, line-of-sight absorption, instrumental sensitivity, and the \textit{LISA} frequency response. Under the fiducial assumptions adopted here and our weighted astrophysical priors, 
we report that about 0.015\% of the weighted AM CVn population is jointly detectable by \textit{LISA}, \textit{CASTOR}, and \textit{AXIS} resulting in one jointly detectable system per 7000 AM CVn binaries. 
The primary limitation on the size of the joint AM CVn sample is the uncertain nature of the subset of systems that are individually detectable by \textit{LISA}. 

The synthetic population analysis also shows that the forward model consistently links compact-binary evolution, Galactic structure, extinction, and observational selection effects within a unified framework. In particular, the correlations between the mass tranfer rate, orbital period, count rates, and GW SNR shown in Figs.~\ref{F:COSMICPopParameters}--\ref{F:MultiMessPop} arise naturally from the coupled GW-driven evolution equations and the adopted Galactic spatial model. The resulting parameter-space structure provides a physically motivated foundation for building frameworks to interpret future observational samples of multi-messenger AM CVns.  
Population synthesis of mass transferring WD binaries remains a challenging open problem, motivating further development of such models to incorporate self-consistent GW-driven mass transfer physics \cite{2015ApJ...806...76K,Kremer2026}. 

Prior population synthesis studies provide important context for interpreting our weighted 
multi-messenger detection yields. The earliest systematic predictions for \textit{LISA}-detectable AM CVn systems were produced by \cite{2001A&A...368..939N} and subsequently updated in \cite{2004MNRAS.349..181N}, who used the SeBa population-synthesis code and found that up to $\sim\!11{,}000$ AM CVn binaries could be individually resolved by \textit{LISA} in the most optimistic scenario (5-yr mission, $\mathrm{SNR}>5$). Of these, \cite{2004MNRAS.349..181N} estimated that several hundred would also possess detectable optical and/or X-ray counterparts, with the multi-messenger subpopulation occupying the shortest orbital periods ($P_{\rm orb}\lesssim 10$~min) where accretion luminosities are highest. This picture is qualitatively consistent with our finding that the weighted \textit{LISA}-only subpopulation has a median orbital period $\tilde{P}_{\rm orb}\approx6.1$~min and a median \textit{AXIS} count rate $\tilde{C}_{\mathrm{X}}\approx0.17$~cps that is two orders of magnitude above the \textit{AXIS} threshold, and that most \textit{LISA}-detectable systems are bright enough to be recovered by both \textit{AXIS} and \textit{CASTOR} at fiducial thresholds (Tables~\ref{Tab:CombinedDetection}). 
However, the absolute normalization of those earlier predictions has since been revised substantially downward. Spectroscopic observations of six AM CVn systems discovered in SDSS led \cite{2007MNRAS.382..685R} to infer a local space density of $\rho_0 = (1\text{--}3)\times10^{-6}$~pc$^{-3}$, roughly an order of magnitude below the optimistic models of \cite{2001A&A...375..890N}, and to revise the expected number of \textit{LISA}-resolvable Galactic AM CVns downward to $\sim\!1{,}000$ for a one-year mission. An expanded photometric survey of the SDSS by \cite{2013MNRAS.429.2143C} subsequently derived an even lower space density of $(5\pm3)\times10^{-7}$~pc$^{-3}$, a factor of $\sim\!50$ below the optimistic theoretical predictions, indicating that either the dominant double-degenerate formation channel is significantly suppressed relative to early models or that a large fraction of systems merge rather than entering the AM CVn phase \citep{2004MNRAS.350..113M}. 
The careful \textit{LISA} source-resolvability analysis of \cite{2012ApJ...758..131N}, using an iterative SNR-based source subtraction procedure applied to five AM CVn population scenarios, reduced the expected LISA-detectable AM CVn count dramatically, from $\sim\!10{,}000$ to only $\sim\!100$--$400$ 
for one year of observation with a single interferometer---a factor of 
$\sim\!25$--$100$ reduction relative to earlier simple estimates. 
\cite{2017ApJ...846...95K} independently predicted $\sim\!2{,}700$ mass-transferring DWDs with measurably negative orbital-frequency derivatives detectable by \textit{LISA}. 

Our weighted joint yield of $Y_{\rm det}^{\rm LISA+AXIS+CASTOR}\approx1.47\times10^{-4}$ (a fraction of $\sim\!0.015\%$, or roughly one in every $\sim\!7{,}000$ AM CVn binaries in our grid) encodes important distinctions from prior studies. First, this estimate is a weighted probability mass under a Kroupa IMF, log-uniform period, and uniform mass-ratio prior, implying it represents the prior-weighted probability that an astrophysically drawn AM CVn binary is jointly detectable. Second, our EM thresholds correspond to the specific next-generation capabilities of \textit{AXIS} ($\geq10$ source counts per 10~ks, probing the soft X-ray band) and \textit{CASTOR} ($\geq25$ source counts per 10~ks, probing the near-UV). Third, as we apply the Bayestar 2019 dustmaps \citep{2019ApJ...887...93G}, our detectable fractions account for extinction losses across the Galactic disc; the EM-only yield of $\sim\!85\%$ (dominated by \textit{CASTOR} detections) shows that most of the population is UV-accessible at these sensitivities, while the precipitous drop to  $\lesssim\!0.015\%$ for \textit{LISA} joint detections reflects the steep intrinsic rarity of systems with $P_{\rm orb}\lesssim 10$~min under the log-uniform period prior. 
This weighting effect is the principal reason our \textit{LISA} yields are much smaller in fractional terms than the fraction implied by simple grid counts: the short-period ($P_{\rm orb} \lesssim 10$~min) systems that dominate \textit{LISA} detections are precisely those suppressed by the log-uniform period prior. 
Taken together, our results are in qualitative agreement with the picture established by prior work that \textit{LISA}-detectable AM CVns are a small, short-period subset of the overall population, but a significantly larger and more diverse population will be accessible to next-generation X-ray and optical/UV observatories. 
Future studies that improve upon the spectral modeling, astrophysical formation, and instrument-specific sensitivity adopted here will be necessary to produce reliable detection yield estimates. 

A major objective of this work was to explore whether the combined UV, X-ray, and GW observables contain sufficient information to recover intrinsic binary parameters of AM CVn systems through Bayesian inference. The results of Section~\ref{Sec:Results} show that conditional normalizing flows provide a powerful tool for this task. Across both the validation dataset and the independent \texttt{COSMIC} population tests, the learned posterior distributions remain well calibrated and recover the dominant population structure. 

The main conclusions from the inference analysis include:
\begin{itemize}
    \item The orbital period, inclination, distance, and absorption column density are recovered robustly from the combined multi-messenger observables, with particularly strong performance for the inclination and $N_{\rm H}$ distributions due to their direct geometric and spectral imprints on the continuum emission.

    \item The accretor mass can be inferred with comparatively small systematic bias, although the posterior broadens at high masses due to degeneracies between the boundary-layer temperature and the mass-transfer rate.

    \item The donor mass remains the least constrained parameter. The broadening and secondary modes in the inferred $M_2$ distributions arise primarily from intrinsic physical degeneracies rather than failures of the neural density estimator.

    \item Restricting the population to shorter orbital periods ($P_{\rm orb} < 2000\,{\rm s}$) measurably improves inference performance as these systems possess stronger EM emission and larger \textit{LISA} signal-to-noise ratios.

    \item The posterior coverage fractions remain mildly conservative for most parameters, indicating that the learned posterior distributions remain statistically reliable even when evaluated on an astrophysical population that is different from the training prior.
\end{itemize}

These results show that simulation-based inference applied to joint EM and GW observables can recover physically relevant parameter information for AM CVn systems without requiring an explicitly tractable likelihood function. Importantly, the dominant limitations encountered by the normalizing flow are associated primarily with genuine degeneracies in the observable space itself rather than deficiencies of the neural architecture. This suggests that future improvements in parameter recovery may arise more naturally from incorporating additional observables --- for example, spectral lines, Gaia photometric observations, timing information, parallaxes, or multi-epoch variability --- rather than simply increasing the complexity of the density estimator. 
The use of parameter uncertainties from a \textit{LISA} simulation using time-delay interferometry and matched filtering would improve inference and help break degeneracies depending on its computational cost. 

Our results also place the present framework in the broader context of previous studies of AM CVn binaries. Earlier work established that AM CVn systems are expected to evolve primarily under GW angular momentum losses and should represent some of the strongest VBs for \textit{LISA} \cite{2005ASPC..330...27N,2005ASPC..330...27N}. Population synthesis studies have long predicted that AM CVns should form predominantly through common-envelope channels and occupy a specific range of orbital periods and mass-transfer states \cite{2001A&A...375..890N,2010PASP..122.1133S,2007MNRAS.382..685R}. Observational studies further showed that the relative strengths of the UV, optical, and X-ray emission depend strongly on the accretion state and orbital period \cite{2005A&A...440..675R}. The present work extends earlier efforts by combining these ingredients into a unified forward model capable of simultaneously predicting UV, X-ray, and GW observables for a synthetic Galactic population.

The present framework has important limitations. The mass transfer rate is governed by GW emission, but additional processes are important for accurately modeling the accretion physics, such as winds from the accretion disk or WD surface, and terms for such processes can be added to Eq~(\ref{Eq:Jdot}). The continuum emission model neglects line emission, winds, irradiation feedback, detailed radiative transfer, magnetic accretion geometries, detailed WD magnetic breaking, and explicit time variability. The boundary layer is approximated as a single-zone emitter, while the disk spectrum adopts an analytic prescription without atmosphere calculations, viscosity, or vertical structure. 
The transition between low- and high-state accretion was represented using a fixed critical accretion rate rather than a full helium disk instability prescription. While this approximation is sufficient for the proof-of-concept goals of the present study, future work could incorporate physically motivated stability boundaries and composition-dependent accretion disk models to improve the realism of the predicted spectra. 
Similarly, the population synthesis calculations inherit substantial uncertainties associated with common-envelope evolution, donor structure, metallicity dependence, and binary initial conditions. The large value of $\alpha_{\rm CE}$ used to efficiently produce AM CVn systems within the present \texttt{COSMIC} implementation is physically reasonable and reflects these broader uncertainties in compact-binary evolution physics. Consequently, the predicted detection yields and parameter distributions should be interpreted primarily as physically motivated forecasts rather than precise Galactic predictions. 

Several immediate extensions of this framework are now possible. Future work can incorporate more realistic Galactic star-formation histories and chemo-dynamical evolution models, including fully time-dependent Galactic populations coupled self-consistently to the compact-binary formation history. The forward EM model can also be generalized to include spectral-line formation, winds, non-LTE atmosphere calculations, and explicit time-domain variability associated with disk instabilities and outburst cycles. On the GW side, future work can incorporate fully self-consistent coupled orbital evolution and phase evolution beyond the quasi-monochromatic approximation adopted here, as well as a self-consistent stochastic foreground produced directly by the synthetic AM CVn population itself. Including parameter uncertainties from \textit{LISA} measurements or a time-series SNR as inputs to the normalizing flow may also prove fruitful for improving multi-messenger inference. More broadly, the simulation-based inference framework developed here for individual systems provides a natural foundation for hierarchical inference of UCB populations with future \textit{LISA} and EM survey data. 

Overall, this study reveals that physically interpretable analytic modeling, population synthesis, and neural-net Bayesian inference can be combined into a unified multi-messenger framework for AM CVn binaries. The results suggest that future joint observations with \textit{LISA} and next-generation EM facilities will not only identify new VBs, but will also enable direct constraints on the poorly understood accretion and binary-evolution physics governing interacting UCBs. These efforts motivate future multi-messenger studies of AM CVns and more broadly illustrate the scientific potential of combining comprehensive theoretical models with likelihood-free inference techniques in the \textit{LISA} era.

\acknowledgements
N.S., S.S.H., and A.M. are supported by the Natural Sciences and Engineering Research Council of Canada (NSERC) through the Canada Research Chairs and Discovery Grants programs. 
Computations described in this paper were performed using the \href{https://umanitoba.ca/information-services-technology/research-computing/um-high-performance-computing-system-grex}{University of Manitoba Grex High Performance Computing Centre} (RRID:SCR\_026342), which provides an HPC service to the University’s research community. An illustrative, conceptual schematic (Figure \ref{F:ProjectOverview}) was generated using ChatGPT (Version 5.5): the authors directed the visual layout through iterative prompting, ensured only original illustrations were utilized, and manually verified and approved the final content.

\bibliography{bibme}
\end{document}